\documentclass[showpacs,aps,prb,twocolumn,groupedaddress,superscriptaddress]{revtex4-1}

\usepackage{amssymb}
\usepackage{graphicx}
\usepackage{amsmath,color}
\usepackage{ulem,accents,verbatim}
\renewcommand{\emph}[1]{\textit{#1}}
\newcommand{\gv}[1]{{\underline{#1}}}
\newcommand{\s}{\sigma}
\renewcommand{\d}{\delta}
\renewcommand{\o}{\omega}
\renewcommand{\k}{\gv{k}}
\newcommand{\q}{\gv{q} }

\newcommand{\op}[1]{\hat{#1}}
\renewcommand{\c}[2]{\op{#1}^{\phantom{\dagger}}_{#2}}
\newcommand{\ck}[2]{\op{#1}^\dagger_{#2}}

\newcommand{\ew}[1]{\langle #1\rangle}
\newcommand{\upa}{\uparrow}
\newcommand{\doa}{\downarrow}

\newcommand{\chimag}{\chi_{\mathrm{mag}}}

\newcommand{\gm}[1]{\uuline{#1}{}}

\newcommand{\twoPart}[4]{\gm{#1}(#2\,|\,#3,#4)}
\newcommand{\twoPartO}[5]{\gm{#1}^{#2}(#3\,|\,#4,#5)}

\newcommand{\onePart}[3]{\gm{#1}(#2\,|\,#3)}
\newcommand{\onePartAmp}[3]{\gm{#1}^{\mathrm{amp}}(#2\,|\,#3)}

\newcommand{\Pic}[3]{\gm{\Pi}(#1\,|\,#2,#3)}
\newcommand{\PicAmp}[3]{\gm{\Pi}^{\mathrm{amp}}(#1\,|\,#2,#3)}
\newcommand{\Chi}[3]{\gm{\chi}(#1\,|\,#2,#3)}

\newcommand{\um}[1]{\underaccent{\approx}{#1}}

\newcommand{\T}{\mathcal{T}}

\begin{document}

\title{ %Dynamics screening of local magnetic moment: 
  Itinerant and local-moment magnetism in strongly correlated electron systems}

\author{Sebastian Schmitt}
\affiliation{Lehrstuhl f\"{u}r Theoretische Physik II, Technische Universit\"{a}t Dortmund,
Otto-Hahn Str.\ 4, D-44227 Dortmund, Germany}
\author{Norbert Grewe}%\email[]{Your e-mail address}
\affiliation{Institut f\"{u}r Festk\"{o}rperphysik, Technische Universit\"{a}t Darmstadt,
Hochschulstr.\ 6, D-64289 Darmstadt, Germany}
\author{Torben Jabben}
\affiliation{Institut f\"{u}r Festk\"{o}rperphysik, Technische Universit\"{a}t Darmstadt,
Hochschulstr.\ 6, D-64289 Darmstadt, Germany}

\date{\today}

\begin{abstract}
  Detailed analysis of the magnetic properties of the Hubbard model within dynamical
  mean-field theory (DMFT) is presented.  
  Using a RPA-like decoupling of two-particle propagators we derive an universal 
  form for susceptibilities, which captures essential aspects of localized and itinerant pictures. This expression
  is shown to be quantitatively valid whenever 
  long-range coherence of particle-hole excitations can be neglected, as is the case in 
  large parts of the phase diagram where antiferromagnetism is dominant.
  The applicability of an interpretation in terms of  the two archetypical 
  pictures of magnetism is investigated for the Hubbard model on a body-centered cubic lattice
  with additional next-nearest neighbor hopping  $t'$. 
  For large values of the Coulomb interaction, local-moment magnetism is found to
  be dominant, while for weakly interacting band electrons
  itinerant quasiparticle magnetism prevails.
  In the intermediate regime and for finite $t'$ an  re-entrant behavior is discovered, where 
  antiferromagnetism  only exists in a finite temperature interval.
\end{abstract}

% 71.27.+a  Strongly correlated electron systems; heavy fermions
% 71.10.Fd  Lattice fermion models (Hubbard model, etc.)
% 71.10.-w  Theories and models of many-electron systems 
% 72.10.Fk  Scattering by point defects, dislocations, surfaces, and other imperfections (including Kondo effect)
% 71.45.Gm  Exchange, correlation, dielectric and magnetic response functions, plasmons
% 75.10.-b  General theory and models of magnetic ordering (see also 05.50.+q Lattice theory and statistics)
% 75.20.Hr  Local moment in compounds and alloys; Kondo effect, valence fluctuations, heavy fermions (for Kondo effect and scattering mechanisms in electronic conduction, see 72.15.Qm and 72.10.Fk)
% 75.10.Lp  Band and itinerant models
% 75.30.Kz  Magnetic phase boundaries (including classical and quantum magnetic transitions, metamagnetism, etc.) (for ferroelectric phase transitions, see 77.80.B-; for superconductivity phase diagrams, see 74.25.Dw)


\maketitle

% Check first occurrence of abbrevs: QMC ENCA AFM FM IC DMFT BCC

\section{Introduction}
\label{sec:introduction}

The magnetic properties of solids are typically described
in terms of two archetypical and opposing viewpoints. On the one hand, 
the picture of weakly interacting itinerant electron magnetism is usually 
implemented for metallic systems. On the other hand, 
for small overlap between sites or for strong local interactions the 
valence electrons can form localized moments and behave effectively like spin 
degrees of freedom of a Heisenberg model (for introductory texts, see, for example, 
Refs.~\onlinecite{fazekas:LectureNotes99} and
\onlinecite{yosidaMagnetism}).

%\cite{ramakrishnanSCES08}
In strongly correlated electron systems, such as cuprate high-temperature 
superconductors\cite{plakidaHighTCBook10} or heavy fermion systems,\cite{greweSteglichHF91,*colemanHeavyFermionMag07} 
such a clear distinction is often obscured since aspects of both pictures appear.
Due to the strong Coulomb interaction the electrons are rather localized  
furnishing large magnetic moments. However, at low temperatures usually a bandstructure
of heavy but itinerant quasiparticles around the Fermi level forms,
giving rise to the screening of local moments.
Central to the magnetic properties in these systems is the 
competition between quasiparticle-band formation, possibly spin-polarized, and ordering 
tendencies of large local moments. 

This competition yields especially interesting and novel 
physics in the case of frustrated systems.\cite{lacroixFrust2010}
In the recently studied  frustrated heavy fermion systems, 
such as LiV$_2$O$_4$, frustrated local moments dominate at elevated temperatures, while 
% do not  order magnetically due to the large geometric frustration, but 
at low temperatures
strongly correlated quasiparticles emerge 
(see, for example, Ref.~\onlinecite{hopkinsonLiV2O402,*joenssonLiV2O407,*aritaLiV2O407}). 
Similarly, a competition between frustrated  magnetic moments 
and a low temperature Fermi liquid   has been proposed as an explanation for the 
re-entrant Mott-transition found in the highly frustrated organic compound 
$\kappa$-(ET)$_2$Cu[N(CN)$_2$]Cl.\cite{kagawaRe-entrantOrganic04,*ohashiTriangularHM08,*ohashiTriangularHM08-2} 
In the context of the  cuprate high-temperature  superconductors there is a longstanding 
question whether  the magnetic properties are best described in the local or itinerant 
picture, see, for example, Ref.~\onlinecite{vojtaMobileItinernantCuprate09}.
Due to the proximity of these systems to the Mott-insulating phase, the localized 
picture is often proposed, whereas recent studies also revealed the presence of 
itinerant spin-fluctuations throughout the whole 
Brillouin zone.\cite{heItinerSpinCuprates11,*taconParamagnonHTC11}

In the Hubbard model, all these aspects can be captured. 
Weakly interacting band electrons can be studied for small values
of the Coulomb repulsion. With increasing interaction strength 
the system undergoes 
a Mott metal-insulator transition above which it is best characterized 
by an effective Heisenberg spin model.
The interesting region is close to the Mott transition where
large magnetic moments and itinerant quasiparticles compete.
Frustration effects can be studied by considering, for example, the influence of 
a next-nearest neighbor hopping in an otherwise bipartite  lattice.

The magnetic properties of the Hubbard model are very well studied.
Even though the model was introduced in order to describe ferromagnetism (FM)
this phase is restricted to very large interaction strength and 
depends on details of the bandstructure.\cite{vollhardtMetallicFM99,fazekas:LectureNotes99} 
Antiferromagnetism (AFM), however, represents a generic phase of the model and is found in 
large regions of phase space.\cite{fazekas:LectureNotes99} 

A very useful method to study the Hubbard model in the strongly
correlated regime  near the Mott transition is provided by the dynamical
mean field theory\cite{pruschke:dmftNCA_HM95,*georges:dmft96,*georgesDMFTreview04} (DMFT).
With this method one can access the single-particle Green function as well as
two-particle quantities like susceptibilities.

We will consider the Bethe-Salpeter 
equations and derive coupled equations for dynamic lattice
susceptibilities which prove equivalent to
the usual expression known from the DMFT,\cite{jarrellQMCHM92,zlatic2PVertexDMFT90,*jarrell:symmetricPAM95}
but lend themselves more directly to our specific investigation of magnetism.
The ultimate goal of this approach is to analytical continue these equations directly to the real axis 
and obtain integral equations for functions of real frequency arguments.

In this work, however,  we employ an additional decoupling scheme for the internal 
Matsubara summations which allows us to derive 
an universal approximation for the susceptibility, which unifies localized and
itinerant aspects of magnetism.
We discuss the range of validity of this formula 
and demonstrate its accurateness by reproducing the DMFT N\'eel  temperature 
for the Hubbard model on a simple cubic (SC) lattice in three dimensions, $D=3$.
The appeal of this approach lies in its simplicity compared to the usual Matsubara 
approach to two-particle properties.\cite{toschiDGA07,kunesTwoDMFT11}
Only functions of real frequencies enter and the inversions of Matsubara-space 
matrices is avoided.

The competition between local-moment and itinerant 
quasiparticle magnetism is studied for a three-dimensional 
body-centered cubic (BCC)  lattice with an additional next-nearest neighbor hopping
$t'$. 
In situations where quasiparticle  magnetism is dominant,
the tendency toward magnetic order should be strongly suppressed
by a finite next-nearest neighbor hopping, since 
the perfect nesting property of the Fermi surface is removed.
On the other hand, local-moment magnetism is less sensitive
since the geometric frustration induced by the next-nearest neighbor hopping
reduces the effective exchange coupling only quadratically in $t'$.
This behavior is indeed found for weak and strong Coulomb repulsions, respectively.  
In the crossover region for intermediate interactions, an interesting 
re-entrant behavior is found where the effective picture to be used 
depends on the temperature of the system.  
Crucial to the occurrence of this behavior is the presence of the frustrating
next-nearest neighbor hopping $t'$.
The findings of this work are in accord with the
common believe that frustration is capable of producing novel and interesting
phenomena.     

The paper is organized as follows.  After a brief introduction of the model,  Sect.~\ref{sec:model}
sketches the derivation of the Bethe-Salpeter equations for lattice susceptibilities.  
We present two different interpretations of the resulting simple formula for the
magnetic susceptibility, one in terms of local moments and the other 
utilizing itinerant quasiparticles.   
In the first part of Sect.~\ref{sec:results} we investigate the validity of our
approach and  compare the 
N\'eel temperature for the SC lattice  to known results from the literature.    
The second part of this section then focuses on the BCC lattice where 
the magnetic properties are investigated in detail.  
We also include two appendices, where more details on the
technicalities are presented. The focus is laid on the similarities and 
differences of the the single-particle and two-particle set-up.

%%%%%%%%%%%%%%%%%%%%%%%%%%%%%%%%%%%%%%%%%%%%%%%%%%%%%%%%%%%%%%%%%%%%%%%%%%%%%%%%%%%
\section{Model and Method}
\label{sec:model}

In metals the long-ranged Coulomb interaction is usually screened and only short-range
components have to be considered. In order to study the effects of electronic correlations
we consider the  Hubbard model, where only the on-site  
Coulomb repulsion is retained. For the sake of simplicity, we assume ionic $s$-shells 
without orbital degeneracy and with remaining spin-degeneracy only, although the formal 
developments described below can easily be generalized.

The Hamiltonian is given by 
\begin{align}
  \label{eq:hubham}
  \op{H}&=
  \sum_{ij,\s}t_{ij}\ck{c}{i\s}\c{c}{j\s}
  +\sum_{i\,\s}  \epsilon \:\ck{c}{i\s}\c{c}{i\s}
  +U\, \sum_i \op{n}_{i\upa}\op{n}_{i\doa}
  ,
\end{align}
where the operator $\c{c}{i\s}$ ($\ck{c}{i\s}$) annihilates (creates) an electron in a 
localized Wannier orbital at lattice  site $i$ with spin $\s$, $\op{n}_{i\s}=\ck{c}{i\s}\c{c}{i\s}$ 
is the number operator and $\epsilon$ is the ionic level position where the chemical 
potential is already taken into account.  The electrons can transfer
between sites $i$ and $j$ with 
an amplitude $t_{ij}$, which accounts for the itinerancy and band formation
of the electrons. The last term in Eq.~\eqref{eq:hubham} implements 
the local Coulomb repulsion with the interaction matrix element $U$.

The structure of the lattice is encoded into the one-particle hopping amplitude $t_{ij}$,
the Fourier transform of which gives the single-particle dispersion relation
\begin{align}
  t_\k&=\frac1N\sum_{jl}e^{i(\gv{R}_j-\gv{R}_l)\k}\:t_{jl}
  .
\end{align}
The noninteracting density of states (DOS) is entirely  determined by the
dispersion relation
\begin{align}
  \label{eq:nonDOS}
\rho_0(\omega)&=\frac1N\sum_{\k}\delta(\omega-t_\k)
.
\end{align}

Despite its simplicity, the exact solution of the Hubbard Hamiltonian of Eq.~\eqref{eq:hubham} is only
possible in one spatial
dimension\cite{liebWu:HubbardModel,*lieb1DHubbard03} (for a recent book, see, e.g., Ref.~\onlinecite{Essler2005})
and 
in infinite spatial 
dimensions, $D\to\infty$ (with  
an appropriate rescaling the hopping parameters),\cite{metznerDInfty89,*muellerHartmannDInfty89}
by means of the DMFT.
In this work we will utilize DMFT in order to extract an approximate solution for three dimensional systems.

In finite dimensions the major approximation of DMFT is to 
treat all spatially nonlocal correlations in a mean-field manner. 
For the single-particle properties, this implies the assumption of 
a momentum independent interaction self-energy,
$\Sigma^{U}(\k,z)\overset{\mathrm{DMFT}}{\to} \Sigma^{U}(z)$. 
Then, the structure and dimensionality of the lattice enters the lattice Green function 
only via the dispersion relation,
\begin{align}
  \label{eq:gfk}
  G(\k,z)=\frac{1}{z-\Sigma^U(z)-t_\k}
  .
\end{align}

However, the dependence on the (complex) energy variable $z$ is not neglected 
and thus dynamic local correlations are fully retained within  DMFT.
The self-energy and the local Green function,
\begin{align}
  \label{eq:lattsum}
  G(z)&=\frac{1}{N}\sum_{\k}\frac{1}{z-\Sigma^U(z)-t_{\k}}\\
  &=\frac{1}{ z-\Sigma^U(z)-\Gamma(z)}
  \label{eq:localGf}
\end{align}
can be obtained from an effective single-impurity Anderson model (SIAM)
embedded in a self-consistent medium characterized by the hybridization function 
$\Gamma(z)$.
Once the effective SIAM is solved for a given $\Gamma(z)$,
a new guess for the self-energy is obtained by inverting Eq.~\eqref{eq:localGf}, 
which in turn is used to get $G(z)$ via Eq.~\eqref{eq:lattsum}.
Only in this last  step the lattice structure enters.  The self-consistency cycle is
closed by reorganizing Eq.~\eqref{eq:localGf} and extracting a new guess for 
$\Gamma(z)$. 

The nontrivial part in this cycle is the solution of the effective SIAM.
But due to its long history, a multitude of different methods for treating 
this model exist, for example,
exact diagonalization,\cite{Caffarel1994,*Si1994} several variations of quantum
Monte-Carlo schemes,\cite{hirschQMC86,*gullCTQMC11} 
and the numerical renormalization group\cite{wilsonNRG75,*bullaNRGReview08} (NRG).
In this work, we employ the enhanced 
non-crossing approximation\cite{pruschkeENCA89,*schmittSus09,*greweCA108, holmFiniteUNCA89,*keiterENCA90} (ENCA),
which has no adjustable parameters and works directly on the
real frequency axis. 

The tendency toward magnetism
is investigated via the magnetic susceptibility of the paramagnetic phase, which
diverges at
a second order phase transition.
The susceptibility is given by 
the nonlocal time-dependent order-parameter correlation function
\begin{align}
  \label{eq:susDef}
  \chi^{\mathrm{mag}}_{ij}(\tau)&=\ew{\T\big[\hat{M}_i^z(\tau)\,\hat{M}_j^z(0)\big] }
  ,
\end{align}
with the total magnetization operator at lattice site $i$ 
\begin{align}
  \label{eq:Mop}
  \hat M^z_i&=\sum_{a}\gamma_a\: \c{n}{ia}
  .
\end{align}
In general, the index $a$ denotes orbital 
and spin quantum numbers. With the assumption of  $s$-shells one
has especially $a=\sigma$ and $\gamma_\sigma=-\frac{g\mu_B}{2}\sigma$, 
where $g$ is the electron Land\'e factor and $\mu_B$ the Bohr magneton.

Equation \eqref{eq:susDef} is already specified to isotropic situations where 
the susceptibility tensor is diagonal and all diagonal elements 
are equal, i.e.\ $\chi_{ij}^{\mathrm{mag};\alpha\beta}=\chi_{ij}^{\mathrm{mag}}\:\delta_{\alpha\beta}$ 
($\alpha,\beta=\{x,y,z\}$). For a paramagnetic situation,  the transverse susceptibility is 
just given by twice this value, 
$\chi_{ij}^{\mathrm{mag};\perp}=2\chi_{ij}^{\mathrm{mag}}$.

The wave-vector and frequency dependent susceptibility is obtained by the Fourier transform
of Eq.~\eqref{eq:susDef},
\begin{align}
\notag
&  \chi_{\mathrm{mag}}(\q,i\nu_n)=\frac{1}{(N)^2}\sum_{ij}\int_0^\beta\!d\tau 
\:e^{i\nu_n\tau} \:e^{i\q(\gv{R}_i-\gv{R}_j)}\chi_{ij}(\tau)
\\  \label{eq:ChiQw}
&=\sum_{ab}
 \gamma_a \gamma_b\Big\{
 \ew{\hat n_{a}}\ew{\hat n_{b}}\:\d_{i\nu_n,0}
 +
 \Big[\gm{\chi}(\q,i\nu_n)\Big]_{abba}
  \Big\}
  .
\end{align}
where $\nu_n=\frac{2\pi}{\beta}n$ ($n\in \mathbb{Z}$) are bosonic Matsubara 
frequencies and $\beta=\frac1{k_\mathrm{B}T}$ the inverse temperature $T$. 
We already separated the static ($i\nu_n=0$) unconnected 
part proportional to a product of local occupation numbers
$\ew{\hat n_a}$
in the first term.

We  introduced a matrix notation in orbital space 
for the Fourier transform of the connected two-particle Green function,
$\gm{\chi}(\q,i\nu_n)  =\{\chi_{a,b;c,d}(\q,i\nu_n)\}$ (see Appendix \ref{app:2Pdmft}).
This is the central quantity of interest, as other particle-hole susceptibilities
can  equally well be calculated from it. 
The charge susceptibility, for example, 
is obtained by using different matrix elements $\gamma_\s = - |e|$
in Eq.~\eqref{eq:Mop}, which amounts to a different sum over the
matrix elements of the susceptibility matrix
in Eq.~\eqref{eq:ChiQw}.
Therefore, we will focus in the following on the susceptibility 
matrix $\gm{\chi}(\q,i\nu_n)$, without the specific 
pre-factors $\gamma_a$ unless needed,  i.e.\ for their signs.

In analogy to the self-energy, the particle-hole irreducible two-particle 
interaction vertex is assumed to be momentum independent within 
DMFT,\cite{jarrellQMCHM92,zlatic2PVertexDMFT90,*jarrell:symmetricPAM95} 
\begin{align}
  \label{eq:PiLoc}
  \gm{\Pi}(i\o_1,\k_1,i\o_2,\k_2;i\o_2',\k_2',i\o_1',\k_1') \\
  \notag \overset{\text{DMFT}}{\to}
  \gm{\Pi}(i\o_1,i\o_2;i\o_2',i\o_1')
  .
\end{align}
As a consequence, sums over crystal momentum only involve the one-particle 
propagators  leading to simple local expressions for closed loops or a geometric 
series for chains through the lattice. At interaction points crystal momentum 
is only conserved on average.
A similar procedure fails for internal Matsubara sums since all
frequency dependencies are retained within DMFT. Therefore, the 
Bethe-Salpeter equations, as  detailed in the appendix,  have 
the structure of coupled integral equations in Matsubara frequency 
space.

The dynamic susceptibility is obtained by summing the appropriate
two-particle Green function over two internal frequencies, 
\begin{align}
  \label{eq:susMagDefG2}
  &\gm{\chi}(\q,i\nu_n)  
  =
  \frac{1}{\beta}\sum_{i\o_1,i\o_2}
  \gm{\chi}(\q,i\nu_n|i\o_1,i\o_2)
  \:e^{(i\o_1\!+i\o_2)\delta}
  .
\end{align}
The exponential incorporates  infinitesimal convergence factors $\delta$
which ensure the correct time ordering of the number-operators in the two-particle Green 
function.

Frequently, in particular   when using QMC as impurity solver,  the 
two-particle lattice susceptibility is obtained 
by interpreting Green functions as matrices in Matsubara frequency 
space, $\gm{\chi}(\q,i\nu_n|i\o_1,i\o_2)=\gm{\hat \chi}_{\q,i\nu_n}|_{\o_1,\o_2}$
(indicated by a hat and the different placement of the external variable in our notation). 
The Bethe-Salpeter equation 
then has the structure of a matrix equation,
where the internal frequency sums are represented by  matrix multiplications,
\begin{align}
  \label{eq:BSlatt}
  \gm{\hat \chi}_{\q,i\nu_n}=-\gm{\hat P}_{\q,i\nu_n}-
  \gm{\hat P}_{\q,i\nu_n}\gm{\hat \Pi}_{i\nu_n}\gm{\hat\chi}_{\q,i\nu_n}
  .
\end{align}
The particle-hole propagator $\gm{\hat P}_{\q,\nu_n}$ is determined 
by the single-particle Green function (cf.\ Eq.~\eqref{eq:PqFinal}),
and explicit two-particle interactions are incorporated 
via  the irreducible vertex $\gm{\hat \Pi}_{\nu_n}$,
which is a priori unknown and hard to calculate directly
for strongly correlated systems.

For a description via effective impurities underlying the DMFT-method, 
an analogous equation  can be formulated,
\begin{align}
  \label{eq:BSloc}
 \gm{\hat \chi}^{\mathrm{loc}}_{i\nu_n}=- \gm{\hat P}^{\mathrm{loc}}_{i\nu_n}
 - \gm{\hat P}^{\mathrm{loc}}_{i\nu_n} \,\gm{\hat \Pi}_{i\nu_n}\, \gm{\hat \chi}^{\mathrm{loc}}_{i\nu_n}.
\end{align}
$\gm{\hat \chi}^{\mathrm{loc}}_{i\nu_n}$ is the dynamic local susceptibility 
of the effective impurity model and can be calculated in principle.
The local particle-hole
propagator is given by the momentum sum of its lattice counter part,
\begin{align}
\gm{\hat P}^{\mathrm{loc}}_{i\nu_n}=\frac 1N \sum_q\gm{\hat P}_{\q,i\nu_n},   
\end{align}
and can be calculated with knowledge of the local Green function (see Eq.~\eqref{eq:PlocFinal}). 

The \textit{same} irreducible local vertex $\gm{\hat \Pi}_{i\nu_n}$ 
is assumed for the impurity and the lattice model.
This makes it possible to eliminate it from 
Eqs.~\eqref{eq:BSloc} and \eqref{eq:BSlatt}
and  the usual DMFT result for the lattice susceptibility is
obtained
\begin{align}
  \label{eq:MatsubaraDMFT}
  & \gm{\hat \chi}_{\q,i\nu_n}= 
%\\ \nonumber   &
  \left[
  -\gm{\hat P}_{\q,i\nu_n}^{-1}
  +{\gm{\hat \chi}^{\mathrm{loc}}_{i\nu_n}}^{-1}
    +{\gm{\hat P}^{\mathrm{loc}}_{i\nu_n}}^{-1}
  \right]^{-1}
  .
\end{align}
The inversions indicate matrix inversions in the space of Matsubara
frequencies and orbital/spin space.
It is  important to realize that all quantities 
entering Eq.~\eqref{eq:MatsubaraDMFT} can be calculated 
directly from the 
effective impurity model and the lattice Green function. 
One thus  has an explicit equation
for the lattice susceptibility. 

However, the shortcoming of the approach sketched above is twofold. 
First, the set of Matsubara frequencies $i\omega_{n}$ is infinite 
as $n\in\mathbb Z$.
The Matsubara matrices are therefore infinite dimensional, 
which is of course not sustainable in practical calculations and
truncations have to be introduced. Second, the calculated 
susceptibilities are functions of complex Matsubara frequencies
which have to be numerically continued to the real axis
which is mathematically an ill-defined problem. There exist
sophisticated techniques, such as 
maximum entropy methods,\cite{jarrellMEM96} 
but uncertainties and uncontrolled errors remain.

Another approach utilizes the dynamic density-matrix
renormalization group technique and a subsequent
deconvolution.\cite{raasSpectralDMRG05,*raasGenSus09}
Apart from being very resource consuming, the deconvolution 
to obtain data on the real frequency axis might 
also introduce artefacts. 

In order to avoid these difficulties we take a different route here. 
We introduce an additional approximation and decouple the Matsubara 
sums in the Bethe-Salpeter equations
in a manner similar to the  random-phase approximation (RPA). 
For technical details see appendix \ref{app:2Pdmft}.
The major advantage is that the analytic 
continuation of all frequency variables to the real axis,
like $i\nu_n\to\nu+i\delta$,
can be done analytically. As derived in the appendix,
the result 
is\cite{schmitt:sus05,schmittPhD08}\footnote{As two-particle quantities --in contrast to single-particle Green functions--
do not depend on the sign of the infinitesimal
imaginary part $\delta$, we omit it here.
}
\begin{align}
  \label{eq:susFinal}
  &\gm{\chi}(\q,\nu)=  \\  \notag  &
  % \hspace*{10mm} 
  \Big[
  -\gm{ P}(\q,\nu)^{-1}
  +{\gm{\chi}^{\mathrm{loc}}(\nu)}^{-1}
  +{\gm{P}^{\mathrm{loc}} (\nu)}^{-1}
  \Big]^{-1}
  .
\end{align}
The structure of this equation is of course the same as 
Eq.~\eqref{eq:MatsubaraDMFT}, but the important difference is that
no Matsubara matrices occur (there are no hats are over the quantities) 
and the external frequency $\nu$ is a real variable. 
The indicated matrix structure and inversions only refer to 
the orbital/spin space.

In the following  we will focus 
on the magnetic susceptibility and on simple $s$-shells.
In that case, all quantities are matrices in spin-space only. 
The particle-hole propagators of the paramagnetic regime are spin-symmetric,
$P_\uparrow=P_\downarrow\equiv P$, and 
only two independent components exist for the susceptibilities,
$\chi_{\uparrow,\uparrow;\uparrow,\uparrow}=\chi_{\downarrow,\downarrow;\downarrow,\downarrow}$
and $\chi_{\uparrow,\downarrow;\downarrow,\uparrow}=\chi_{\downarrow,\uparrow;\uparrow,\downarrow}$.
The magnetic susceptibility is a linear combination of these two components,   
$\chi_\mathrm{mag}(\q,\nu)=\chi_{\uparrow,\uparrow;\uparrow,\uparrow}(\q,\nu)
-\chi_{\uparrow,\downarrow;\downarrow,\uparrow}(\q,\nu)$,
and can be expressed as (neglecting the prefactor $(\frac{g\mu_B}2)^2$),
\begin{align}
  \label{eq:susFin}
  \chi_\mathrm{mag}(\q,\nu)&=
  %\frac{1}{
  \Big[-\frac 1 {P(\q,\nu)}+\frac 1{P^{\mathrm{loc}}(\nu)}+\frac 1{\chi^{\mathrm{loc}}_\mathrm{mag}(\nu)}
  \Big]^{-1}
  % }
  .
\end{align}
\newcounter{FinalEqCounter}
\setcounter{FinalEqCounter}{\value{equation}}
All quantities entering this form are scalar functions of a real variable.
$\chi^{\mathrm{loc}}_\mathrm{mag}(\nu)$ is the dynamic local magnetic susceptibility
of the impurity model, $P^{\mathrm{loc}}(\nu)$ and $P(\q,\nu)$  are the local
and lattice dependent particle-hole propagators, which only depend on the 
fully interacting single-particle Green function 
(see Eqs.~\eqref{eq:PlocFinal} and \eqref{eq:PqFinal}).

This specific form \eqref{eq:susFin} for the susceptibility lends itself directly to
interpretations in terms of the two archetypical physical pictures 
underlying magnetism:
\begin{itemize}
\item[(i)]
  The picture of local-moment magnetism 
is suggested if Eq.~\eqref{eq:susFin} is re-written as
\renewcommand{\theequation}{\arabic{FinalEqCounter}a}%
\begin{align}
  \chi_\mathrm{mag}(\q,\nu)&
  \label{eq:susJq}
  =\frac{\chi^{\mathrm{loc}}_\text{mag}(\nu)}{
    1-J(\q,\nu)\,\chi^{\mathrm{loc}}_\text{mag}(\nu)}
  ,
\end{align}
\addtocounter{equation}{-1}%
\renewcommand{\theequation}{\arabic{equation}}%
where local spins characterized by $\chi^{\mathrm{loc}}_\text{mag}$
interact via a nonlocal and dynamic exchange coupling
\begin{align}
  \label{eq:effJ}
  J(\q,\nu)&=\frac 1 {P(\q,\nu)}-\frac 1{P^{\mathrm{loc}}(\nu)}
  .
\end{align}
This view is substantiated by 
approaching the atomic limit.  
Then, the hopping matrix elements vanish, $t_\k=0$,
and the lattice and the local particle-hole propagators become identical,
$P(\q,\nu)|^{t=0}=P^\mathrm{loc}(\nu)$.
The susceptibility correctly reduces to the fully 
interacting susceptibility of isolated ions,
\begin{align}
  \chi_{\mathrm{mag}}(\q,\nu)|^{t=0}&= \chi_\mathrm{mag}^{\mathrm{loc}}(\nu)
  .
\end{align}
For half-filling and $U>0$ it is just given by that of a free spin
\begin{align}
  \label{eq:susAtomicLimit}
  \chi_{\mathrm{mag}}(\q,\nu)|^{t=0}&  =\frac1T\,\delta(\nu)
  .
\end{align}
This result is easily extended to incorporate the leading order correction
due to the coupling to neighboring ions. Expanding up to 
second order in the hopping around the atomic limit yields the exchange coupling 
$J(\q,\nu)=\frac1U\frac1N\sum_\k t_\k\,t_{\k+\q}+\mathcal O(t^3)$.
The susceptibility then correctly reproduces the 
the mean-field approximation to the Heisenberg model, 
\begin{align}
  \label{eq:LocMomTc}
  \chi_{\mathrm{mag}}(\q,\nu)=\frac1{T-T_C(\q)}\,\delta(\nu) +\mathcal O(t^3)
  .
\end{align}
In case of a simple-cubic lattice with nearest hopping, 
the $T_C(\q)$  is given by  $T_C(\q)=-\frac{2t^2}U\sum_{i=1}^D\cos(q_i)$.
This  reproduces the AFM $\gv{Q}=\pi(1,1,\dots)^T$ mean-field transition
at the well-known  critical temperature $T_C(\gv{Q})=\frac{2Dt^2}U$.

\item[(ii)] The interpretation in terms of the itinerant picture 
of magnetism is obtained by re-writing 
Eq.~\eqref{eq:susFin} as
\renewcommand{\theequation}{\arabic{FinalEqCounter}b}%
\begin{align}
  \label{eq:susIt}
  \chi_\mathrm{mag}(\q,\nu)&=
  \frac{-P(\q,\nu)}{1+\Gamma_\mathrm{mag}(\nu)P(\q,\nu)}
  .
\end{align}
\addtocounter{equation}{-1}
\renewcommand{\theequation}{\arabic{equation}}%
Here, propagating electrons characterized by their particle-hole
susceptibility 
$-P(\q,\nu)$ interact locally with the vertex
\begin{align}
  \label{eq:LocVertex}
  \Gamma_\text{mag}(\nu)&=
  -\frac1 {\chi_\mathrm{mag}^{\mathrm{loc}}(\nu)} - \frac 1{P^{\mathrm{loc}}(\nu)}
  .
\end{align}
This view is substantiated by the noninteracting limit, where the Coulomb
interaction matrix element vanishes, 
$U=0$. Then, Wick's theorem is applicable and the
local susceptibility reduces to the negative local particle-hole propagator
$\chi_\mathrm{mag}^\mathrm{loc}(\nu)|^{U=0}=-P^\mathrm{loc}(\nu)$. The 
susceptibility is then indeed that of noninteracting particles
on a lattice given by the Lindhard-function
\begin{align}
  \label{eq:susULimit}
  \chi(\q,\nu)|^{U=0}&= -P(\q,\nu)|^{U=0}\\ \nonumber
  &
  =-\frac1N\sum_\k\frac{f(t_{\k+\q})-f(t_\k)}{\nu+i0^++t_{\k+\q}-t_{\k}}
  .
\end{align}
Improving this by employing a perturbation theory in $U$ we arrive at 
the usual RPA expression. In the lowest order the local vertex is 
nothing but the bare Coulomb interaction,  $\Gamma_\mathrm{mag}(\nu)=U$,
and then
\begin{align}
  \label{eq:susRPA}
  \chi(\q,\nu)&= \frac{-P(\q,\nu)|^{U=0}}{1+U\,P(\q,\nu)|^{U=0}}
  +\mathcal O (U^2).
\end{align}
 
\end{itemize}

The two opposing cases of a weakly interacting electron gas and the atomic limit
are exactly incorporated in the approximate form for the susceptibility, Eq.~\eqref{eq:susFin}.
This substantiates the hope 
to be  able to correctly describe the regime of 
intermediate coupling strength and especially
the transition from itinerant to localized forms 
of magnetism.

At this point a comment on the advantages and shortcomings of
the final expression, Eq.~\eqref{eq:susFin}, is in place. The major advantage
of this formula lies in its simplicity in combination with 
the correct incorporation of the weak and strong coupling limit.
And as it will  be shown in the following, the qualitative 
and quantitative description of  correlation
effects even for  intermediate coupling is very good.

%%%%
In the usual weak-coupling RPA~\cite{doniachGreenFunc74,*schaeferDysonRPAHM99}
or extensions thereof, such as the fluctuation exchange 
approximation\cite{bickers:ConservingFLEX89b,*bickers:ConservingFLEX91}
and the two-particle self-consistent approach\cite{vilk:SpinChargeFlucHM94}, 
the particle-hole propagator is either calculated with bare propagators, i.e.\ without interactions, 
or interaction processes are included only at a very crude level. This also applies to  
auxiliary boson approaches\cite{khaliullinDesityResponsetJ96,*saikawa:DynamicSusHM98}
and equation of motion decoupling 
schemes.\cite{ereminDynChargeHM01,*uldryTwoParticleMagSusHM05}
In contrast, the particle-hole propagator employed here is calculated
with the fully interacting  single-particle
Green function obtained from DMFT, which  incorporates life-time and many-body effects 
and can already lead to considerable modifications.
Furthermore, in the above mentioned approximation schemes the two-particle interaction 
vertices are given by weighted linear combinations
of bare interaction matrix elements, i.e.\  by numbers,
and the frequency dependence of two-particle vertices is usually ignored. 
In the present treatment  nontrivial
\textit{dynamic} interaction vertices
$\Gamma(\nu)$ or $J(\q,\nu)$ are incorporated into the susceptibilities.
Finite order cumulant expansions in the hopping $t$ as done, for example, in 
Ref.~\onlinecite{moskalenko:dynamicSusHM97,*sherman:MagSusHM07,*shermanChargeSusHM08}
do respect the leading frequency dependence of the two-particle vertex and include leading
order nonlocal effects. Lifetime broadening and Kondo physics, 
however, are not captured in these treatments.

Our approach requires a thorough assessment of the quality of the
RPA-like decoupling of frequency sums, 
which could possibly cause errors.
%in addition  to the neglect of certain nonlocal correlations, which are inherent in the DMFT-approximation. 
Virtues and limitations of our result 
contained in Eq.~\eqref{eq:susFin} will be borne out by 
the discussion in the next section. 
%
\begin{comment}
As a consequence of local 
approximations like DMFT, in general, the momentum dependence of susceptibilities will 
exclusively be determined  by 
the particle-hole propagator $P(\q,\nu)$.
If the particle-hole propagator shows an enhanced response 
at some wave vector, any susceptibility will show this very tendency
as well. As an example, it will be impossible to obtain an enhanced 
antiferromagnetic spin susceptibility ($\q=\pi(1,1)$-component large)
and a simultaneous tendency toward phase separation in
the charge susceptibility ($\q=0$-component large). 
The difficulties mentioned do not neccessarily cause serious 
errors in most parts of the phase diagram, and even the RPA-like 
decoupling proves able to capture dominant physical effects.
%
I turns out that the decoupling proves able to capture dominant physical effects. 
Propagating modes are strongly influenced
by the lattice structure, i.e.\ the Fermi surface, and  lifetime effects, which
are both included into the particle-hole propagator. The local two-particle 
interactions are  correctly captured by the local
susceptibility. So generally any trend in one of these quantities, such as
an enhanced $\gm{P}$ due to a large phase space volume for some
scattering process, will also be present in the exact (i.e.\ non-decoupled)
lattice susceptibility.
\end{comment}
However, we can already formulate the expectation, that 
in situations, where coherent two-particle  propagations through the lattice 
lead to the build-up of nonlocal correlations the
decoupling is likely to fail. 
In the general set-up, the coherence is maintained at each interaction vertex
due to its energy dependence, whereas in the decoupled version  
averages are taken at each local vertex. Therefore, coherence is  maintained
only between consecutive local two-particle vertices, so that 
long-ranged correlations are affected. 
The most prominent
example where such excitations are crucial is FM, as
it is clear from the conditions favoring Nagaoka-type FM.\cite{fazekas:LectureNotes99,hanisch:LattDepFMHM97,*hanisch:InstFMHM95,*hanisch:FerroHubbard}
As will be demonstrated below, this leads to a very accurate description
of AFM phase transitions, while FM
transition temperatures are overestimated, as in the usual Stoner theory.

Another source of inaccuracies  stems from the method chosen to solve the effective
impurity model. In most cases, the drawbacks of the impurity solver 
pose the  strongest limitations on the accuracy and validity of the 
calculated Green functions.

%%%%%%%%%%%%%%%%%%%%%%%%%%%%%%%%%%%%%%%%%%%%%%%%%%%%%%%%%%%%%%%%%%
\section{Results}
\label{sec:results} 
The following investigation of 
the Hubbard model on two different lattices uses the DMFT scheme, 
applied to various one- and two particle properties furnishing, e.g.,
spectral properties, local moments, susceptibilities, effective interactions 
and transition temperatures. 
We concentrate on the 
approach from the paramagnetic regime in order to identify phase transitions 
via diverging magnetic susceptibilities.
We set $g\mu_\mathrm{B}=\hbar=c=k_\mathrm{B}=1$ and 
use the noninteracting half bandwidth $W$ as unit of energy.

For the results presented in this Section, we used the ENCA\cite{pruschkeENCA89,*schmittSus09,*greweCA108}
as impurity solver in order to calculated the local self-energy and susceptibility.
The three-dimensional momentum integrations of the lattice particle-hole propagator
were performed using the  CUBA-library.\cite{hahn:cubaLibrary05}
%%%%%%%%%%%%%%%%%%%%%%%%%%%%%%%%%%%%%%%%%%%%%%%%%%%%%%%
\subsection{Simple-cubic lattice}
\label{sec:simpleCube}

In this section the magnetic transitions of the Hubbard 
model on a simple-cubic (SC) lattice in three dimensions will be examined. 
The magnetic phase diagram of the Hubbard model has 
been studied with a multitude of methods
for various lattices. These include 
perturbation theory in the interaction 
$U$,\cite{pennStabilityMag66,*dongen:extHMHighDim91}
direct QMC,\cite{staudt3dHM00,hirsch:HM87I,*3DphaseDiagHMScalettar89,*ulmke3DHM96}
diagrammatic approaches,\cite{dare:HM00,vilk:NonPerturbHM97,*dare:ItinAFM96,*arita:MagFlexHM00}
DMFT,\cite{jarrellQMCHM92,jarrellPruschke:HM93a,ulmkeAndersonHM95,zitzler:PhaseFrustratedHM04,freericks:magPhase_HM95,*tahvildar-zadeh3dHM97,*petersBetheHM09,aryanpourTriangular06,*deLeoThermo3dHM11} 
and cluster extensions thereof.\cite{kent3dHM05,*rohringer3dHM11,*fuchs3dHM11,*fuchs3dHMprl11}

The reason we redo such an analysis here, is 
to investigate possible shortcomings
of the present approach by comparing its results to the 
literature.
The impurity solvers based on the hybridization expansion are known
to have some limitations in the Fermi liquid regime at too low 
temperatures.\cite{pruschkeENCA89,*schmittSus09,*greweCA108,Grewe:siam83,*kuramoto:AnalyticsNCA85,*Kuramoto:ncaIII84,*muellerhartmann:NCAgroundstate84}
Additionally, the RPA-like decoupling of the Bethe-Salpeter equations 
introduces a further approximation as envisaged above. As such decoupling schemes 
are known to overestimate the tendency toward 
phase transitions, we investigate the quality of this approximation.

The single-particle dispersion for the SC lattice
with nearest-neighbor hopping in three dimensions 
is given by
\begin{align}
  \label{eq:3dscDisp}
  t_{\k}=&-2t  \big[\cos(k_x)+\cos(k_y)+\cos(k_z)\big]
%  \\ & \notag
%  -4 t' \big(\cos(k_x)[\cos(k_y)+\cos(k_z)] +\cos(k_y) \cos(k_z)\big)
  .
\end{align}
The half bandwidth is $W=6t$ and will be 
used as the unit of energy in the following. 

\begin{figure}
  \includegraphics[width=0.9\linewidth]{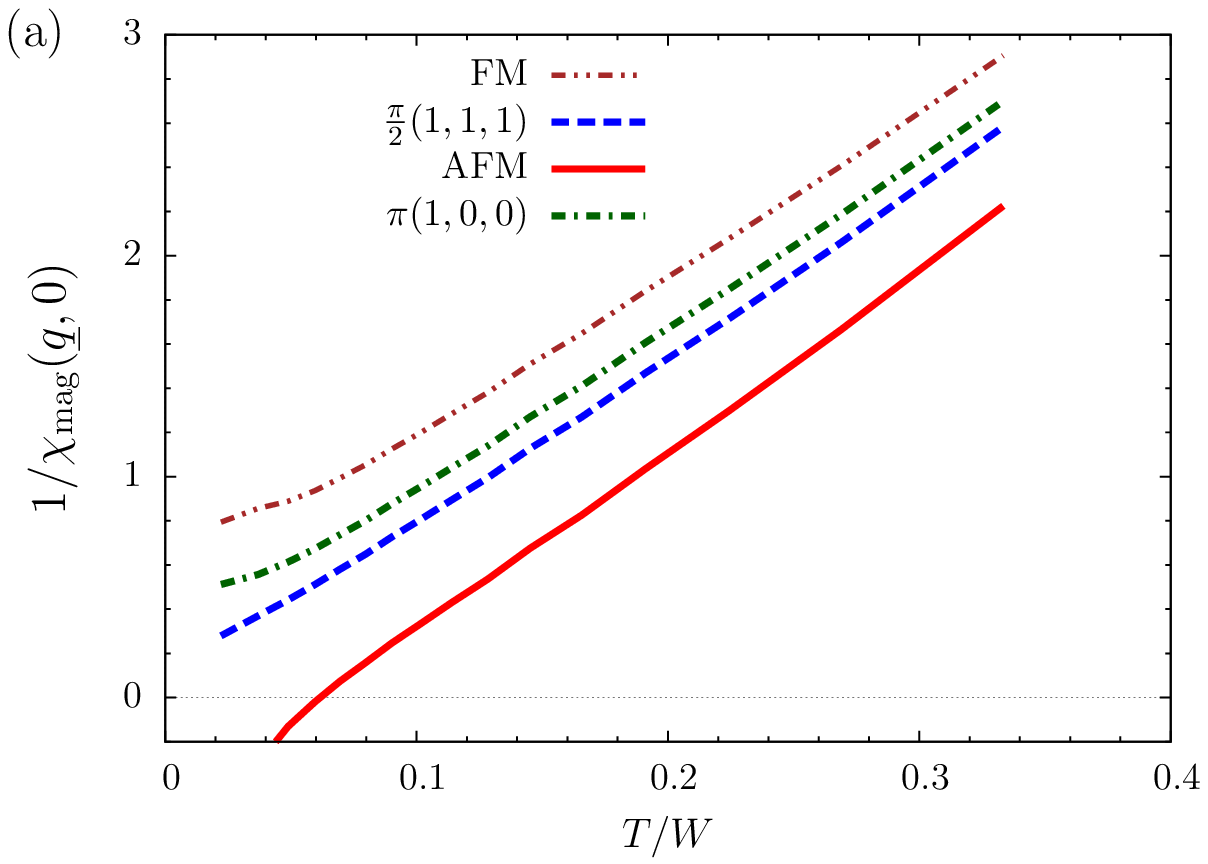}
  \includegraphics[width=0.9\linewidth]{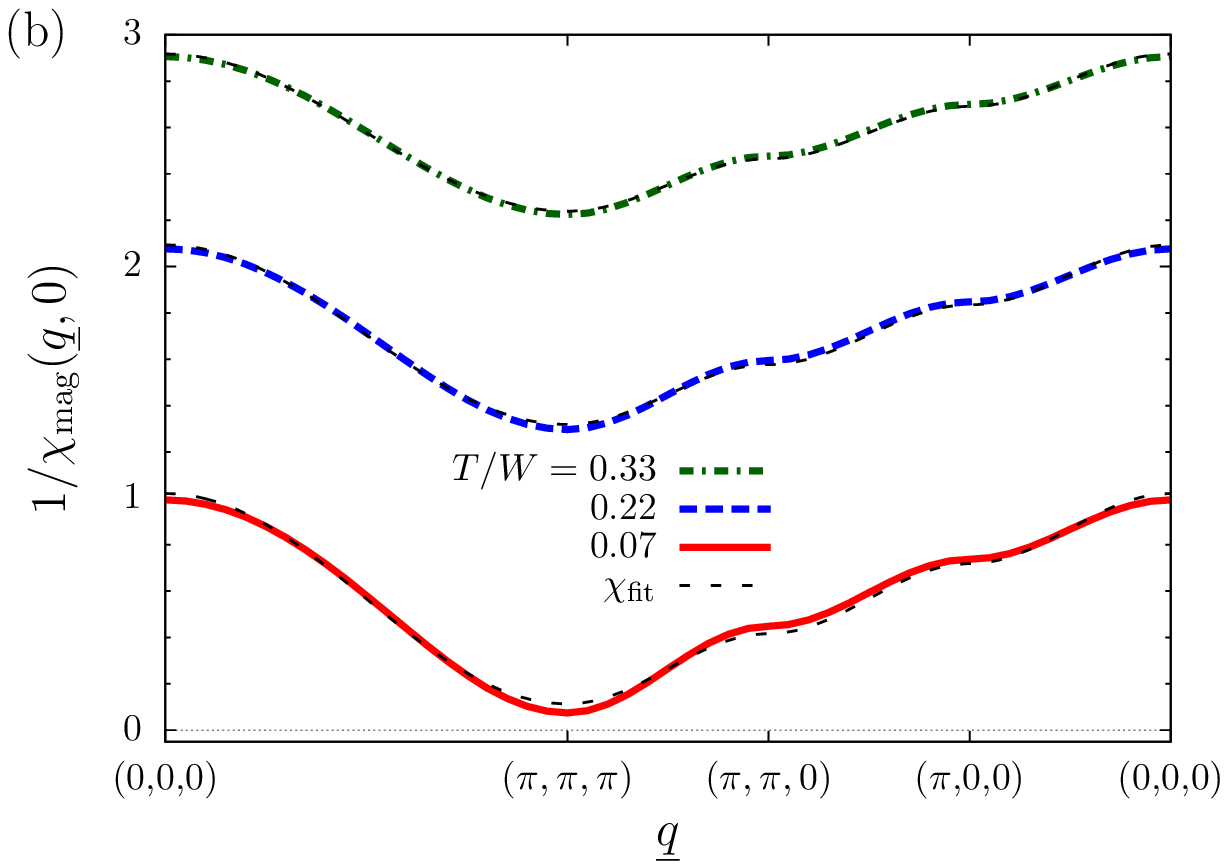}
  \caption{  \label{fig:chiInvSC}
    (Color online) Inverse static susceptibility (a) as function of temperature for various values of the 
    wave vector $\q$ and (b) as function of $\q$ for three  different temperatures. Both 
    plots are calculated for the SC lattice with nearest neighbor hopping
    only ($t'=0$), 
    $U/W=1.5$, and at half-filling $n=1$. }
\end{figure}
Figure \ref{fig:chiInvSC}(a) displays the temperature dependent inverse static 
susceptibility for a half-filled  SC lattice in three dimensions 
and  various wave vectors $\q$.  $\chimag(\q,\nu=0)^{-1}$ depends 
almost linearly on $T$  over the whole range of temperatures, which is expected 
from the mean-field character of DMFT.\cite{jarrellQMCHM92,ulmke:FMQMCHubbard98,byczukCurieDMFT02}
The wave-vector dependence of  $\chimag(\q,\nu=0)^{-1}$ is shown in Fig.~\ref{fig:chiInvSC}(b) for various 
temperatures. The $\gv{Q}=\pi(1,1,1)^T$ component of the susceptibility  is always largest,
indicating the strong tendency toward AFM. 
The dashed lines are fits to an expansion of the inverse susceptibility up to second
order in the hopping,
\begin{align}
  \label{eq:chiFit}
  \chi_{\mathrm{fit}}(\q,\nu=0)&=\frac{1}{a(T)+b(T)\ew{ t^2}_{\q}}
\end{align}
where $\ew{t^2}_{\q}=\frac 1N\sum_{\k} t_{\k+\q}t_{\k}$.
For the simple-, body-centered, and face-centered cubic lattice with nearest and
next-nearest neighbor hopping, $t$  and $t'$ respectively, this elementary 
two-particle dispersion is just given by the negative single-particle dispersion,
where all  hopping-matrix elements are replaced with their squares,     
\begin{align}
  \ew{t^2}_{\q}&=-t_{\q}\big|_{t\to t^2, t'\to {t'}^2}
  .
\end{align}
(The sign stems from the definition of the single-particle dispersion 
with an overall negative sign, see Eq.~(\ref{eq:3dscDisp}.))
This form approximates the $\q$-dependence of the actual susceptibility very well, as
it is visible in Fig.~\ref{fig:chiInvSC}(b). 

This result provides a justification of the RPA-like 
decoupling of frequency sums in the 
Bethe-Salpeter equations.  The reasoning is as follows:
Comparing the  $\q$-dependency of the approximate form Eq.~(\ref{eq:chiFit})
to  Eq.~(\ref{eq:susFin}), we can infer  that the 
particle-hole propagator -- as the only $\q$-dependent quantity entering the 
susceptibility --  must be of the form 
\begin{align}
  P(\q)&=\frac{p_0}{1-\alpha \ew{t^2}_{\q}}=p_0\sum_{n=0}^\infty \big[\alpha \ew{t^2}_{\q}\big]^n
  .
\end{align}
The geometric series, used  in the last equality, can be interpreted as the result of an RPA-like decoupling
of the frequency sums appearing in the expansion of 
the exact particle-hole propagator, see Eq.~(\ref{eq:Pexpanded}).
Given the quality of the fit,  we are led to the conclusion that the RPA-like decoupling
works very well for the particle-hole propagator. 
It therefore seems reasonable to assume that it is also a good approximation
in the Bethe-Salpeter equations.

The physical reason lies in the nature of antiferromagnetic correlations. 
They favor neighboring spins to point in opposite directions, which does 
not require a precise adjustment in the time-domain for a propagating 
particle-hole pair. Therefore, decoupling in frequency space cannot 
induce a qualitative error (in contrast to the ferromagnetic case, see below). 
We always found
good agreement between the form of 
Eq.~(\ref{eq:chiFit}) and the calculated susceptibility, whenever the tendency toward AFM was dominant.

\begin{figure}
  \includegraphics[width=0.9\linewidth]{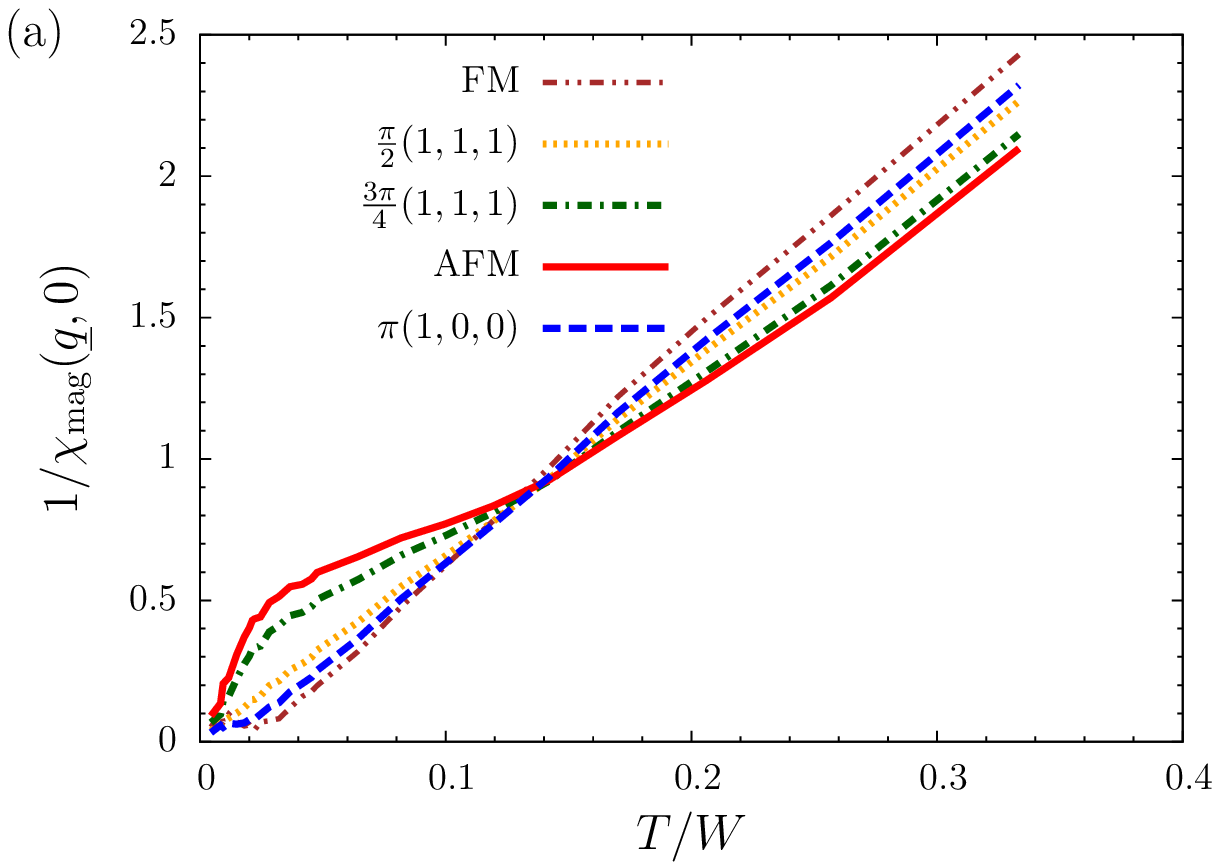}
  \includegraphics[width=0.9\linewidth]{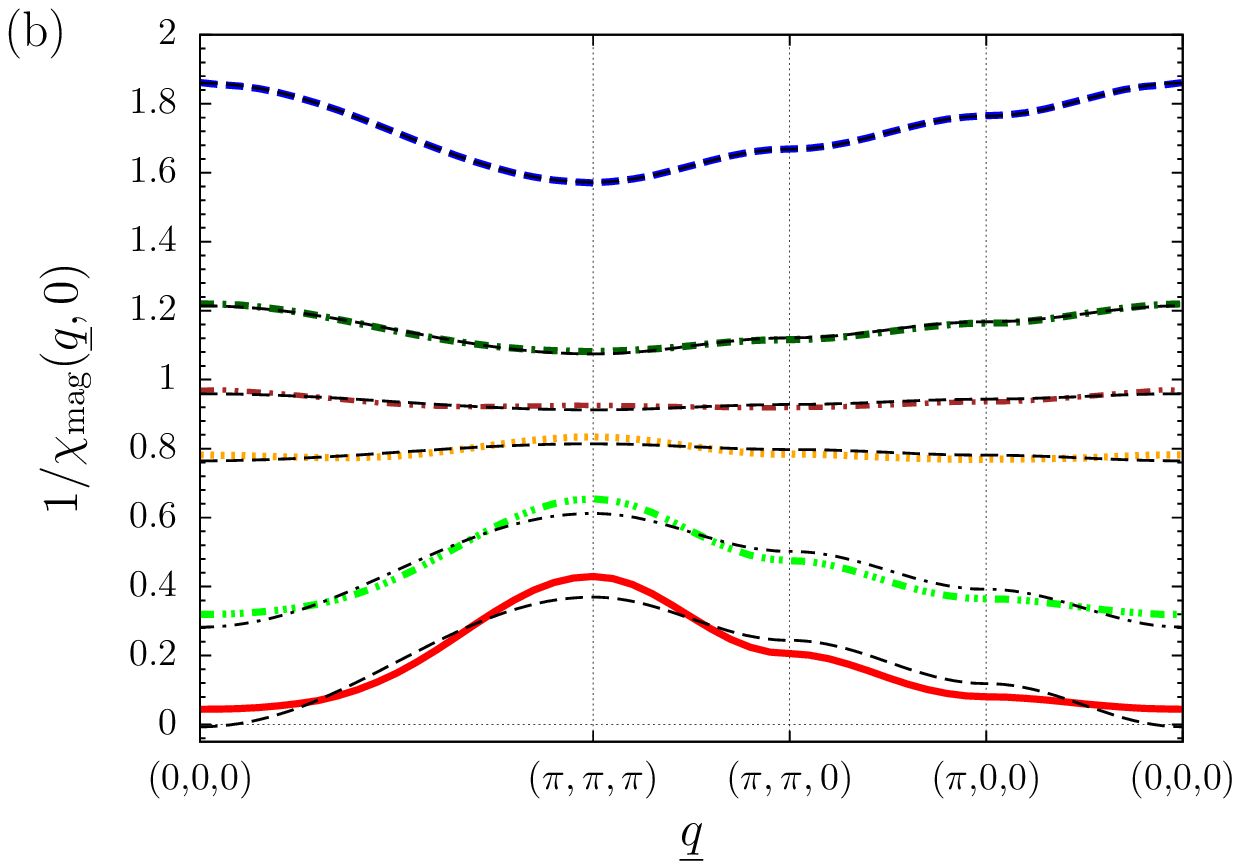}
  \caption{ 
    \label{fig:Tc3dChiU10}
    (Color online) (a) Inverse static susceptibility as function of temperature for various values of the 
    wave vector $\q$.  (b) $1/\chi_{\mathrm{mag}}(\q,0)$ as function of $\q$ for the temperatures (top to bottom):
    $T/W=0.26$, $0.17$, $0.14$, $0.12$, $0.065$, and $0.021$.
    Both plots are calculated for the SC lattice with nearest neighbor hopping
    only ($t'=0$),     $U/W=3.3$, and  $n=0.97$.
  }
\end{figure}
For a larger Coulomb interaction of $U=3.3W$ and at finite doping, $\delta=3\%$ ($n=0.97$),
the AFM susceptibility is largest only at high temperatures. This can be observed 
in Fig.~\ref{fig:Tc3dChiU10}(a) where $\chi(\q,0)^{-1}$ decreases linearly with 
temperature only for $T/W\gtrsim 0.2$ and changes qualitatively at low $T$. There, 
the FM component becomes dominant. 
The wave-vector dependency as shown in Fig.~\ref{fig:Tc3dChiU10}(b) is well approximated 
by Eq.~(\ref{eq:chiFit}) for the AFM-dominated region at high $T$, but 
below $T\lesssim 0.12W$ the fit does not work well and strong deviations occur.

\begin{figure}
  \includegraphics[width=\linewidth]{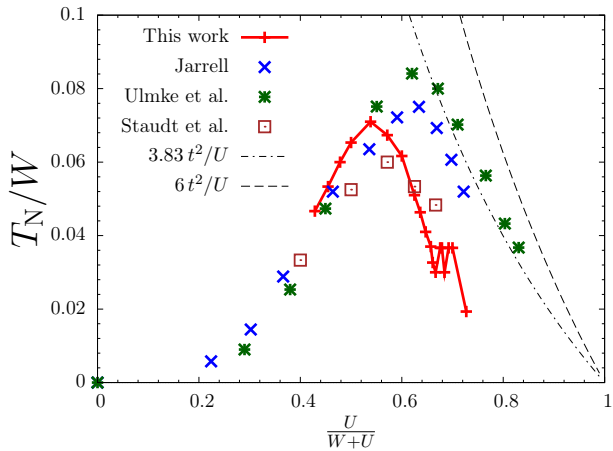}
  \caption{(Color online)  The N\'eel temperature of the half-filled 
    Hubbard model for the three dimensional SC lattice  
    as function of $U$ (red line with points).
    For comparison, results from other studies are included as well: 
    DMFT+QMC\cite{jarrellQMCHM92} (blue crosses, labeled Jarrell), 
    DMFT+QMC\cite{ulmkeAndersonHM95} (green stars, labeled Ulmke et al.), 
    QMC\cite{staudt3dHM00} (brown squares, labeled Staudt et al.),
    high temperature expansion\cite{rushbrooke:HeisenberM} (dash-dotted line, labeled $3.83 t^2/U$),
    and mean-field (dashed-line, labeled $ 6 t^2/U$). 
    \label{fig:Tc3dscD0}
  }
\end{figure}

The AFM transition temperature $T_\mathrm{N}$ (N\'eel temperature) is shown  in Fig.~\ref{fig:Tc3dscD0} 
as function of the Coulomb repulsion
for the half-filled ($\delta=0$) model. Results from other studies are also shown
for comparison. The present result should be close to the data of
Jarrell (Ref.~\onlinecite{jarrellQMCHM92}) 
or Ulmke \textit{et al.} (Ref.~\onlinecite{ulmkeAndersonHM95}) as these results 
were also obtained with DMFT, but for a Gaussian (Jarrell) and semi-circular (Ulmke \textit {et al.})
free DOS. For $U\leq W$ the agreement is quite satisfactory, while for larger $U$
the N\'eel temperature from the present approach is substantially smaller. 
It is, however, roughly in  agreement with data from QMC simulations of the 
three-dimensional model (Staudt \textit {et al.}, Ref.~\onlinecite{staudt3dHM00})
but this may be viewed as a coincidence.

\begin{figure}
  \includegraphics[width=1\linewidth]{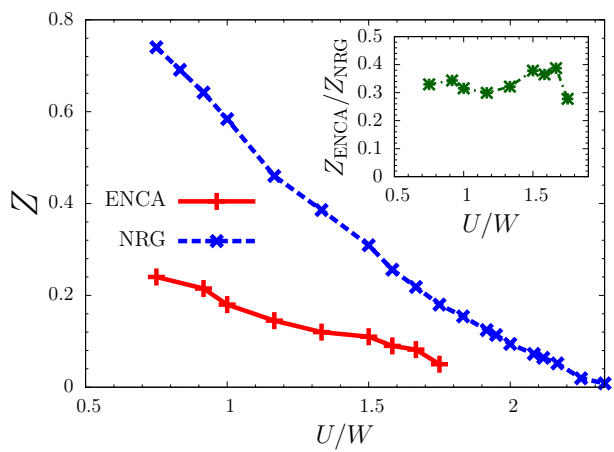}
  \caption{  \label{fig:Z3dsc}
    (Color online) The quasiparticle weight for the half-filled Hubbard model as function of the Coulomb
    repulsion $U$ extracted from the DMFT with ENCA and NRG as impurity solvers.
    The inset shows the ratio of both.
  }
\end{figure}
The  exchange interaction decreases as $1/U$ in leading perturbative order for large $U$.
Moreover, it is renormalized  by small quasiparticle 
weights ($Z$-factors) when  quasiparticle bands emerge. 
So a decrease of $T_\mathrm{N}$ at larger $U$ is to be expected, and its
steepness depends on the size of the $Z$-factors, 
\begin{align}
  \label{eq:QPZ}
  Z&=\frac{1}{1-\frac{\partial \mathrm{Re}\Sigma}{\partial \omega}(0)}
  .
\end{align}

While the ENCA as our impurity solver gives the correct order of magnitude and parameter dependence for
the low-energy scale of the SIAM,
it underestimates its actual value.\cite{pruschkeENCA89,*schmittSus09,*greweCA108}
Within the self-consistent treatment of DMFT this tendency is
retained\cite{greteKink11} as it can be observed in  
Fig.~\ref{fig:Z3dsc} where $Z$ is shown as  function of $U$.
For comparison, the result from  DMFT calculations with the 
NRG\cite{petersNewNRG06} as impurity solver 
for the same parameter values is also shown. While the ENCA quasiparticle weight is 
smaller than the one extracted from the NRG, both follow the same trend.
The inset displays the ratio of both and reveals the ENCA to yield 
a low-energy scale which is roughly a factor of $3$ too small.
One may argue, that the ENCA-calculation is performed 
for an effectively larger value of the interaction $U$. Therefore,
the reduction of the critical temperature compared to the other 
DMFT calculations as observed in Fig.~\ref{fig:Tc3dscD0} is
a result of the too small low-energy scale of the ENCA.

\begin{figure}
  \includegraphics[width=1\linewidth]{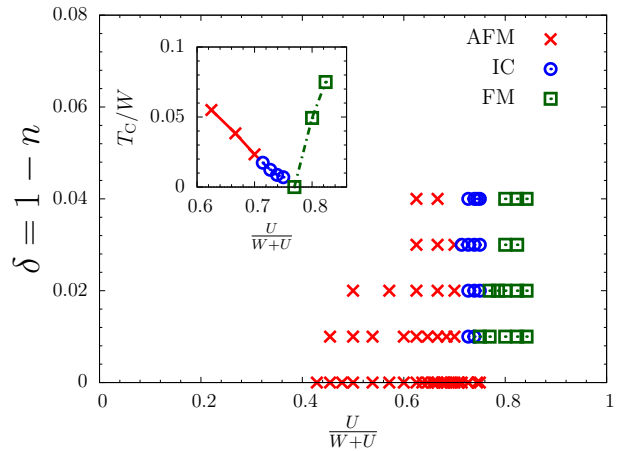}
  \caption{(Color online) Magnetic phase diagram in the $\delta$-$U$ plane. 
    Each point indicates a magnetic phase transition. 
    (The regions without points were unaccessible with the ENCA as impurity solver).
    The inset shows the transition temperature as function of $U$ for a fixed doping of
    $\delta=1-n=0.03$.
    In both graphs, the red crosses indicate AFM, the blue circles IC, and 
    the green squares FM transitions. 
    \label{fig:Tc3dsc}
  }
\end{figure}
Figure \ref{fig:Tc3dsc} displays the regions in the $\delta-U$ phase-space where 
magnetic phase transitions are observed. The AFM region extends from half-filling $\delta=0$ 
to finite doping for Coulomb interactions $U\lesssim 2.3W$ ($U/(W+U)\lesssim 0.7$).
The investigation of regions in phase space with larger doping or smaller Coulomb 
repulsion was not possible, as there the pathology
of the ENCA\cite{Grewe:siam83,*kuramoto:AnalyticsNCA85,*Kuramoto:ncaIII84,*muellerhartmann:NCAgroundstate84,pruschkeENCA89,*schmittSus09,*greweCA108}
prevented the access of low-enough temperatures. 

In accord with earlier studies\cite{pennStabilityMag66,freericks:magPhase_HM95,*tahvildar-zadeh3dHM97,*petersBetheHM09,igoshevMagnetismHM10,kunesTwoDMFT11}
we also find incommensurate (IC) spin-density wave transitions 
away from half filling at the edge of the AFM region.  
As a function of $U$ (see inset) or doping (not shown), 
the transition  temperature follows the trend foreshadowed 
by the N\'eel temperature, but the ordering vector 
shifts away from the AFM $\gv{Q}=\pi (1,1,1)^T$.\cite{freericks:magPhase_HM95,*tahvildar-zadeh3dHM97,*petersBetheHM09}
\begin{figure}
  \includegraphics[width=1\linewidth]{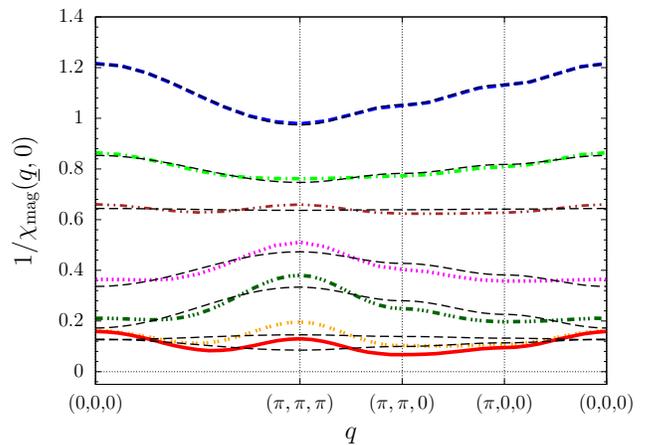}
  \caption{(Color online) Inverse static susceptibility as function of wave vector $\q$
    for various values of temperature (top to bottom):
    $T/W=0.17$, $0.12$, $0.093$, $0.056$, $0.033$, $0.016$, and $0.012$.
    The curves are calculated for the SC lattice with nearest neighbor hopping
    only ($t'=0$),     $U/W=3$, and  $n=0.97$.
    \label{fig:3dIC}
  }\label{fig:ICQ}
\end{figure}

This can be observed in Fig~\ref{fig:ICQ}, where the inverse magnetic susceptibility is
shown as function of $\q$ for various temperatures. There the transition vector 
probably lies in the region around $\q=(1,1,0)^T\pi$ but a precise location 
is not possible since we did not scan a sufficient part of the Brillouin zone
but only along selected axes. 

For finite doping, the nesting property of the Fermi 
surface is lost and incommensurate ordering is expected with increasing 
doping.\cite{freericks:magPhase_HM95,*tahvildar-zadeh3dHM97,*petersBetheHM09}
At a fixed doping, as presented in the inset of Fig.~\ref{fig:Tc3dsc}, the transition is induced
by increasing the interaction. We attribute this to the temperature dependent 
formation of quasiparticles, whose characteristic energy scale is larger
at lower $U$. There, the transition takes place at higher $T$ where the details
of the Fermi surface are washed out and the nesting is approximately 
better fulfilled which leads to a AFM transition. Increasing $U$ 
decreases the transition temperature and the details  of the Fermi surface become increasingly,
making the system more sensitive to the lack of nesting and leading to the IC transition.

As depicted in the inset of Fig.~\ref{fig:Tc3dsc}, the AFM and FM 
phases seem to be separated by a 
quantum critical point (QCP), where the transition temperature vanishes
and the systems goes from AFM to FM ground states at $T=0$.  
This QCP could in principle be physical as was speculated in the literature\cite{zitzlerPruschke:MagPhaseSepHM02} 
but this question can not be addressed within the present study.
The ENCA  as impurity solver does not allow the investigation of very low temperatures 
and consequently whether or not the transition temperature vanishes cannot be decided.
As FM is overestimated within the present approach (see below) it could well be
shifted to larger values of $U$ in more accurate treatments
and thus removing the apparent QCP. 

At very large $U\gtrsim 3W$ we observe FM at finite doping, which is the region
where it is in principle expected and observed for various lattices.\cite{ulmke:FMQMCHubbard98,nagaoka:FerroHMI,*obermeierPruschke:FMHM97,*fazekasHM89,*merino:FMtriangular06,*petersFrustHM09,*guentherFMHM10}
However, for the unfrustrated ($t'=0$) SC lattice studied here,
the transition temperatures $T_\mathrm{C}$ are too large (see inset). 

We attribute this overestimation of the tendency toward FM to the RPA-like 
decoupling.
Time-coherent propagation of particle-hole pairs over long 
distances is known to play a major role 
for FM\cite{hanisch:LattDepFMHM97,*hanisch:InstFMHM95,*hanisch:FerroHubbard} 
(ordering vector $\gv Q=0$). FM correlations 
favor equal spin-alignment, and a transfer of electrons between sites does 
require a time-correlation with accompanying holes due to the 
Pauli principle. But such correlations in the time domain are not conserved 
by the RPA-frequency decoupling at interaction vertices. 
This conclusion is supported by the fact, that the FM transition temperatures as shown in the inset 
are roughly in accord with the Stoner-like criterion,
$\rho(0,T_\mathrm{C}) U\stackrel != 1$, 
where $\rho(0,T)$ is the fully interacting and temperature dependent
DOS at the Fermi level.

We have now established that the present approach yields reliable results whenever 
short-ranged particle-hole excitations dominate the magnetic response.
In and close to the AFM regime, the transition temperatures are even reproduced
quantitatively, apart from a reduction of $T_\mathrm{N}$ due to the too small 
low-energy scale produced by the ENCA.
For very large Coulomb repulsions, the tendency towards FM
is found to be overestimated which is attributed to the RPA-like decoupling 
of frequency sums in the Bethe-Salpeter equations.

%%%%%%%%%%%%%%%%%%%%%%%%%%%%%%%%%%%%%%%%%%%%%%%%%%%%%%%
\subsection{Body-centered cubic lattice}
\label{sec:bcc}

In this section we focus on the 
three-dimensional body-centered cubic (BCC) lattice with
nearest and next-nearest neighbor hopping, $t$ and $t'$, respectively.
The dispersion is given by
\begin{align}
  \label{eq:3dbccDisp}
  t_{\k}=&-8t\cos(k_x)\cos(k_y)\cos(k_z) 
  \\ & \notag
  -2 t' \big( \cos(2 k_x)+\cos(2 k_y)+\cos(2 k_z)\big)
  ,
\end{align}
where the half bandwidth is $W=8t$ as long as $|t'|\leq \frac 43 |t|$, 
which is always the case in the following. 
As for the SC lattice,  the BCC lattice
is bipartite and exhibits perfect nesting
for vanishing next-nearest neighbor hopping, $t'=0$.
The AFM nesting vectors $\gv{Q}$
satisfy $t_{\k+\gv{Q}}=-t_{\k}$
and are of the type $\gv{Q}=\pi(1,0,0)^T$ and $\gv{Q}=\pi(1,1,1)^T$.

\begin{figure}
  \includegraphics[width=0.85\linewidth]{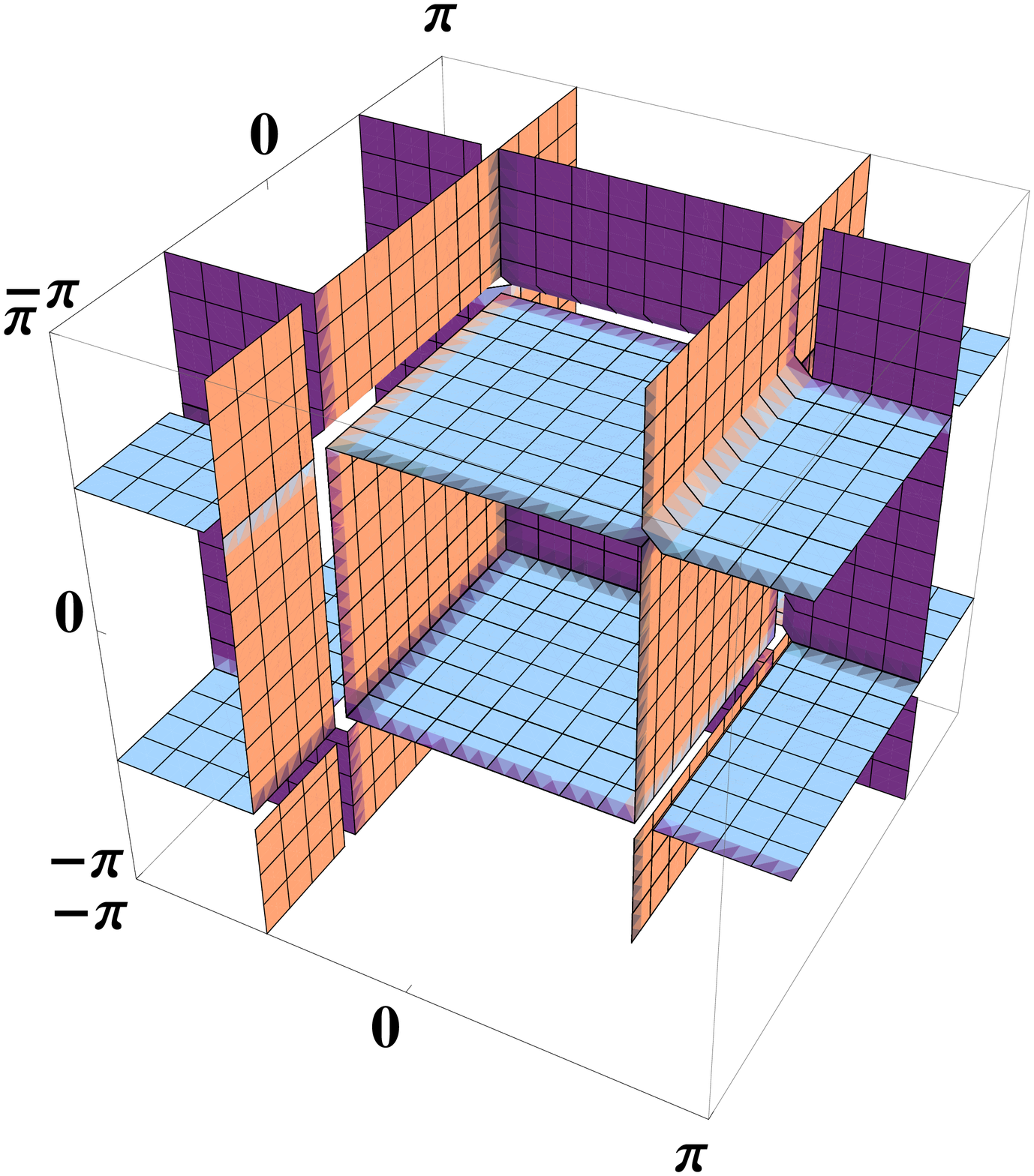} 
  \includegraphics[width=0.85\linewidth]{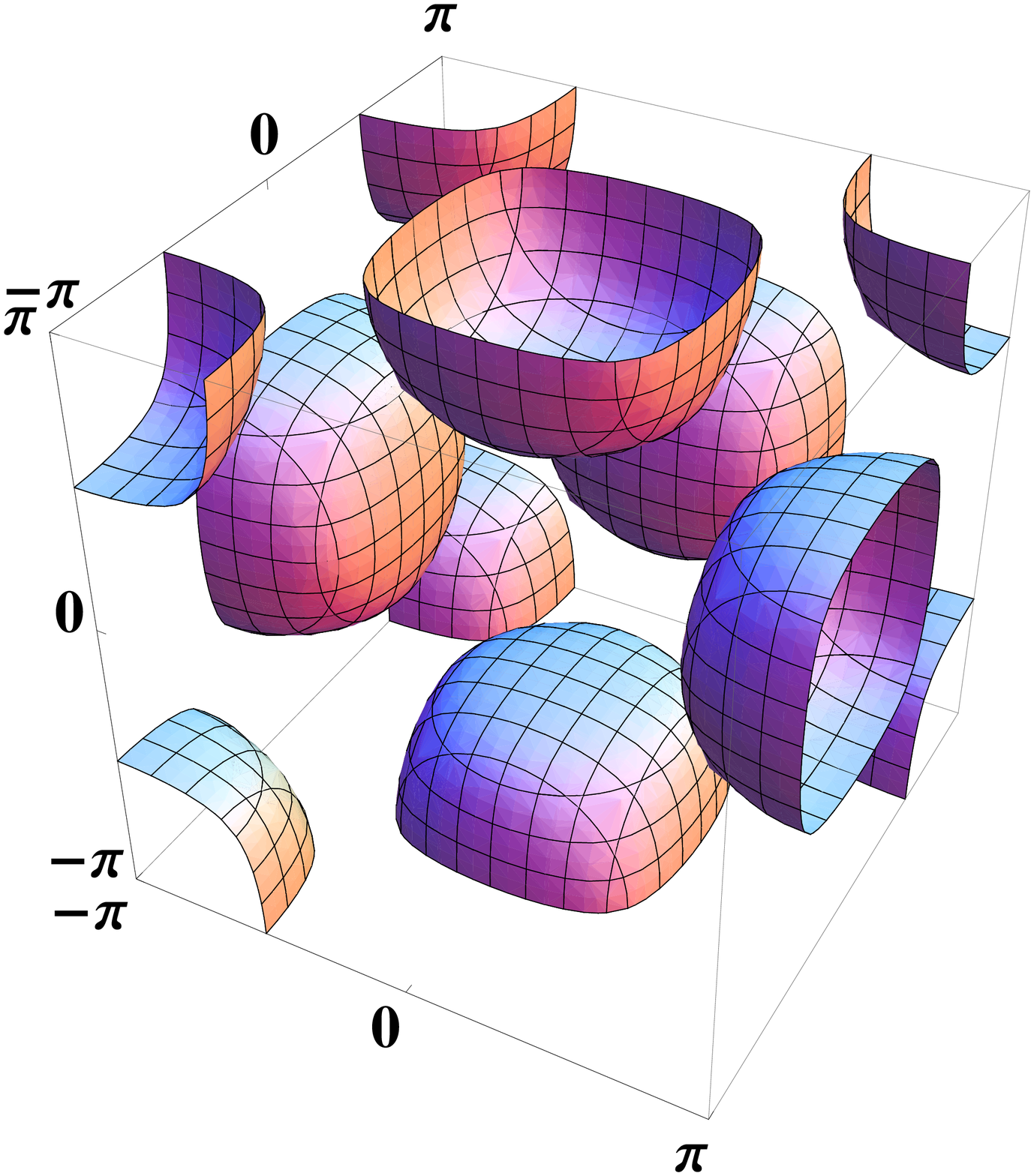}
  \caption{(Color online) 
    Noninteracting Fermi surface for the BCC lattice for (a) $t'=0$ and (b) $t'=-0.2t$  at
    half-filling. (In the front quadrant some parts are omitted in order to get a better view.)  
    \label{fig:bbcFS}
  }
\end{figure}

The perfect nesting property is illustrated in Fig.~\ref{fig:bbcFS}, where the Fermi surface 
of the noninteracting system at half-filling is shown for two values
of the next-nearest neighbor hopping, $t'=0$ and $t'=-0.2t$. 
The flat and parallel sections of the
Fermi surface are clearly visible for $t'=0$  in panel (a).
For finite $t'$ (see panel (b)) the Fermi surface acquires a substantial 
curvature so that the nesting 
vector no longer connects large parts of the Fermi surface. 

\begin{figure}
  \includegraphics[width=1\linewidth]{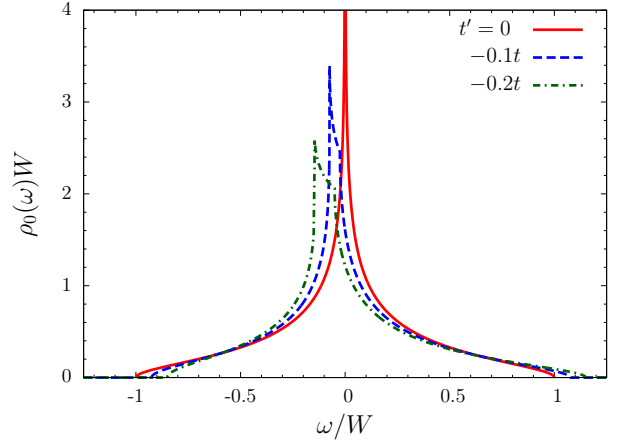}
  \caption{(Color online) 
    Noninteracting DOS for the BCC lattice at half-filling and 
    three different next-nearest neighbor hoppings, $t'/t=0$, $-0.1$, and $-0.2$.  
    \label{fig:bccRho0}
  }
\end{figure}

The noninteracting DOS as shown in Fig.~\ref{fig:bccRho0}
displays the characteristic van-Hove singularities due to the extrema 
and saddle points of the dispersion relation $t_\k$ in the Brillouin zone.
Most prominent is the 
logarithmic divergence near the Fermi level due to the maxima at
$\k=\frac{\pi}{2}(1,0,0)^T$.
Such a van-Hove singularity can induce profound changes in the low temperature, 
low-energy Fermi liquid properties, even in DMFT.\cite{schmittNFLvHove10}
But here we do not study this possibility 
but focus on magnetic transitions which generally occur at higher temperatures.

The reason we are considering the BCC lattice instead of the more 
common SC lattice is found in the more pronounced 
influence of the frustrating next-nearest neighbor hopping $t'$
in the former lattice. 
The shape of the Fermi surface changes much more strongly with increasing $t'$ for the BCC
lattice as, for example, in the SC lattice. 
In particular, results pointing to a competition between local-moment 
and quasiparticle magnetism 
are found  to be more pronounced for the BCC lattice than for the SC case, 
even though we also observed them in the latter case as well.

\begin{figure}
  \includegraphics[width=1\linewidth]{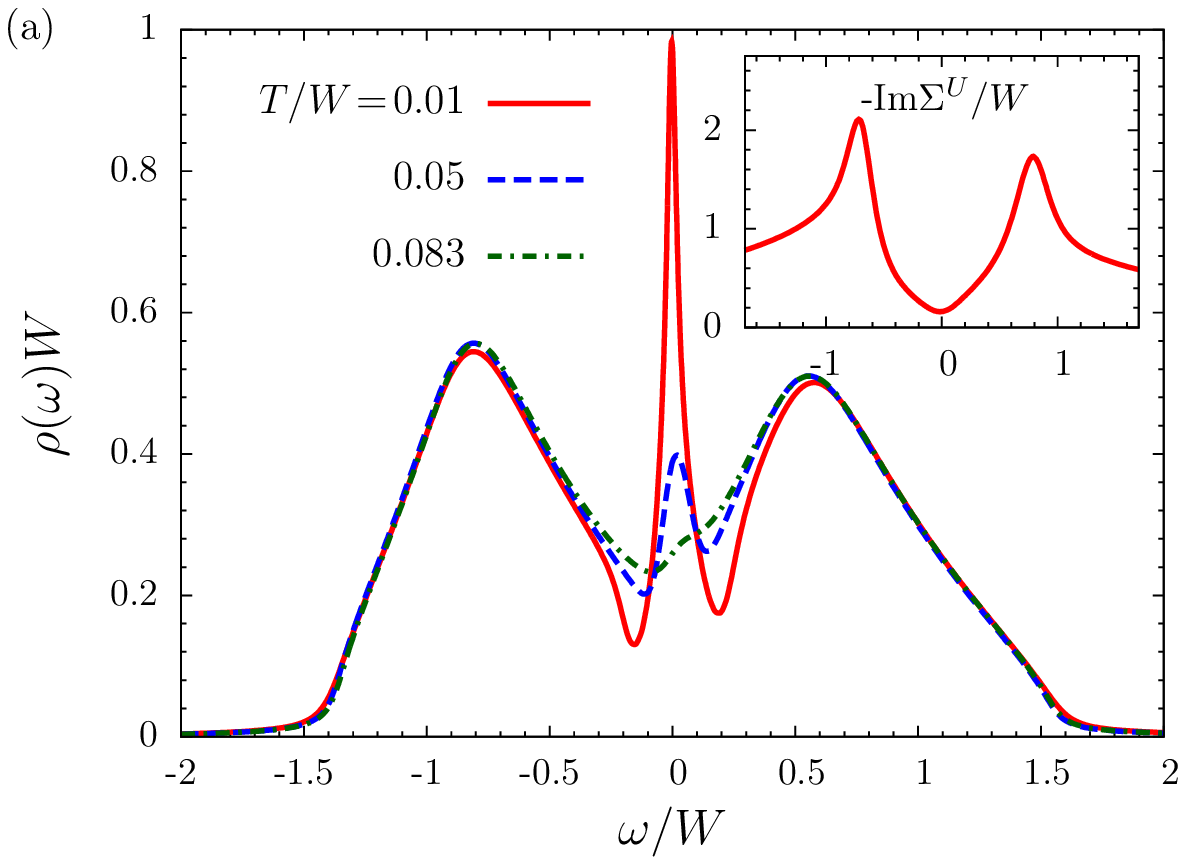}
  \includegraphics[width=1.1\linewidth]{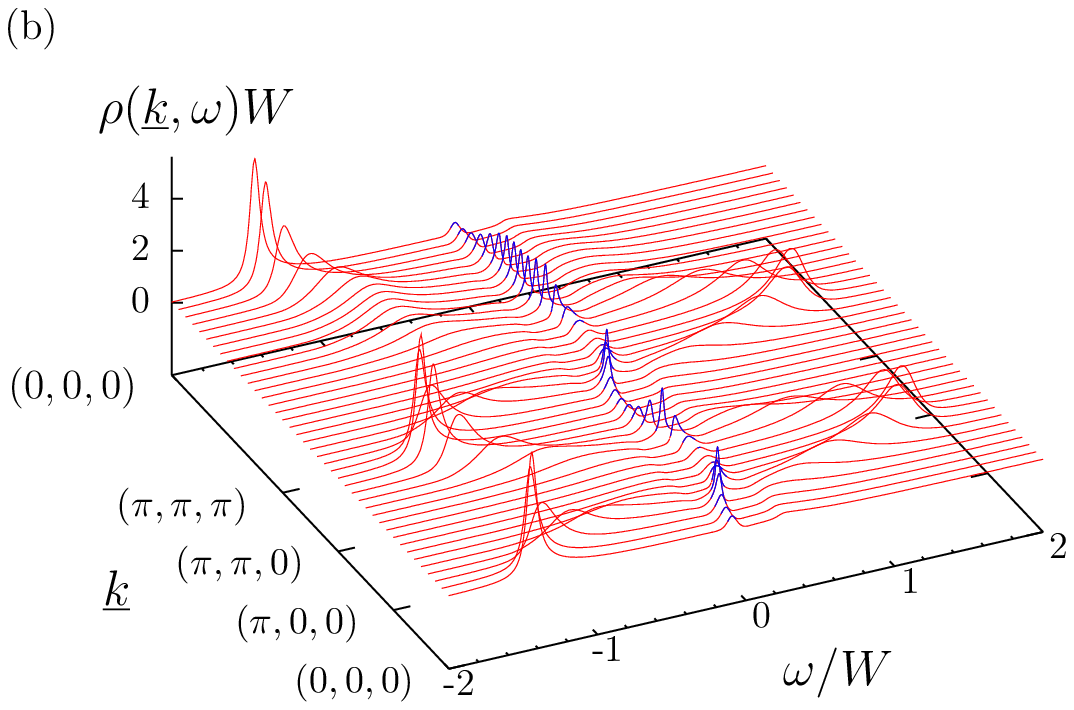}
 \caption{(Color online) (a) Local spectral function $\rho(\omega)$ obtained with the DMFT for the BCC lattice for three different 
   temperatures. The curves are calculated for half-filling with a Coulomb interaction $U/W=1.375$ and a 
   finite next-nearest neighbor hopping $t'/t=-0.2$ is used.  
   The inset shows the (negative) imaginary part of the self-energy -Im$\Sigma(\omega+i \delta)$ for the lowest temperature
   $T/W=0.01$. (b) The momentum dependent lattice spectral function as function of $\k$ along high symmetry directions in 
   the Brillouin zone for the same parameters as in (a) and $T/W=0.01$. 
\label{fig:bccRhoDMFT}
  }
\end{figure}
A DMFT-spectral function for the BCC lattice with finite next-nearest neighbor hopping 
is shown in Fig.~\ref{fig:bccRhoDMFT}.
As a consequence of the rather large Coulomb interaction the lower and upper Hubbard band 
are clearly visible.
Upon lowering the temperature a quasiparticle peak emerges at the Fermi level 
indicating the formation of a correlated metallic state. 
The inset shows the imaginary part of the self-energy. 
The quadratic minimum  at the Fermi level reveals this state to be a Fermi 
liquid as it is expected within DMFT.\cite{pruschke:dmftNCA_HM95,*georges:dmft96,*georgesDMFTreview04,muellerHartmannDInfty89-2,schmittNFLvHove10}
The momentum resolved spectral function $\rho(\k,\omega)$ shown in panel (b)
exhibits the dispersive low-energy quasiparticle band around the Fermi
level.
   
The formation of this low-temperature Fermi liquid is associated with a 
screening of local magnetic moments due to Kondo-correlations.
\begin{figure}
 \includegraphics[width=0.9\linewidth]{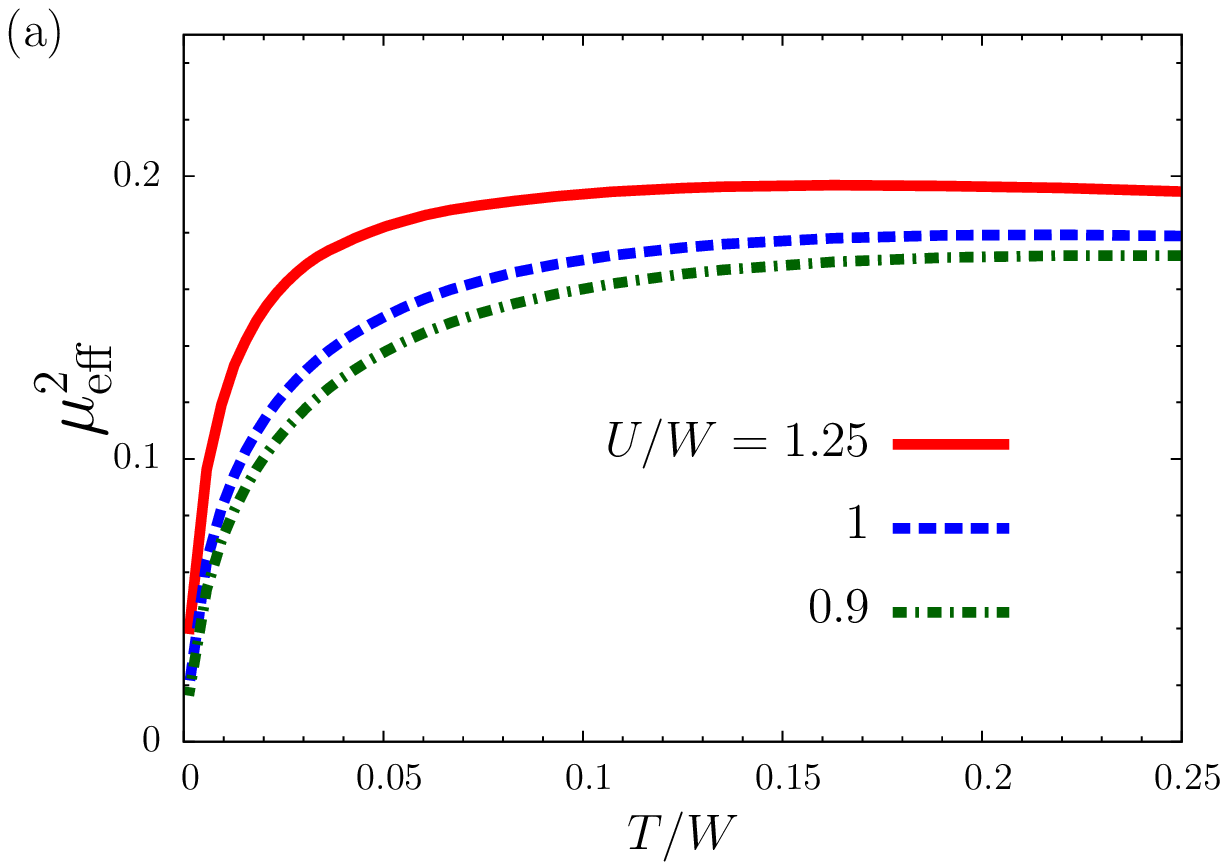}
 \includegraphics[width=0.9\linewidth]{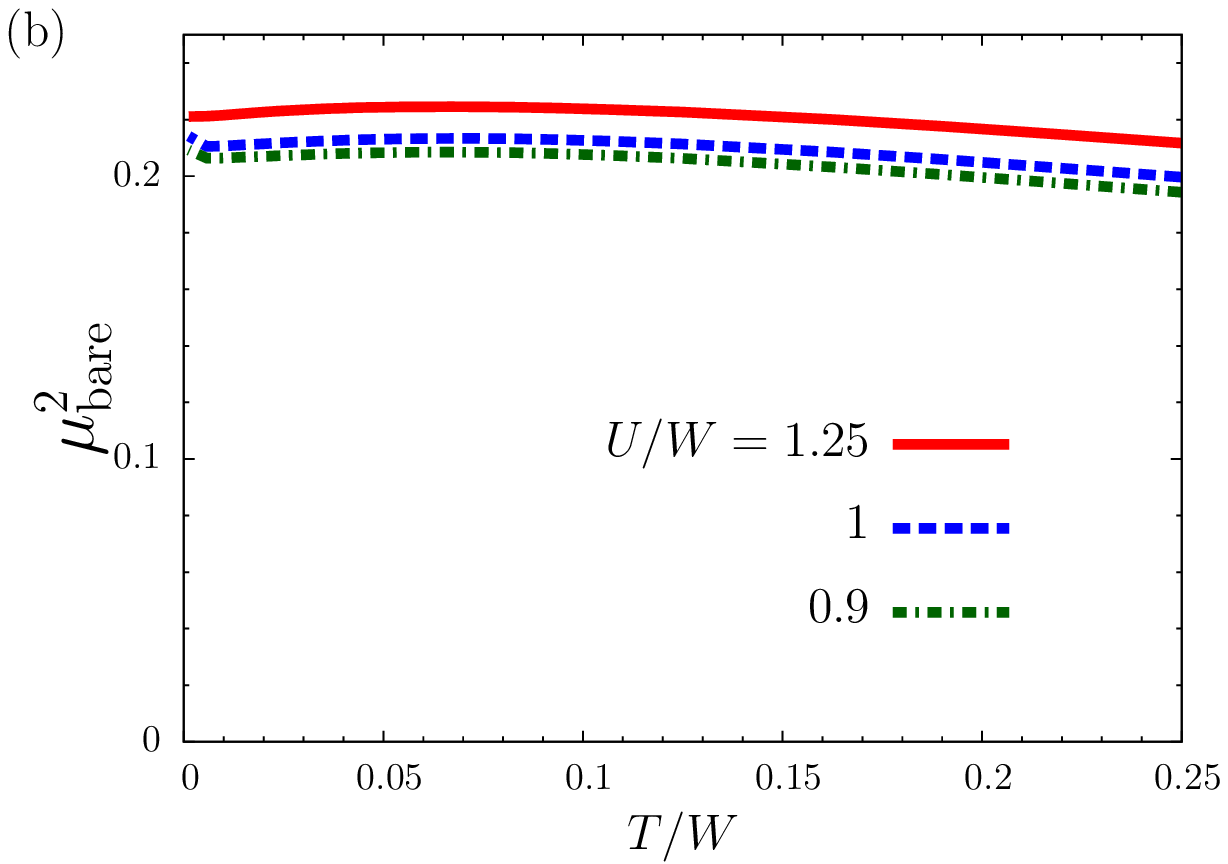}
 \caption{(Color online) The screened (a) and unscreened (b) local moment,  $\mu_{\text{eff}}^2$
   and   $\mu_{\text{bare}}^2$, respectively,  as function of temperature 
   for $t'/t=-0.15$ and 
   three different values of the Coulomb interaction. 
   % Other parameters are as in Fig.~\ref{fig:bccRhoDMFT}.
   \label{fig:bccMuEff}
 }
\end{figure}
Figure \ref{fig:bccMuEff}(a) displays the screened local moment,
\begin{align}
  \mu_{\text{eff}}^2=T\chi^{\mathrm{loc}}_{\mathrm{mag}}(T),
\end{align}
as function of temperature for three  
different values of the Coulomb interactions. Due to the Coulomb interaction the moments 
at high temperatures are larger than
the induced moment $\mu_{\text{eff},U=0}^2=\frac 18$ of the noninteracting 
band electrons (where empty and doubly occupied local states fully contribute), 
and increases with $U$ toward the value $\mu_{\text{eff},\mathrm{spin}}^2=\frac 14$
of a free spin.
At temperatures of the order of the characteristic low-energy scale $T^*=WZ$, where
$Z$ is the quasiparticle weight of Eq.~\eqref{eq:QPZ}, 
the moments decrease due to the buildup of Kondo-correlations. 

In contrast, the unscreened moment
\begin{align}
  \mu_{\text{bare}}^2=\frac 14 \big[\ew{\hat n_\uparrow}+\ew{\hat n_\downarrow}-2\ew{\hat n_\uparrow\hat n_\downarrow}\big]
\end{align}
which is essentially 
determined by the double occupancy
is almost independent of  temperature, see Fig.~\ref{fig:bccMuEff}(b). 
Upon lowering the temperature it slightly increases due to the 
decrease in the thermally induced double occupancy. The formation of the 
Fermi liquid quasiparticle band at low temperatures then leads to a slight
increase of the double occupancy\cite{georges:halfFilledHMInfDim93,dare:HM00}
and consequently the moment decreases again.

\begin{figure}
  \includegraphics[width=0.9\linewidth]{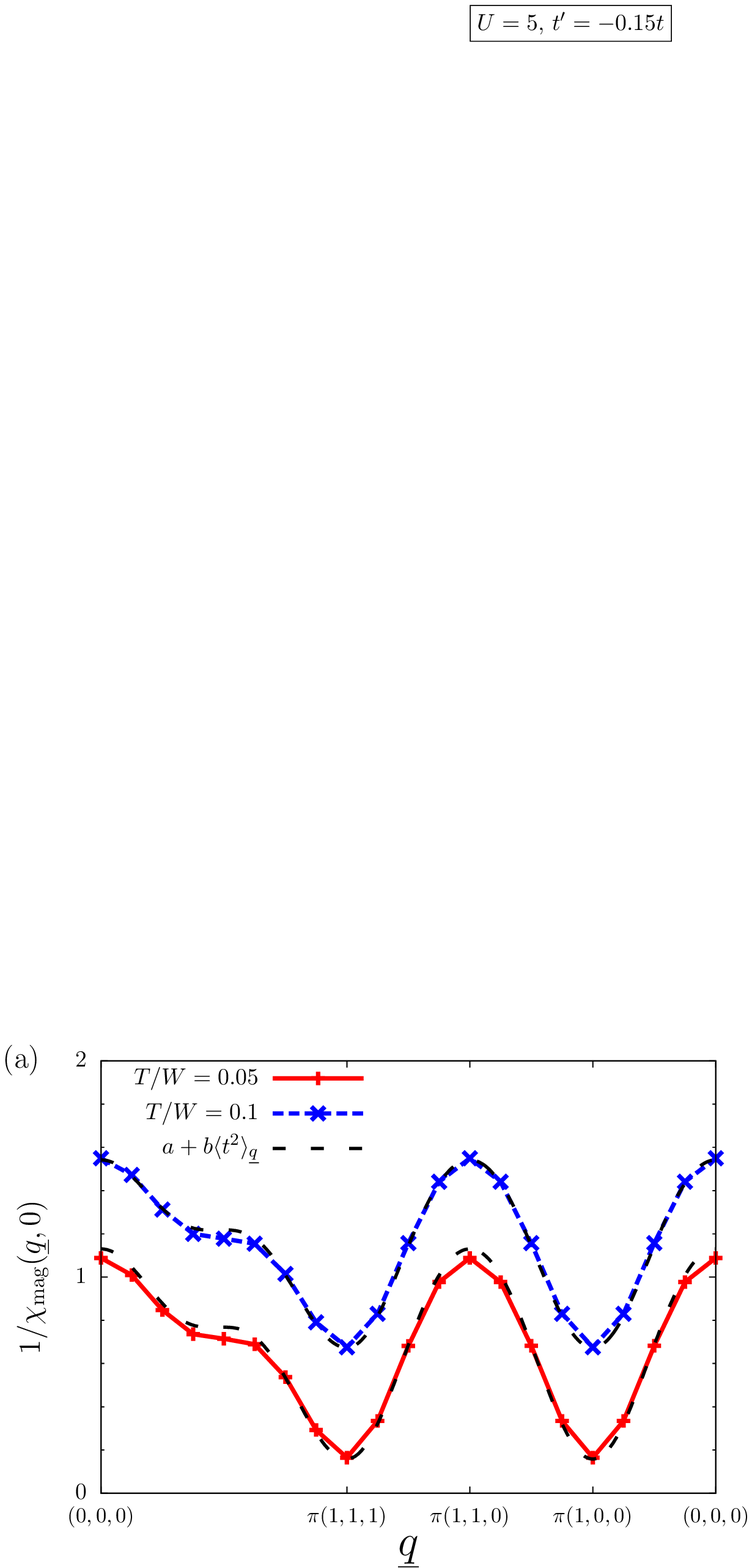}
  \includegraphics[width=0.9\linewidth]{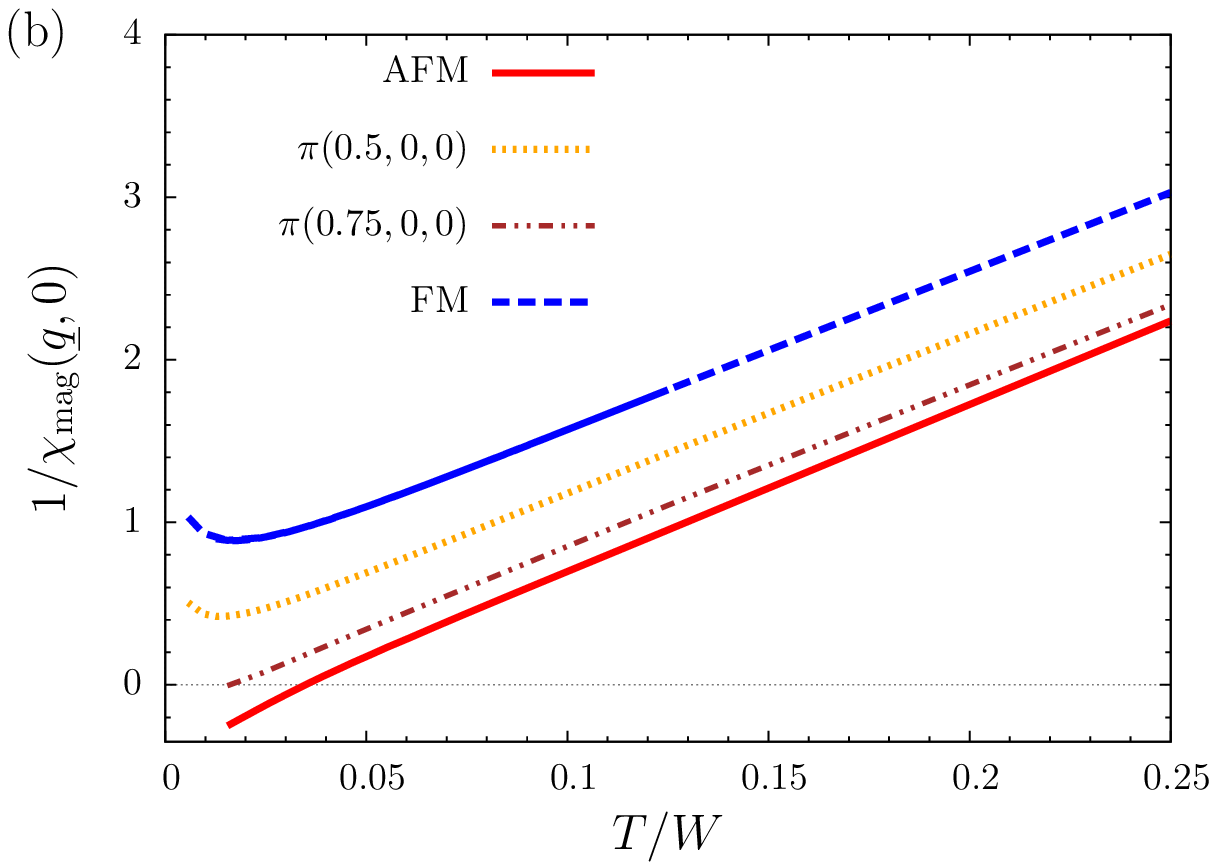}
  \caption{(Color online) (a)  The inverse magnetic susceptibility for
    different  temperatures  as function 
    of the wave vector along with fits of the form $a+b\ew{t^2}_\q$,
    see Eq.~\eqref{eq:chiFit}. 
    (b) The inverse magnetic susceptibility for different wave vectors 
    as function of temperature.
    The curves are calculated 
    for half-filling with a Coulomb interaction $U/W=1.25$ and a 
    next-nearest neighbor hopping of $t'/t=-0.15$.   }
  \label{fig:ChiBCCU5}
\end{figure}
AFM should prevail for not too 
large next-nearest neighbor hopping $t'$
due to the near-nesting property of the Fermi surface.
Figure \ref{fig:ChiBCCU5}(a) displays the inverse magnetic 
susceptibility for $U/W=1.25$ and $t'/t=-0.15$ 
as function of the wave vector along a path through the Brillouin 
zone. As in the previous section for the SC lattice, the AFM components
($\gv{Q}=\pi(1,0,0)^T$ and $\gv{Q}=\pi(1,1,1)^T$) are enhanced. Additionally, 
fits with  the approximate function of Eq.~\eqref{eq:chiFit}   
agree very well with the susceptibility, supporting the view
developed in the previous section. The temperature dependence
of the susceptibility is depicted in  Fig.~\ref{fig:ChiBCCU5}(b).
The AFM component exhibits a Curie-Weiss behavior
$\frac 1 {T-T_\mathrm{N}}$ with a divergence at $T_\mathrm{N}/W\approx 0.035$,
which supports an interpretation in terms of local-moment magnetism.

\begin{figure}
  \includegraphics[width=0.9\linewidth]{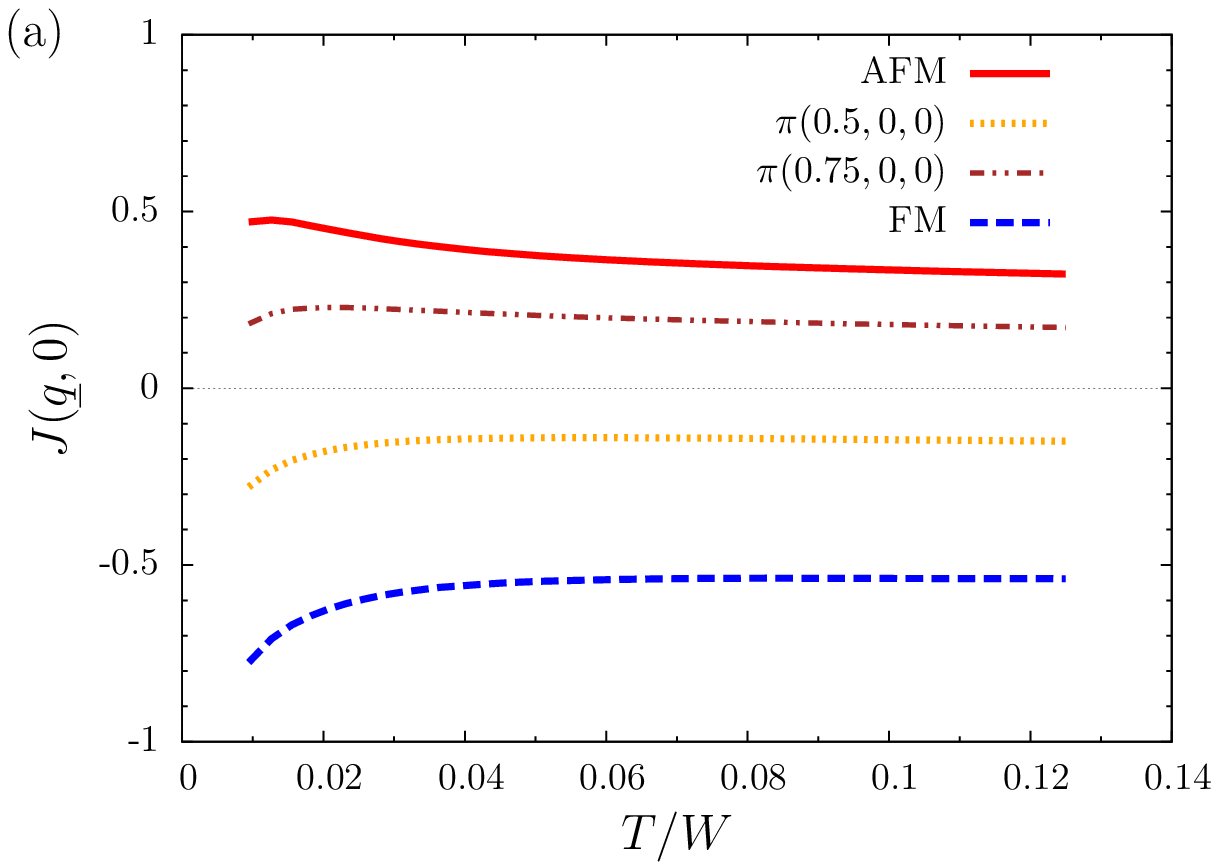}
  \includegraphics[width=0.9\linewidth]{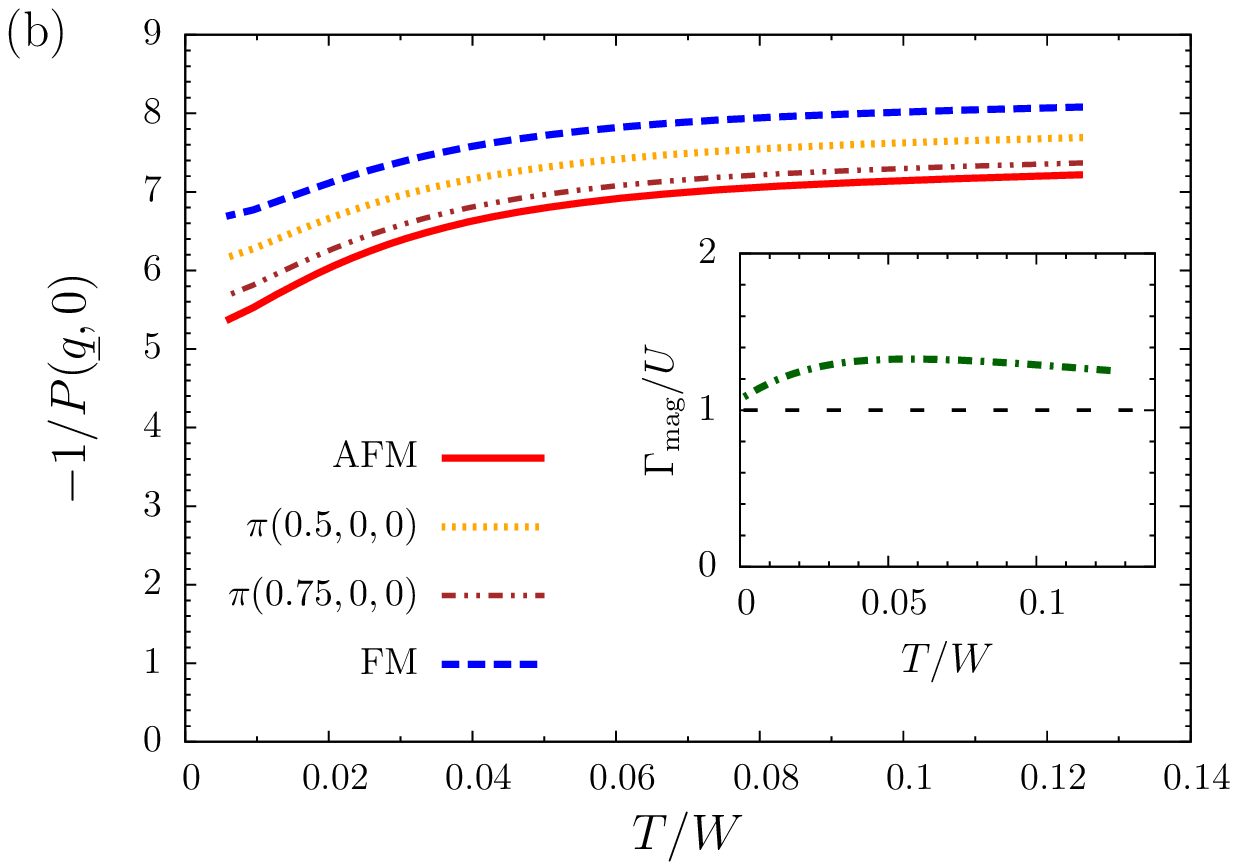}
  \caption{(Color online) (a) The effective static exchange coupling $J(\q,\nu=0)$
    of Eq.~\eqref{eq:effJ} 
    for different wave vectors  as function of temperature for 
    $U/W=1.25$ and $t'/t=-0.15$. (b) 
    The inverse negative static particle-hole propagator  
    for different wave vectors  as function of temperature.
    The inset displays the effective local vertex of Eq.~\eqref{eq:LocVertex}.
  }
  \label{fig:JGammaBCC}
\end{figure}

However, this perspective becomes less obvious when the constituents 
are analyzed in detail. The effective local moment which
enters  Eq.~\eqref{eq:susJq} is temperature dependent
due to screening, see Fig.~\ref{fig:bccMuEff}. Additionally, the 
static exchange coupling $J(\q,\nu=0)$ also displays a temperature 
dependency. This can be observed in Fig.~\ref{fig:JGammaBCC}(a), where
the AFM coupling $J(\gv{Q},0)$  slightly increases toward lower temperatures. 
The increase in $J(\gv{Q},0)$  is compensated by the reduced moment
and together both yield the observed Curie-Weiss behavior characteristic of 
local-moment magnetism.

Figure \ref{fig:JGammaBCC}(b) shows the same data but organized 
in terms of  the 
itinerant picture of magnetism. The ``noninteracting'' susceptibility, i.e.\
the particle-hole propagator calculated without explicit two-particle
interactions but with the full single-particle Green functions, 
increases with decreasing temperature for all wave vectors.
This is a consequence of the accumulation of spectral weight 
near the Fermi level, cf.\ Fig.~\ref{fig:bccRhoDMFT}(a). 
The effective magnetic two-particle vertex $\Gamma_\mathrm{mag}(\nu=0)$
is of the order but slightly larger then the bare interaction $U$ (see inset). 
It slightly decreases for low temperatures indicating the quasiparticle 
excitations to experience somewhat less scattering.

Thus, the buildup of Kondo-like resonances in the lattice 
has two opposing effects, which nearly compensate each other here:
The screening of local moments and the accumulation of additional 
quasiparticle weight near the Fermi level.

\begin{comment}
  The local susceptibility which is also shown in the plot seems to behave like 
  a pure local moment, i.e.\ $\chi^{\mathrm{loc}}_{\mathrm{mag}}\propto \frac 1 T$.
  The very slight upturn at low temperatures due to screening is hardly visible in this
  plot and indicates that screening is not very active for these parameter values.
\end{comment}

\begin{figure}
  \includegraphics[width=0.9\linewidth]{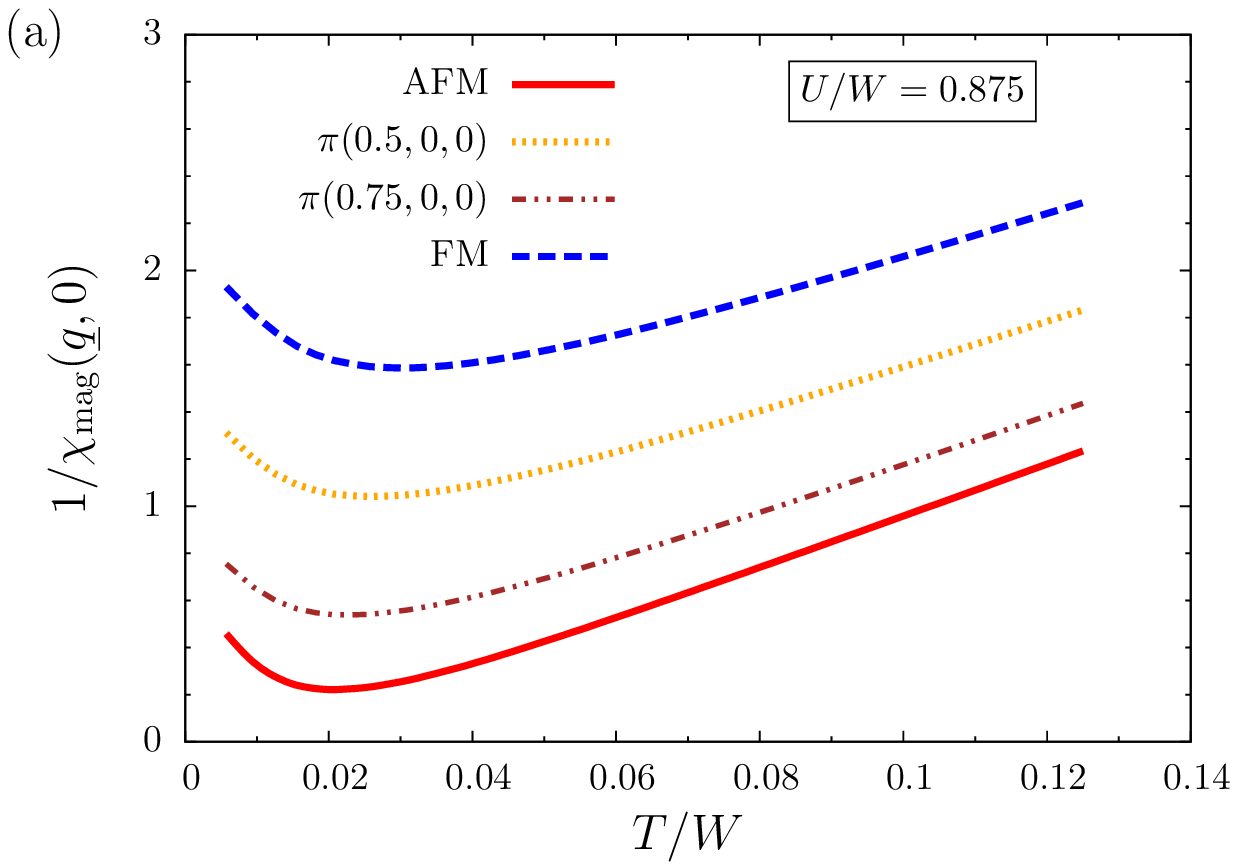}
  \includegraphics[width=0.9\linewidth]{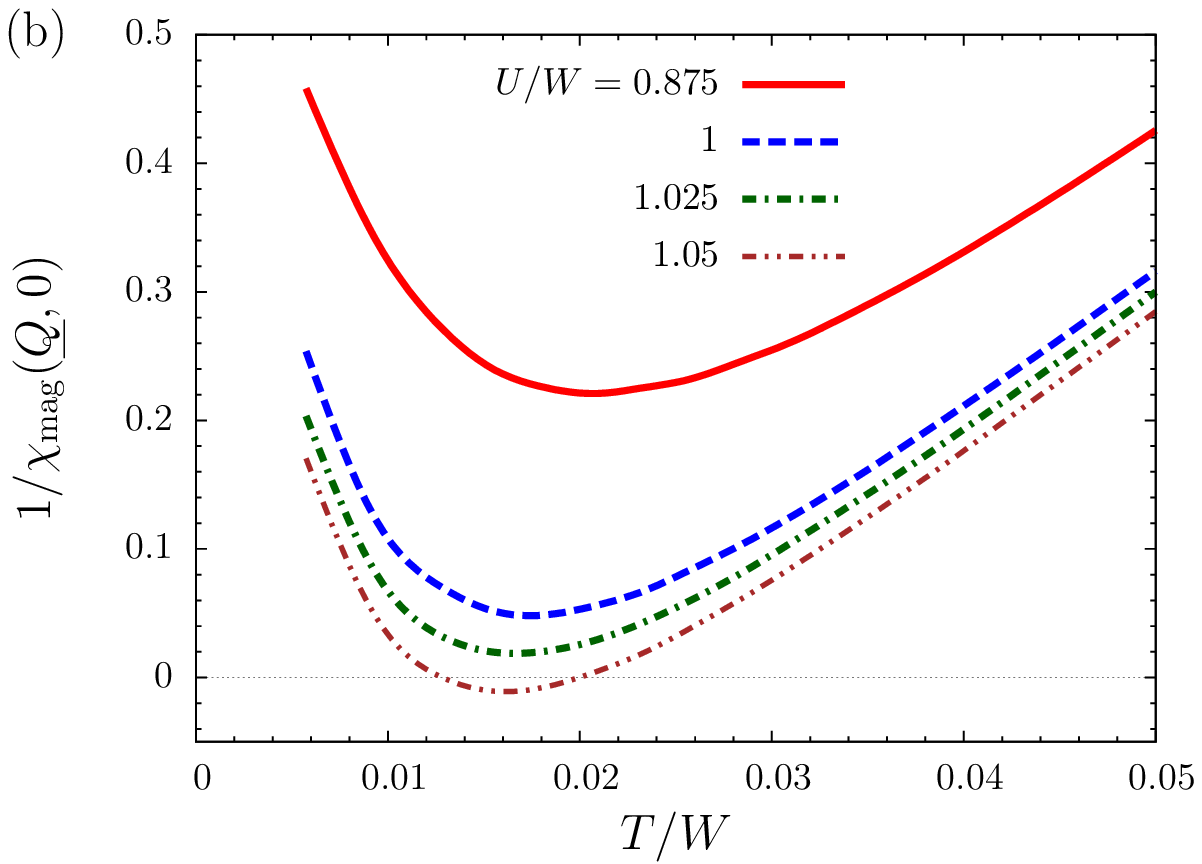}
  \caption{(Color online) (a) The inverse magnetic susceptibility 
    for different wave vectors as function of temperature for  $U/W=0.875$.
    (b) The inverse AFM component of the susceptibility for small temperatures
    and various values of $U$. In both plots is $t'/t=-0.17$.
  }
  \label{fig:ChiBCCtp085Urun}
\end{figure}
The balance between itinerant and localized contributions to the susceptibility 
is rather fragile.  
Already for moderate  changes of  parameter values the Curie-Weiss behavior of the 
susceptibility is lost. This is the case for a slightly larger next-nearest neighbor 
hopping of $t'/t=-0.17$ and $U/W=0.875$ as shown in Fig.~\ref{fig:ChiBCCtp085Urun}(a). 
AFM is still dominant but no magnetic transition occurs. All components of the 
susceptibility exhibit a maximum (minimum in inverse susceptibility) and decrease 
toward lower temperatures. The decrease occurs at a temperature of the order of the 
low-energy scale indicating that the coherent quasiparticles 
do not favor any tendency toward magnetic order.

This maximum in the susceptibility opens up the possibility of an unusual
phase-transition scenario. 
Increasing the Coulomb repulsion
enhances the AFM susceptibility as can be seen in Fig.~\ref{fig:ChiBCCtp085Urun}(b).
For large enough $U$, the minimum in $1/{\chi_{\mathrm{mag}}(\gv{Q},0)}$ even reaches  zero
indicating a divergent susceptibility at a finite temperature 
($T_\mathrm{N}/W\approx 0.016$ for the case shown in the graph with 
a Coulomb repulsion between $U/W=1.025$ and  $U/W=1.05$).

Fixing the temperature at some value and increasing $U$ can lead to
the peculiar situation that the system stays paramagnetic even though
at a \textit{higher} temperature it would undergo an AFM
phase-transition. In that regime, there exists
a finite temperature interval $T_\mathrm{N,1}\leq T\leq T_\mathrm{N,2}$
in which  AFM order is expected  to occur, while outside this interval 
the system stays paramagnetic. 
%For decreasing temperature and  a fixed value of $U$, 
%the system would  first encounter an AFM transition at 
%$T_\mathrm{N,2}$ and at the lower $T_\mathrm{N,1}$ re-enter the 
%paramagnetic state.
Unusually, for $T<T_{\mathrm{N},1}$ the system can be 
driven into the  AFM ordered phase by heating it.

It should be noted, that similar re-entrant behavior 
was already found for the AFM transition
in the Hubbard model within a weak-coupling  
treatment for an infinite-
and two- dimensional lattice,\cite{halvorsenRentranftAFMHM94,hongReentrantHM96}
as well as within an equation of motion technique.\cite{manciniHMEQM04}
In these cases, the frustration was brought about by  
finite hole-doping, instead of a finite 
next-nearest neighbor hopping implemented here.
The extended Hubbard model with an additional nearest-neighbor
Coulomb interaction also displays a re-entrant behavior
for the charge-ordering  transition.\cite{pietigReentrantChargeOrderHM99}

Another peculiarity feature is connected to the critical exponent.
The theory we employed has mean-field 
character regarding spatial correlations. Thus, we expect
a mean-field critical exponent $\gamma=1$ for the nonlocal 
magnetic susceptibility,  $\chi(\gv{Q},T)\sim (T-T_C)^{-\gamma}$.
This exponent is indeed always found, except for that  value of $U$ where
the minimum in $\chi^{-1}$ touches the zero-axis. There, the exponent is 
is larger and of the order  $\gamma= 2$ as this minimum can be
fit with a parabola.

The re-entrant behavior is a result of
the competition between the ordering  of local moments and the 
formation of low-temperature quasiparticles, i.e.\
the competition between local moment and itinerant quasiparticle magnetism.
The exact shape of the Fermi surface has
a strong influence on the quasiparticle scattering which determines the 
particle-hole propagator.
% and therefore generates the effective exchange coupling $J$. 
At low temperatures, the quasiparticles
are fully formed and the frustration induced by $t'$ leads to a 
suppression of the particle-hole propagator, essentially eliminating 
magnetism (similar to  what happens at weak coupling).

\begin{figure}
  \includegraphics[width=0.9\linewidth]{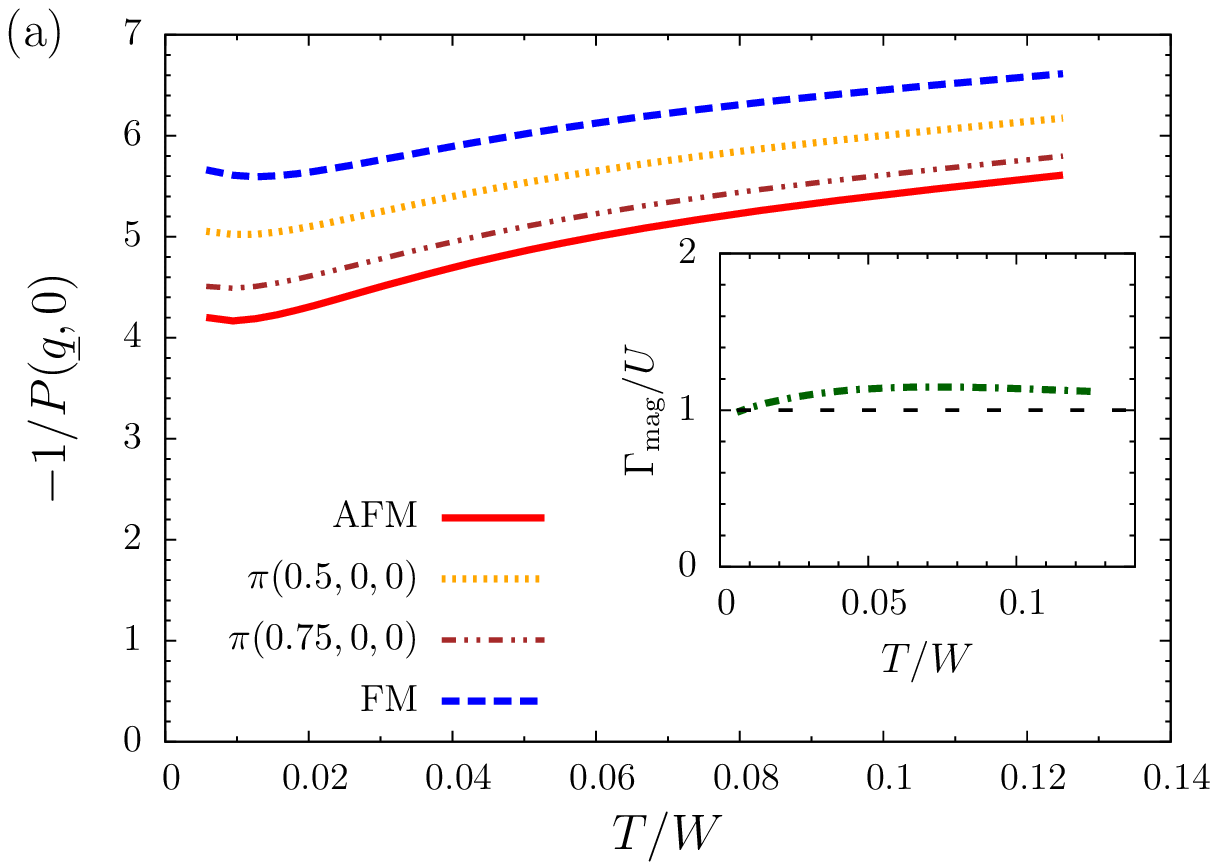}
  \includegraphics[width=0.9\linewidth]{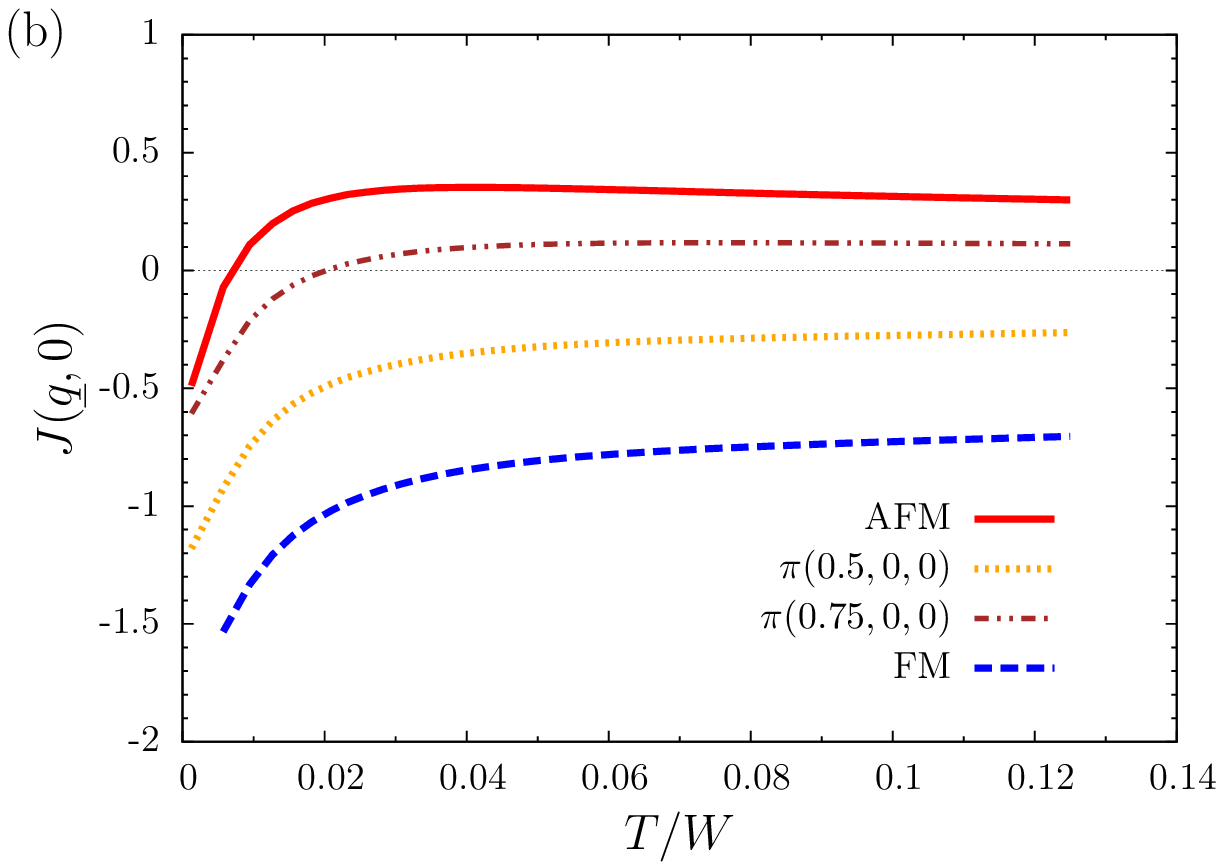}
  \caption{(Color online)
    (a) The inverse negative static particle-hole propagator  
    for different wave vectors  as function of temperature.
    The inset displays the effective local vertex of Eq.~\eqref{eq:LocVertex}.
    (b) The effective static exchange coupling $J(\q,\nu=0)$
    of Eq.~\eqref{eq:effJ} 
    for different wave vectors  as function of temperature. for 
    In both plots $U/W=1$ and $t'/t=-0.17$ is used. 
  }
  \label{fig:BubU4t085}
\end{figure}

This can be observed in Fig.~\ref{fig:BubU4t085}(a),
where the inverse particle-hole propagator shows a saturation and 
even a slight upturn toward low temperatures
(compare Fig.~\ref{fig:JGammaBCC}(b) where this is not the case),
even though the system is metallic with large spectral weight at the 
Fermi level, cf.\ Fig.~\ref{fig:bccRhoDMFT}(a).
The effective vertex (see inset) is not sensitive to the Fermi surface and consequently 
displays a qualitatively similar behavior to the case shown
in Fig.~\ref{fig:JGammaBCC}(b).

A local moment  perspective can be employed above $T/W\approx 0.02$
where the effective exchange coupling $J$ 
is almost  temperature independent  (see Fig.~\ref{fig:BubU4t085}(b))
and very similar to the one shown in Fig.~\ref{fig:JGammaBCC}(a).
Below this temperature, however, coherent  quasiparticles 
emerge and the coupling changes dramatically.
The AFM component even changes sign and becomes ferromagnetic.

\begin{figure}
  \includegraphics[width=0.9\linewidth]{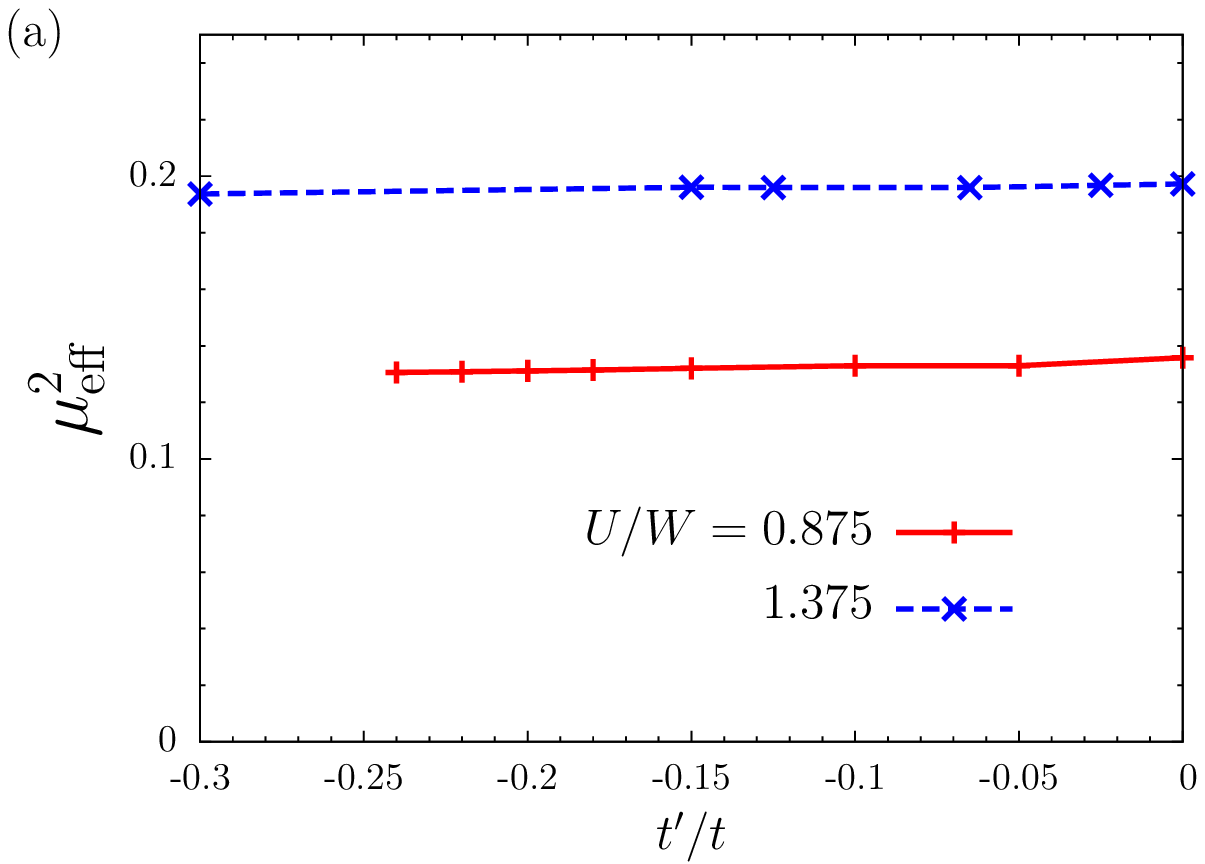}
  \includegraphics[width=0.9\linewidth]{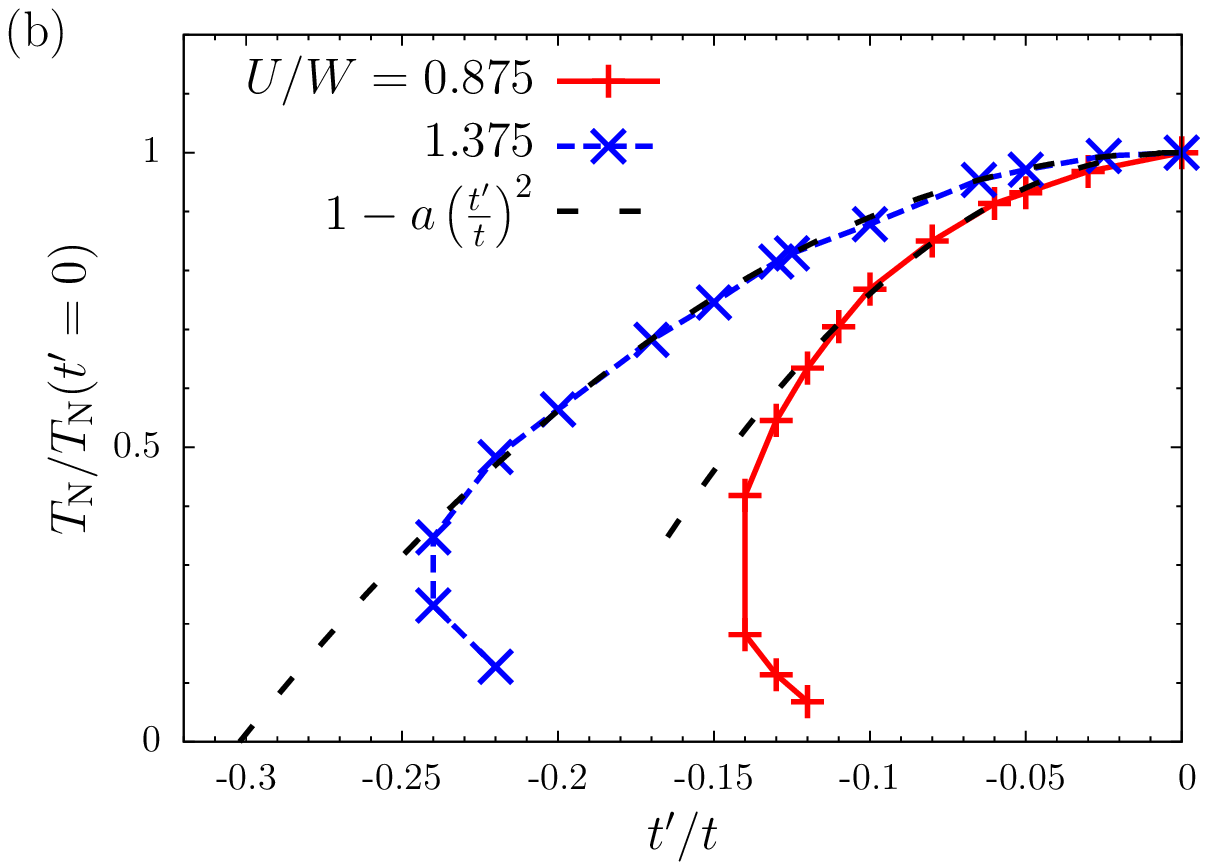}
  \caption{(Color online) (a) The effective local moment
    as function of the next-nearest neighbor hopping $t'$
    for a fixed temperature $T/W=0.046$ and two values of the Coulomb repulsion. 
    (b) The N\'eel temperature as function of $t'$
    normalized to its value for $t'=0$ for two values of $U$. 
    Also shown are fits with a quadratic decrease,
    where the fit parameters are $a=24$ and $a=10.95$ for $U/W=0.875$ and $U/W=1.375$, respectively.   }
  \label{fig:TC_3dBCC_tpRun}
\end{figure}
Whether the local moments  order at higher temperature 
is determined by their
size and the geometry of the lattice. Both aspects are much less sensitive to 
the frustration generated by a finite $t'$ than the magnetic quasiparticle scattering.
The value of the screened effective moments is not influenced much 
by $t'$ as can be seen in Fig.~\ref{fig:TC_3dBCC_tpRun}(a). 
$\mu_\mathrm{eff}^2$ is shown for a fixed temperature 
as function of $t'$ for two different values of $U$. 
The  decrease with increasing $t'$ is barely visible 
confirming the screening to be independent of the details 
of the Fermi surface or the shape of the spectral functions
(as long as the system is a metal, of course).

In order to get an estimate for the $t'$ dependence of the 
N\'eel temperature from the local moment perspective, it is 
instructive to start  from the atomic limit. The effective exchange
coupling  can be estimated in second order
perturbation theory in the hopping,
\begin{align}
J_{\mathrm{eff}}(\q)&=- \frac{8 t^2}{U} \cos(q_x)\cos(q_y)\cos(q_z) 
\\ & \notag
- \frac{2 {t'}^2}U \Big[ \cos(2 q_x)+\cos(2 q_y)+\cos(2 q_z)\Big]
,
\end{align}
which yields the effective AFM coupling for $\q=\gv{Q}$,
\begin{align}
\label{eq:JeffBccAFM}
% J_{\mathrm{eff}}^{\mathrm{AFM}}&= \frac{8 t^2}{U} - \frac{6 {t'}^2}U 
J_{\mathrm{eff}}^{\mathrm{AFM}}&= \frac{8 t^2}{U} \Big(1- \frac 34 \frac{{t'}^2}{t^2}\Big) 
\equiv J_{t'=0}^{\mathrm{AFM}}\Big(1- \frac 34 \frac{{t'}^2}{t^2}\Big) 
.
\end{align}
This coupling directly determines the AFM transition temperature as it was
discusses around Eq.~\eqref{eq:LocMomTc}. 
The Hubbard model for the parameter values under consideration 
is far from the atomic limit as it is a metal and itinerant 
electronic excitations are  present. 
However, for the local moment picture to be 
applicable the N\'eel temperature should follow the quadratic
decrease with $t'$ of Eq.~\eqref{eq:JeffBccAFM}, but renormalized 
pre-factors would be expected. Figure \ref{fig:TC_3dBCC_tpRun}(b) displays 
$T_\mathrm{N}$ as function of $t'$ for two 
different values of $U$. The transition temperature indeed decreases in accordance 
with that expectation for small to moderate values of $|t'/t|$.
The rather large values for the fit parameter $a$ compared to $3/4$ of Eq.~\eqref{eq:JeffBccAFM} indicate 
strong renormalizations due to the formation of heavy quasiparticles.
At too large $t'$, the N\'eel temperature comes in regions where
the forming quasiparticles begin to dominate and magnetism is suppressed
quite abruptly.  For smaller values of $U$, the local moments
are generally less pronounced and consequently this suppression occurs 
faster.  

The N\'eel temperature does not
go to zero continuously, but rather remains finite at the
maximal $t'$ where a transition occurs. In the vicinity of this
value the re-entrant behavior can be observed, and the system 
tends to exhibit AFM only in a finite temperature interval.

\begin{figure}
  \includegraphics[width=\linewidth]{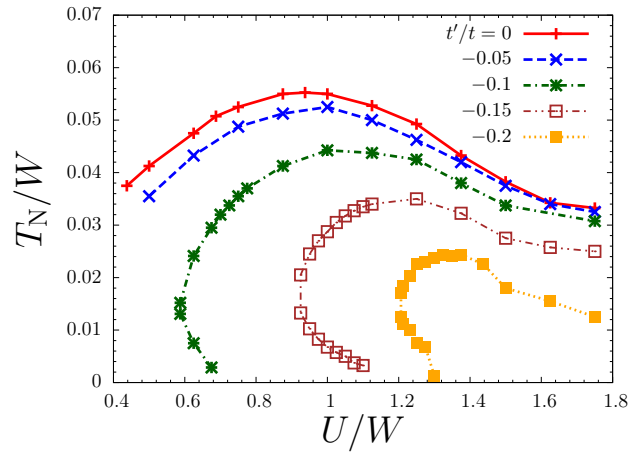}
  \caption{(Color online) The N\'eel temperature of the half-filled 
    Hubbard model on BCC lattice as function of $U$ for various $t'$.  
  }
  \label{fig:TnBCCSC}
\end{figure}
The N\'eel temperature as function of 
the Coulomb repulsion $U$ for various values of the next-nearest neighbor hopping $t'$
is shown in Fig.~\ref{fig:TnBCCSC}.
With increasing $t'$ the transition temperature is generally reduced
especially in the weak coupling regime where AFM
is completely eliminated. For finite $t'$, there exists a   
nonzero critical value for the Coulomb  repulsion, $U_\mathrm{C}>0$, only 
above which AFM  can occur.  
Similar behavior is  found for the SC lattice with
finite next-nearest neighbor hopping $t'$.\cite{zitzler:PhaseFrustratedHM04}
The perturbation expansion in $U$ shows that for 
vanishing next-nearest neighbor hopping, $t'=0$,
AFM extends to $U=0$, i.e.\ $U_\mathrm{C}=0$, due to perfect  nesting (see, for example, 
Ref.~\onlinecite[Chap.\ 7]{fazekas:LectureNotes99}). 

For large $U$ AFM is found to be very robust with respect to $t'$ and   
$T_\mathrm{N}$ is only reduced. The dependence of the N\'eel temperature on 
$t'$ is in agreement with the quadratic decrease expected from a 
local-moment picture, and the quadratic coefficient of the $t'$-term 
decreases with increasing $U$,
cf.\ Fig~\ref{fig:TC_3dBCC_tpRun}(b).
In the crossover regime at intermediate $U/W$ 
the re-entrant behavior is observed. For a given $U$, there exist two
critical temperatures, $T_\mathrm{N,2}$ and $T_\mathrm{N,1}$,
between which the system exhibits AFM, while below $T_\mathrm{N,1}$ and 
above $T_\mathrm{N,2}$ paramagnetism prevails.

%%%%%%%%%%%%%%%%%%%%%%%%%%%%%%%%%%%%%%%%%%%%%%%%%%%%%%%
\section{\label{sec:concl} Conclusion and Outlook}
%%%%%%%%%%%%%%%%%%%%%%%%%%%%%%%%%%%%%%%%%%%%%%%%%%%%%%%

We have derived an universal and physically intuitive form of 
the magnetic susceptibility of strongly correlated electron systems 
within DMFT using an additional RPA-like decoupling in the 
Bethe-Salpeter equations. It was thus possible to work completely 
in the real-frequency domain and to write this final expression 
in terms of functions of real variables, which all are obtained 
from an effective impurity model. The static and dynamic magnetic 
response could readily be calculated from there.
 
We analyzed the validity of this form and found the RPA-like 
decoupling to be quantitatively correct whenever short-ranged
particle-hole excitations dominate. This is the case in and 
close to the antiferromagnetic region in phase space of the 
Hubbard model where the N\'eel temperature was 
correctly reproduced (modulo shortcomings 
of the ENCA impurity solver). 

The tendency toward ferromagnetism, on the other hand, 
is overestimated by the approach as a direct consequence of 
the decoupling.  
However, this shortcoming could be circumvented in 
the future by employing a different decoupling strategy,
where averages are taken over  products of a particle-hole propagator 
and a local vertex.  Leading order coherence effects 
during a two-particle interaction should then be retained.

The applicability of the two archetypical pictures of magnetism
was then investigated in case of the Hubbard model on a three-dimensional 
body-centered cubic lattice with next-nearest neighbor hopping.
For small interactions strength the antiferromagnetism  can be 
understood in terms of itinerant quasiparticles. The magnetic
response is determined by particle-hole excitations across 
the Fermi surface. With increasing
next-nearest neighbor hopping $t'$,
the perfect-nesting property of the  Fermi surface
is lost rapidly. As a consequence, AFM is almost
entirely  removed from the weak coupling phase diagram.

At large values of the local Coulomb interaction $U$
the local-moment picture is more 
appropriate, as the antiferromagnetism is rather
robust against geometric frustration induced by $t'$. The expansion around
the atomic limit provides an suggestive form for the $t'$-dependence of
the N\'eel temperature which is in agreement with the DMFT results.

At intermediate $U$  both types of magnetism compete. 
The many-body effects leading to the formation of the quasiparticle bands 
and the screening of local moments are strongly temperature dependent.
This produces an unusual re-entrant behavior. 
Local moments still sizable at elevated temperatures 
order antiferromagnetically at a characteristic temperature $T_{\mathrm{N},2}$.
But at even lower temperatures the coherent quasiparticles are dominating and 
due to the lack of nesting of the Fermi surface, magnetism is 
suppressed below  $T_{\mathrm{N},1}$.

Beyond the scope of magnetism, the interplay of 
localized and itinerant effective pictures is of  
fundamental interest. 
For example, the explanation of the re-entrant Mott-transition found in 
the highly frustrated organic compound 
$\kappa$-(ET)$_2$Cu[N(CN)$_2$]Cl\cite{kagawaRe-entrantOrganic04,*ohashiTriangularHM08,*ohashiTriangularHM08-2} 
is in direct analogy to the mechanism presented here. 
The temperature dependency of the effective picture
and the re-entrant behavior is generated by a 
competition between frustrated  magnetic moments and a
low temperature Fermi liquid.

It is an interesting prospect and needs further investigations 
whether similar temperature dependencies could play a role 
in other strongly correlated systems. 
In the cuprate superconductors,\cite{plakidaHighTCBook10} for example, 
the single-particle excitations are believed
to be of a more itinerant character along the nodal 
direction while of more localized nature
along the anti-nodal.\cite{shenCuprateReview08}
The frustration brought about by the next-nearest neighbor hopping $t'$ 
could play a decisive role, at least for some compounds and parameter
values.

\begin{acknowledgments}
  The authors acknowledge fruitful discussions with E.\ Jakobi, F.~G\"uttge, and F.B.\ Anders. 
  We also thank F.B.\ Anders for providing us with his NRG-code with which the data shown
  in Fig.~\ref{fig:Z3dsc} was calculated.
  We acknowledge financial support from the Deutsche Forschungsgemeinschaft under grant No.\ 
  AN  275/6-2.
\end{acknowledgments}

\appendix

\section{DMFT for single particle Green functions}
\label{app:1Pdmft}
In this section we sketch the DMFT self-consistency which will
turn out to be useful for the formulation of
the two-particle properties in appendix \ref{app:2Pdmft}.
The Dyson-equation 
can be written in terms of the effective local single-particle cumulant Green function
\begin{align}
  \tilde G(z)=\frac{1}{ z-\Sigma^U(z)}
  .
\end{align}
This quantity represent dressed atoms\cite{metzner:cumulantsHM91}
between which single-particle transfers $t_\k$ are possible. 
This view is established within the linked cluster expansion around the
atomic limit utilizing local cumulants\cite{metzner:cumulantsHM91,greweCA108,grewe:LokaleTheorie,schmittPhD08}
and yields the Dyson equation
\begin{align}
  \label{eq:GAtomLatt} 
  G(\k,z) &=\tilde{G}(z)+\tilde{G}(z)\: t_\k\: G(\k,z) 
  .
\end{align}
Solving this equation for the Green function just gives Eq.~\eqref{eq:gfk}.
 
The local Green function is obtained by a Fourier transform
\begin{align}
    G(z)&=\tilde{G}(z)+\tilde{G}(z)\frac1N\sum_\k t_\k\: G(\k,z) \\
  &=\tilde{G}(z)+\tilde{G}(z)\: T(z)\: \tilde G(z) 
  \label{eq:LocGF}
  .
\end{align}
It was used that $\sum t_\k=0$, which holds for all inversion symmetric lattices
and we introduced the local scattering matrix
\begin{align}
  \label{eq:appTdef}
T(z)&=\frac1N\sum_\k t_\k\: G(\k,z)\:t_\k
.
\end{align}

\begin{figure}[t]
  \begin{center}
    \includegraphics[width=\linewidth]{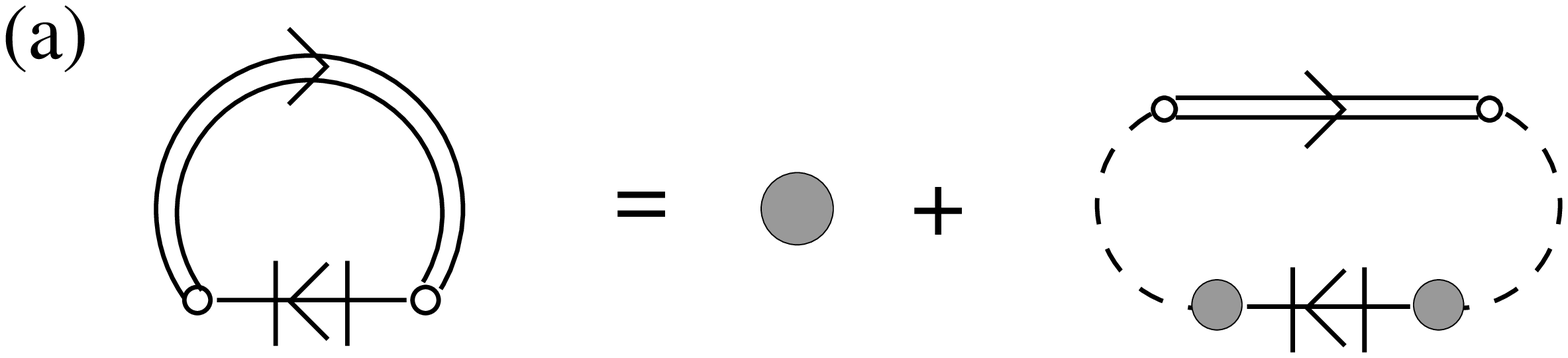}
    \includegraphics[width=\linewidth]{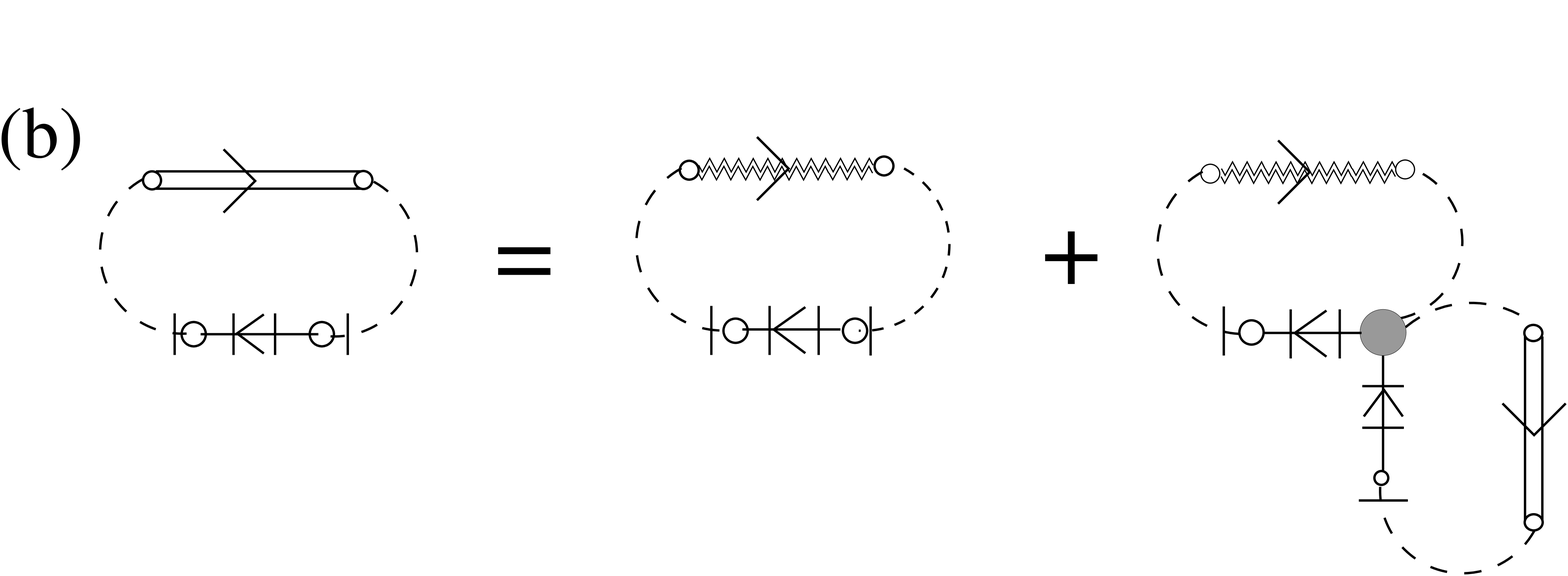}
  \end{center}
  \caption{(a) Dyson equation for the local single-particle Green function in terms of the local
    cumulant $\tilde G$ (grey shaded circles) and the local scattering matrix $T(z)$. 
    The double line with the arrow and the  small circles at the end
    indicate a full Green functions $G(\k,z)$
    and a  dashed line an elementary  hopping $t_\k$.
    (b) The local scattering matrix $T(z)$ in terms of the irreducible loops, i.e.\ 
    the effective medium, $\Gamma(z)$ as in Eq.~\eqref{eq:Tmatrix}. 
    The full scattering matrix is already graphically
    composed of two elementary hoppings and a full propagation, see Eq.~\eqref{eq:appTdef}.
    The effective medium is represented in a similar way, only the propagation 
    through the lattice without visits to the local site is indicated by
    a zig-zag line. In both quantities the elementary hoppings are included 
    and,  therefore, do not appear in Eq.~\eqref{eq:Tmatrix0} - \eqref{eq:Tmatrix}.
    The start and end points of all propagations are connected to
    indicate local i.e.\ $\k$-summed quantities and the 
    small horizontal bars indicate that these lines are not 
    accounted for in an analytical expression. 
    The small circles are drawn only to indicate a 
    lattice site at which a propagation starts/ends and do not appear in any 
    analytical expressions. In the present context these are
    unnecessary, but they are helpful for two-particle quantities which are discussed 
    in the next section.  
  }
  \label{fig:dyson1PLoc}
\end{figure}
A graphical representation of this equation is depicted in Fig.~\ref{fig:dyson1PLoc}(a). 
$T(z)$ incorporates all processes where an electron leaves the site under consideration, 
propagates through the entire lattice ---including that local site---   
and then returns. These processes can be reorganized in terms of re-visits to
the local site,\cite{craco:atomicLimitHM95}
\begin{align}
  \label{eq:Tmatrix0}
  T(z)&=\Gamma(z)+\Gamma(z)\:\tilde G(z)\:\Gamma(z)+\dots
  \\ &=\Gamma(z)+\Gamma(z)\:\tilde G(z) \:T(z)
  \\
  \label{eq:Tmatrix}
  &=\frac1{\Gamma(z)^{-1}-\tilde G(z)} 
  .
\end{align}
The effective medium $\Gamma(z)$ now has the clear interpretation of being
the irreducible loop for a propagation of an electron from a specific site
through the lattice back to that site, without returning during the propagation,
see Fig.~\ref{fig:dyson1PLoc}(b). 
This makes it very obvious to interpreted $\Gamma$ as an noninteracting
effective medium and thus establishes the DMFT mapping
of the lattice problem to an SIAM.

Formally, this is obtained by inserting Eq.~\eqref{eq:Tmatrix} into  Eq.~\eqref{eq:LocGF} which 
gives the local Green function
\begin{align}
  \label{eq:GlocG}
  G(z)&=\frac{1}{\tilde{G}^{-1}-\Gamma(z)}
\end{align}
which is nothing but  Eq.~\eqref{eq:localGf}.

%%%%%%%%%%%%%%%%%%%%%%%%%%%%%%%%%%%%%%%%%%%%%%%%%%%%%%%%%%%%%%%%%%
\section{Bethe-Salpeter equations and dynamic susceptibilities}
\label{app:2Pdmft}
%%%%%%%%%%%%%%%%%%%%%%%%%%%%%%%%%%%%%%%%%%%%%%%%%%%%%%%%%%%%%%%%%%

The physical susceptibility matrix depends on one external bosonic Matsubara 
frequency $i\nu_n$ and wave vector $\q$. It  is obtained 
from a more general two-particle Green function by summing 
over two internal fermionic frequencies and wave vectors
(omitting the convergence factors $e^{(i\o_1\!+i\o_2)\delta}$),  
\begin{align}
  \label{eq:susMagDefG2A}
  \gm{\chi}(\q,i\nu_n)  
%  \\
%    &=  \frac{1}{\beta N}\sum_{i\o_1,i\o_2}
%    \sum_{\k_1,\k_2}
%    \gm{G}^{2C}(\k_1,i\o_1,\k_2\!+\!\q,i\o_2\!+\!i\nu_n;
%    \\
%   \notag
%   &\phantom{MMMMMMM} \k_2,i\o_2,\k_1\!+\!\q,i\o_1\!+\!i\nu_n)
%  %\:e^{(i\o_1\!+i\o_2)\delta}
%  \\
%%\end{comment}
% \label{eq:susMagDefG2a}
&=
  \frac{1}{\beta N}\sum_{i\o_1,i\o_2}
  \sum_{\k_1,\k_2}
  \gm{\chi}(\q,i\nu_n|\k_1,\k_2|i\o_1,i\o_2)
%  \:e^{(i\o_1\!+i\o_2)\delta}
  \\  \label{eq:susMagDefG2b}
  &\equiv
  \frac{1}{\beta}\sum_{i\o_1,i\o_2}
  \gm{\chi}(\q,i\nu_n|i\o_1,i\o_2)
 % \:e^{(i\o_1\!+i\o_2)\delta}
  .
\end{align}
Here, $i\o_1$ and $i\o_2$ are short-hand notations for the sets of 
fermionic Matsubara frequencies  $i\o_{n_1}$ and  $i\o_{n_2}$ ($n_1,n_2\in \mathbb{Z}$).
The notation is introduced 
to discriminate between external arguments
($i\nu_n$, $\q$)  and internal momenta ($\k_1$, $\k_2$) and frequencies
($i\o_1$, $i\o_2$) which are separated by vertical bars.
In the last equation we absorbed the momentum sum into the susceptibility
as indicated by the missing internal momentum variables.  

We also introduced a matrix notation for two-particle quantities
in orbital/spin space indicated by the straight double underline, i.e.\
$A_{a,b;c,d}=\{\gm{A}\}_{a,b;c,d}$. The  matrix multiplication is defined as
\begin{align}
  \{\gm{A}\,\gm{B}\}_{a,b;c,d}& =\sum_{ef}A_{f,b;c,e}B_{a,e;f,d}.
\end{align}

\begin{figure}
  \includegraphics[width=7cm]{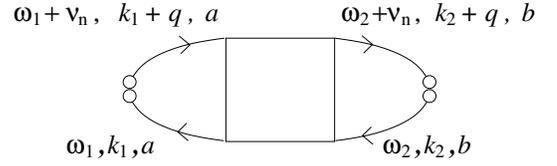}
  \caption{Graphical representation of the susceptibility matrix  $\chi_{a,b;b,a}(\q,i\nu_n|\k_1,\k_2|i\o_1,i\o_2)$.}
  \label{fig:chiDef}
\end{figure}

The graphical representation of the susceptibility matrix 
$\gm{\chi}(\q,i\nu_n|\k_1,\k_2|i\o_1,i\o_2)$ is 
shown in Fig.~\ref{fig:chiDef}.   The two external pairwise contracted 
lines signal the specific choice of orbital
matrix elements $\chi_{a,b;b,a}$, as well as
frequency and momentum arguments.

\begin{figure}
  \begin{center}
    \includegraphics[width=1\linewidth]{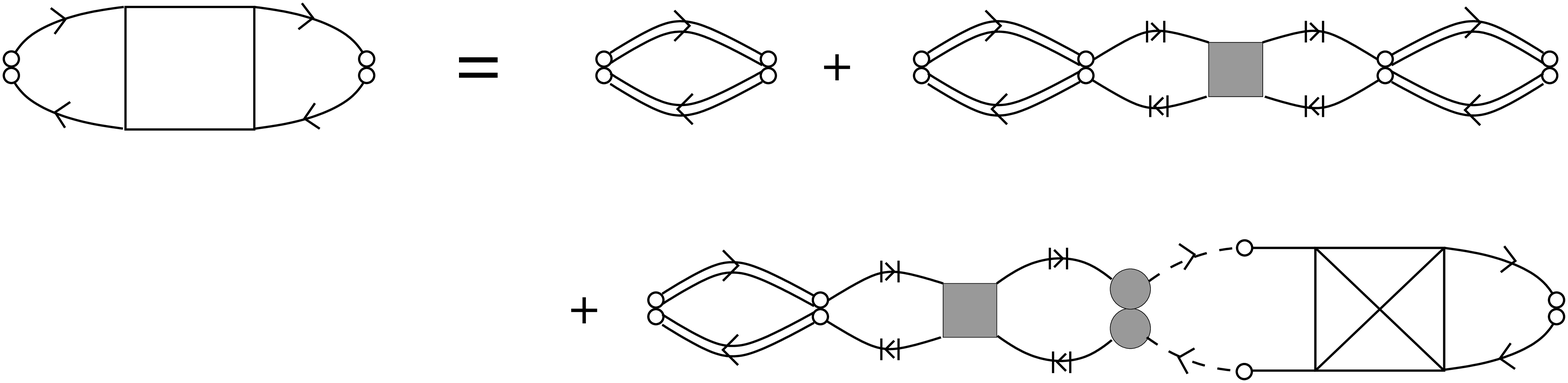}
  \end{center}
  \caption{Bethe-Salpeter equation for the susceptibility matrix. 
    The shaded square represents the irreducible vertex $\Pi$.  }
  \label{fig:bethePC}
\end{figure}
The Bethe-Salpeter equation for the susceptibility matrix is obtained
by summing all particle-hole \textit{reducible} diagrams not  
accounted for in the vertex $\gm{\Pi}$. 
This is shown graphically in Fig.~\ref{fig:bethePC}.

Within DMFT the full particle-hole irreducible two-particle 
vertex $\gm{\Pi}$ is assumed to be momentum independent
and momentum conservation is ignored at internal vertices.
This allows all sums over internal momenta to be absorbed into effective quantities
only depending on $\q$. The particle-hole excitation of such an effective
quantity thus always propagates between two sites only.  
Graphically this is indicated  in Fig.~\ref{fig:bethePC} by the endpoints  
of the particle-hole excitation always being right on top of each other.  

Analytically, the Bethe-Salpeter equation depicted in Fig.~\ref{fig:bethePC} is
\begin{align}
  \label{eq:betheLattChi}
  &\Chi{\q,i\nu_n}{i\o_1}{i\o_2}=
  \Big[-\delta_{\o_1,\o_2}
  \\ \notag
  & - \sum_{i\o_3}\twoPartO{S}{c}{\q,i\nu_n}{i\o_3}{i\o_2}\:\PicAmp{i\nu_n}{i\o_1}{i\o_3}  \\ \notag
  & %\phantom{=}
+  \onePart{P}{\q,i\nu_n}{i\o_2}\:\PicAmp{i\nu_n}{i\o_1}{i\o_2} 
% \:\onePart{P}{\q,i\nu_n}{i\o_1}
  \Big]
  \onePart{P}{\q,i\nu_n}{i\o_1}
  .
\end{align}

The particle-hole propagator $\gm{P}$ is constructed from
the single-particle Green function 
\begin{align}
  \label{eq:PDef}
  \onePart{P}{\q,i\nu_n}{i\o_1}
  &\equiv
  \frac{1}{N}\sum_\k \gm{GG}(\q,i\nu_n|\k|i\o_1)
 % \equiv \frac{1}{N}\sum_\k \um{G}(\k,i\o_1)\otimes\um{G}(\k\!+\!\q,i\o_1\!+\!i\nu_n)
 .
\end{align}
The matrix  notation on the right-hand side indicates a two-particle matrix 
% indicated in equation~\eqref{eq:PDef}
build from one-particle quantities by means of  a tensor product
\begin{align}
&\left[\gm{G G}(\q,i\nu_n|\k|i\o_1)\right]_{a,b;c,d} =
\\ \notag
& \phantom{MMMMM}G_{ac}(\k,i\o_1)\: G_{bd}(\k+\q,i\o_1+i\nu_n)
\\
&\phantom{MMMM}= \um{G}(\k,i\o_1)\otimes \um{G}(\k+\q,i\o_1+i\nu_n)
\end{align}
In the last line we introduced the double under-wave as indicator
for a one-particle matrix with two indices only.

%%%%%%%%%%%%%%%%%%%%%%%%%%%%%%%%%%%%%

\begin{figure}
  \begin{center}
    \includegraphics[width=6cm]{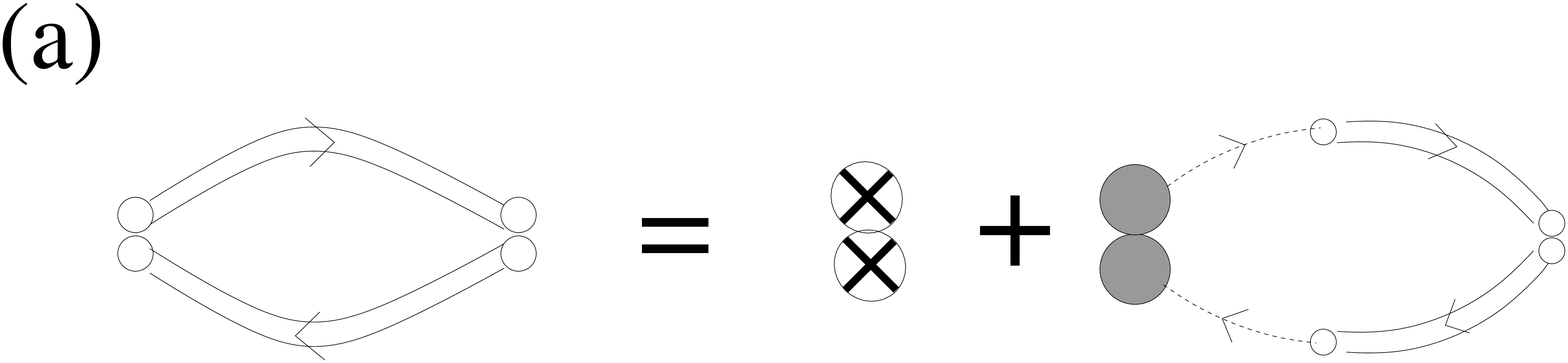}
    \includegraphics[width=6cm]{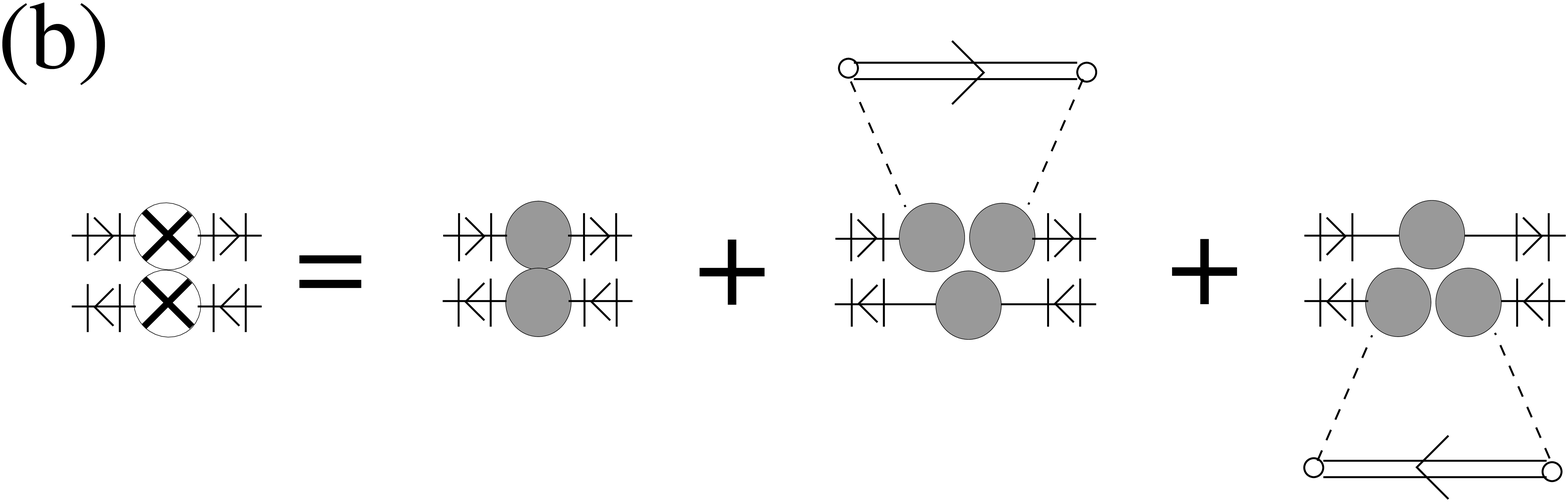}
  \end{center}
  \caption{(a) The particle-hole propagator expressed through 
    $\gm{Q}$ and $\gm{\Lambda}\:\gm{\tilde G\tilde G}$. (b) The quantity
    $\gm{\Lambda}\:\gm{\tilde G\tilde G}$ 
    where the external lines are made explicitly for clarity.
  }
  \label{fig:Pexpanded}
\end{figure}
% The matrix elements 
Inserting  the 
Dyson equation~(\ref{eq:GAtomLatt})
and using  $\sum_\k \um{t}_\k=0$,
the particle-hole propagator can be separated into a momentum-dependent
and  momentum-independent part
\begin{align}
  \label{eq:Pexpanded}
&  \onePart{P}{\q,i\nu_n}{i\o_1} 
  % &=\onePart{\tilde G\tilde G}{i\nu_n}{i\o_1}\left(\onePart{\Lambda}{i\nu_n}{i\o_1}+\onePart{I}{\q,i\nu}{i\o_1}\right)
  % \\
 \\ \nonumber & =
  \onePart{\Lambda}{i\nu_n}{i\o_1}\,\onePart{\tilde G\tilde G}{i\nu_n}{i\o_1}
  +\onePart{Q}{\q,i\nu_n}{i\o_1},
\end{align}
with the momentum dependency fully accounted for by the function
\begin{align}
  \label{eq:JDef}
  &  \onePart{Q}{\q,i\nu_n}{i\o_1}
\\ \nonumber &=
  \frac{1}{N}\sum_\k 
  \gm{GG}(q,i\nu_n|\k|i\o_1)\:\:\onePart{t t}{\q}{\k}\:
   \onePart{\tilde G\tilde G}{i\nu_n}{i\o_1}
 \end{align}
 and
 \begin{align}
  \label{eq:LambdaDef}
    \onePart{\Lambda}{i\nu_n}{i\o_1}&= 
  \gm{1}
  +\um{1}\otimes\big[\um{\tilde G}(i\o_1\!+\!i\nu_n)\:\um{T}(i\o_1\!+\!i\nu_n)\big]
  \nonumber
  \\ &
    +\big[\um{T}(i\o_1)\:\um{\tilde G}(i\o_1)\big]\otimes\um{1}
    .
\end{align}
This decomposition is show in Fig.~\ref{fig:Pexpanded}. 
The function $\gm{\Lambda}\,\gm{\tilde G\tilde G}$ is 
represented graphically as two circles with crosses
and incorporates
processes, where only one electron leaves a site and propagates through the lattice,
while the other remains at this site. 

%%%%%%%%%%%%%%%%%%%%%%%%%%%%%%%%%%%%%
The negative signs in front of the particle-hole propagators in 
Eq.~\eqref{eq:betheLattChi}  stem from the closed Fermion loop, i.e.\ from the
permutations necessary to obtain the ordering of the creation- and annihilation 
operators. Such signs are viewed as part of the diagrammatic rules  
and thus do not explicitly occur in the figures.

In the Bethe-Salpeter equation (\ref{eq:betheLattChi}) the amputated version of
the local vertex is used which is obtained by factorizing 
the one-particle  contributions of the local site, 
\begin{align}
  \label{eq:Piamp}
  &\Pic{i\nu_n}{i\o_1}{i\o_2}
  =\\ \notag
  &\onePart{\tilde{G}\tilde{G}}{i\nu_n}{i\o_2}\:\PicAmp{i\nu_n}{i\o_1}{i\o_2}
  \:\onePart{\tilde{G}\tilde{G}}{i\nu_n}{i\o_1}
  .
\end{align}
This is done in order to avoid over-counting of contributions 
from  
the start- and endpoint of the propagation. Graphically, amputation is indicated
by the vertical bars around the arrows.

\begin{figure}[t]
  \begin{center}
    \includegraphics[width=\linewidth]{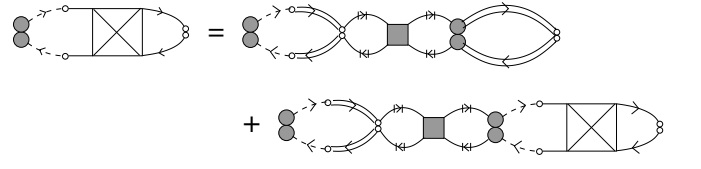}
  \end{center}
  \caption{Equation for the two-particle cumulant  transfer propagator $\gm{S}^c$.  }
  \label{fig:betheSC}
\end{figure}

In Eq.~(\ref{eq:betheLattChi}) an additional quantity
$\gm{S}^c$ had to be introduced
in order to avoid  two local irreducible vertices $\Pi$ occuring 
at the same site right after each other.
This two-particle cumulant transfer-propagator 
describes correlated particle-hole propagations starting with
two elementary transfers. 
A separate  Bethe-Salpeter equation can be formulated
%%% store equation number (=counter) for later use 
%\newcounter{BetheChiCounter}
%\setcounter{BetheChiCounter}{\value{equation}}
% for it 
%%%%this two-particle cumulant transfer-propagator
\begin{align}
  \label{eq:betheLattSC}
  &  \twoPartO{S}{c}{\q,i\nu_n}{i\o_1}{i\o_2}=
\\ \notag
% \onePart{I}{\q,i\nu_n}{i\o_1}\,\delta_{i\o_1,i\o_2-i\nu_n}\,
&  \onePart{P}{\q,i\nu_n}{i\o_2}\:\PicAmp{i\nu_n}{i\o_1}{i\o_2}\:
  \onePart{Q}{\q,i\nu_n}{i\o_1}
  -  \\\notag
  &
  \sum_{i\o_3} \twoPartO{S}{c}{\q,i\nu_n}{i\o_3}{i\o_2}\:\PicAmp{i\nu_n}{i\o_1}{i\o_3}\:\onePart{Q}{\q,i\nu_n}{i\o_1}
  ,
\end{align}
which is shown graphically in Fig.~\ref{fig:betheSC}.
The crossed box signals,
that only terms with at least one local irreducible  interaction vertex $\Pi$
involving all four external lines contributes.

\begin{figure}[t]
  \begin{center}
    \includegraphics[width=\linewidth]{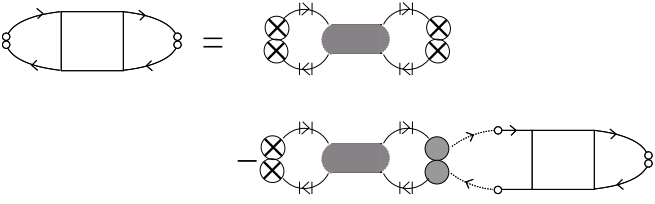}
  \end{center}
  \caption{Alternative Bethe-Salpeter equations for the lattice susceptibility expressed with
    the effective local two-particle cumulant Green functions.}
  \label{fig:betheLatt1}
\end{figure}
Within DMFT, the local irreducible  two-particle vertex $\Pi$
can in principle be determined from the effective impurity model.
But it is more instructive to  
to re-arrange the Bethe-Salpeter equation~\eqref{eq:betheLattChi}
in terms of dressed local two-particle Green functions, analogous to 
the one-particle Dyson equation~\eqref{eq:GAtomLatt}. 
The lattice susceptibility can then be expressed as
%\begin{widetext}
\begin{align}
  \label{eq:betheLattChiAlt}
  &
  \Chi{\q,i\nu_n}{i\o_1}{i\o_2}=
  \sum_{i\o_3}  \Big[ \onePart{\Lambda}{i\nu_n}{i\o_2}\:\delta_{\o_2,\o_3}
  \\\notag
  & \phantom{=}  -\onePart{\tilde G\tilde G}{i\nu_n}{i\o_2}\: \twoPartO{S}{\mathrm{amp}}{\q,i\nu_n}{i\o_3}{i\o_2}\Big]\times 
  \\\notag
  &\phantom{==============}
  \times \twoPartO{\tilde{G}}{2,\Lambda}{i\nu_n}{i\o_1}{i\o_3}
  .
\end{align}
%\end{widetext}
where
the amputated two-particle transfer propagator
$\gm{S}^{\mathrm{amp}}$ is introduced. This equation is depicted in Fig.~\ref{fig:betheLatt1}.
The (un-amputated) two-particle 
transfer propagator  $\gm{S}$  is closely related to the  
two-particle cumulant transfer propagator $\gm{S}^c$ via
\begin{align}
  \label{eq:SbarS}
  \twoPart{S}{\q,i\nu_n}{i\o_1}{i\o_2}&=
  -\onePart{Q}{\q,i\nu_n}{i\o_1}\delta_{i\o_1,i\o_2}
  \\ \notag 
  &+\twoPartO{S}{c}{\q,i\nu_n}{i\o_1}{i\o_2}
  .
\end{align}

 \begin{figure}
  \begin{center}
    \includegraphics[width=8cm]{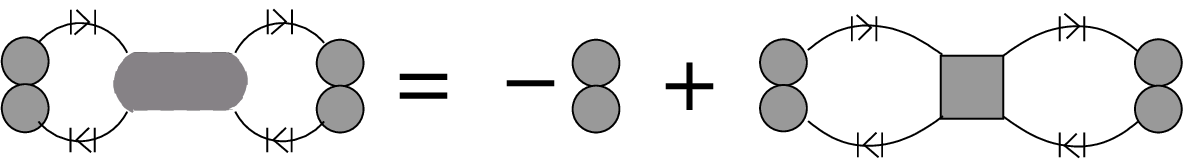}
    \includegraphics[width=8cm]{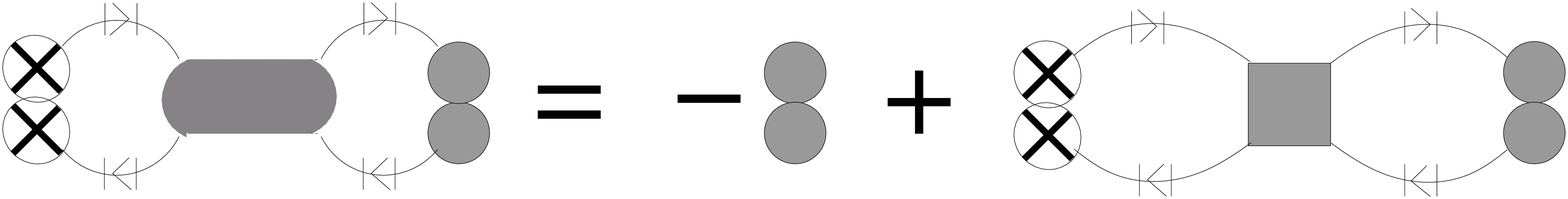}
    \includegraphics[width=8cm]{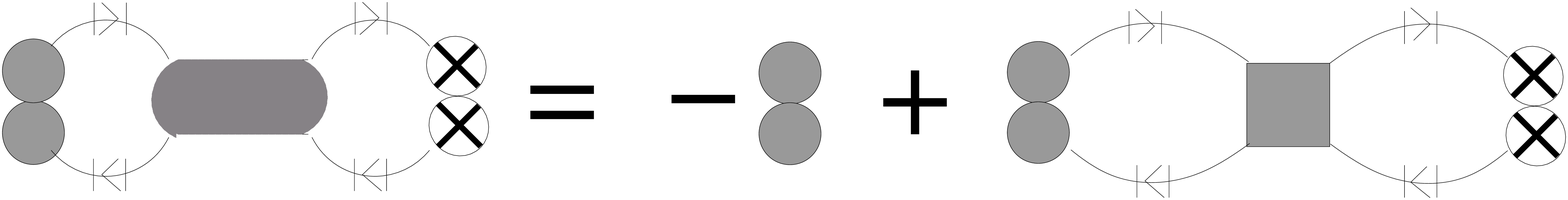}
    \includegraphics[width=8cm]{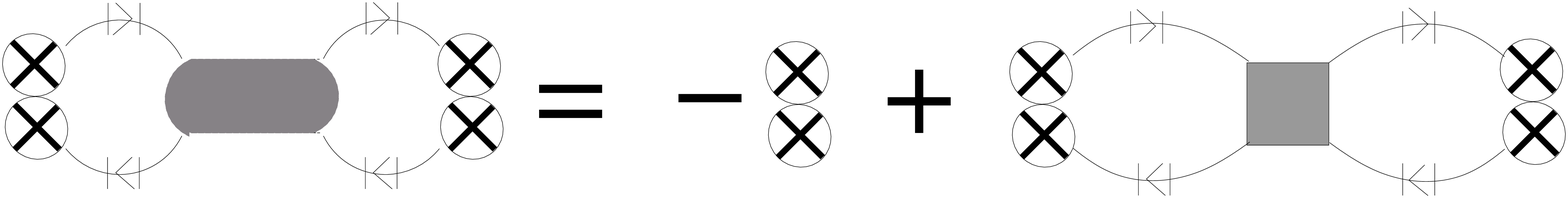}
  \end{center}
  \caption{ The effective local two-particle cumulant Green functions $\gm{\tilde G}^2$, 
    $\gm{\tilde G}^{2,\Lambda}$,  
    $\gm{\Lambda}\gm{\tilde G}^{2,\Lambda}\gm{\Lambda}^{-1}$ (see footnote~\onlinecite{Note2}),
    and $\gm{\Lambda}\gm{\tilde G}^{2,\Lambda}$ (from top to bottom).}
  \label{fig:G2tildeDef}
\end{figure}

The  dressed local two-particle  Green function
entering Eq.~\eqref{eq:betheLattChiAlt} is 
\begin{align}
  \label{eq:G2tildeLamDef}
  \twoPartO{\tilde{G}}{2,\Lambda}{i\nu_n}{i\o_1}{i\o_2}&=
  -\onePart{\tilde{G}\tilde{G}}{i\nu_n}{i\o_1}\:\delta_{i\o_1,i\o_2}
\\ \notag& +
  \Pic{i\nu_n}{i\o_1}{i\o_2}\:\onePart{\Lambda}{i\nu_n}{i\o_1}
  ,
\end{align}
which is shown graphically in Fig.~\ref{fig:G2tildeDef}.
It is essentially determined by the local irreducible vertex but additional
pseudo-local corrections due to one-particle propagations do occur. 
The minus sign  associated with the first term  reflects the fact that this term
originates from the particle-hole propagator. Since the closed fermion loop 
is no longer visible in the diagrammatic representation it is included
explicitly in the graphical representation.

The factor of $\gm{\Lambda}$ appearing at the cumulant vertex in Eq.~\eqref{eq:G2tildeLamDef}
is not valid by the rules of the original cumulant perturbation theory, as two 
local contributions are multiplied at the very same lattice site.  
The reason it appears here lies in the internal momentum sums already
performed in the lattice susceptibility
$\Chi{\q,i\nu_n}{i\o_1}{i\o_2}$ (see Eq.~\eqref{eq:susMagDefG2b})
together with the approximation of a momentum independent vertex. 
Separating pseudo-local and nonlocal contributions
the momentum-independent part of the particle-hole propagator
inevitably leads to the appearance of such terms
which are originally not allowed. 

Equation~\eqref{eq:betheLattChiAlt} is the two-particle analog of the
Dyson equation~\eqref{eq:GAtomLatt} with 1PI diagrams replaced by 2PI
diagrams, along with the resulting factors of $\gm{\Lambda}$, and the complication
of a second equation needed for the two-particle transfer propagator $\gm{S}$.

\begin{figure}
  \begin{center}
    \includegraphics[width=\linewidth]{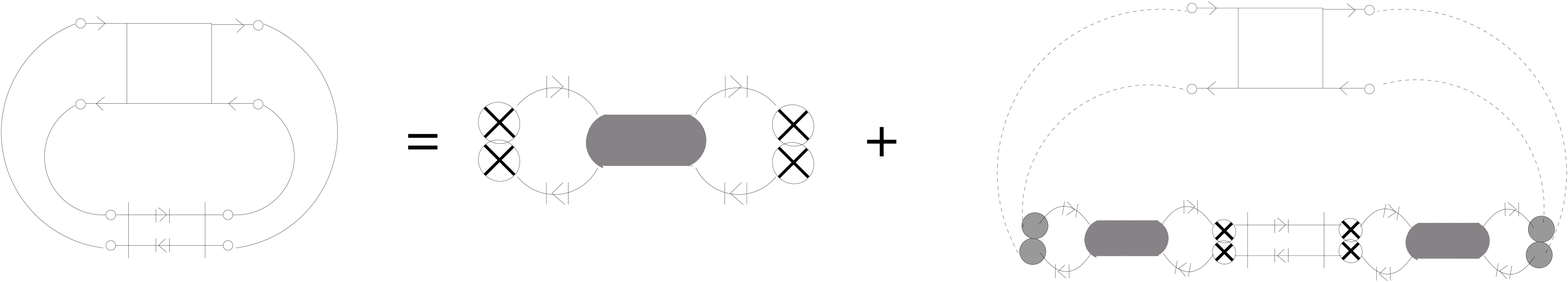}
  \end{center}
  \caption{Bethe-Salpeter equation for the local two-particle Green function $\gm{\chi}^{\mathrm{loc}}$.}
  \label{fig:betheLoc}
\end{figure}

For the mapping onto the effective impurity model  
we construct the local 
two-particle susceptibility by 
iterating Eq.~\eqref{eq:betheLattChiAlt} once
and  summing over $\q$ which gives\footnote{ %\label{footnote:GL} 
  The cumbersome expression 
  $\gm{\Lambda}\,\gm{\tilde{G}}^{2,\Lambda}\gm{\Lambda}^{-1}$
  just represents $\gm{\tilde{G}}^{2,\Lambda}$, but with the factor of $\gm{\Lambda}$
  at the other side of the vertex, i.e.\
  $\gm{\Lambda}\,\gm{\tilde{G}}^{2,\Lambda}\gm{\Lambda}^{-1}=
  -\gm{\tilde{G}\tilde{G}}\:\delta_{i\o_1,i\o_2}+
  \gm{\Lambda}\:  \gm{\Pi}
  $
  (see Fig.~\ref{fig:G2tildeDef}). This form is used to
  avoid another abbreviation.} 
\begin{widetext}
\begin{align}
  \twoPartO{\chi}{\mathrm{loc}}{i\nu_n}{i\o_1}{i\o_2}&\equiv \frac{1}{N}\sum_\q \Chi{\q,i\nu_n}{i\o_1}{i\o_2}
  \notag\\  \label{eq:betheLocChi}
  &=
  \onePart{\Lambda}{i\nu_n}{i\o_2}\,\twoPartO{\tilde{G}}{2,\Lambda}{i\nu_n}{i\o_1}{i\o_2}
  \\\notag
  &\hspace*{-2.4cm}
  +\sum_{i\o_3,i\o_4} 
  \onePart{\Lambda}{i\nu_n}{i\o_2}\twoPartO{\tilde{G}}{2,\Lambda}{i\nu_n}{i\o_4}{i\o_2} \onePart{\Lambda}{i\nu_n}{i\o_4}^{-1}\, 
  \twoPart{TT}{i\nu_n}{i\o_3}{i\o_4}\,
  \twoPartO{\tilde G}{2,\Lambda}{i\nu_n}{i\o_1}{i\o_3}
.
\end{align}
\end{widetext}
This equation is represented in  Fig.~\ref{fig:betheLoc}.
The quantity $\gm{TT}$ is closely related to the $\q$-summed  two-particle transfer propagator 
and represents the two-particle scattering matrix, where both particles
%simultaniously 
leave the local site and propagate through the lattice.    
Equation \eqref{eq:betheLocChi} is the analog of the one-particle equation~\eqref{eq:LocGF}
and establishes the connection between the effective  local two-particle cumulant
Green function $\gm{G}^{2,\Lambda}$ 
(or equivalently the local cumulant vertex $\gm{\Pi}^c$) and the local  physical Green function. 

\begin{figure}[t]
  \begin{center}
    \includegraphics[width=\linewidth]{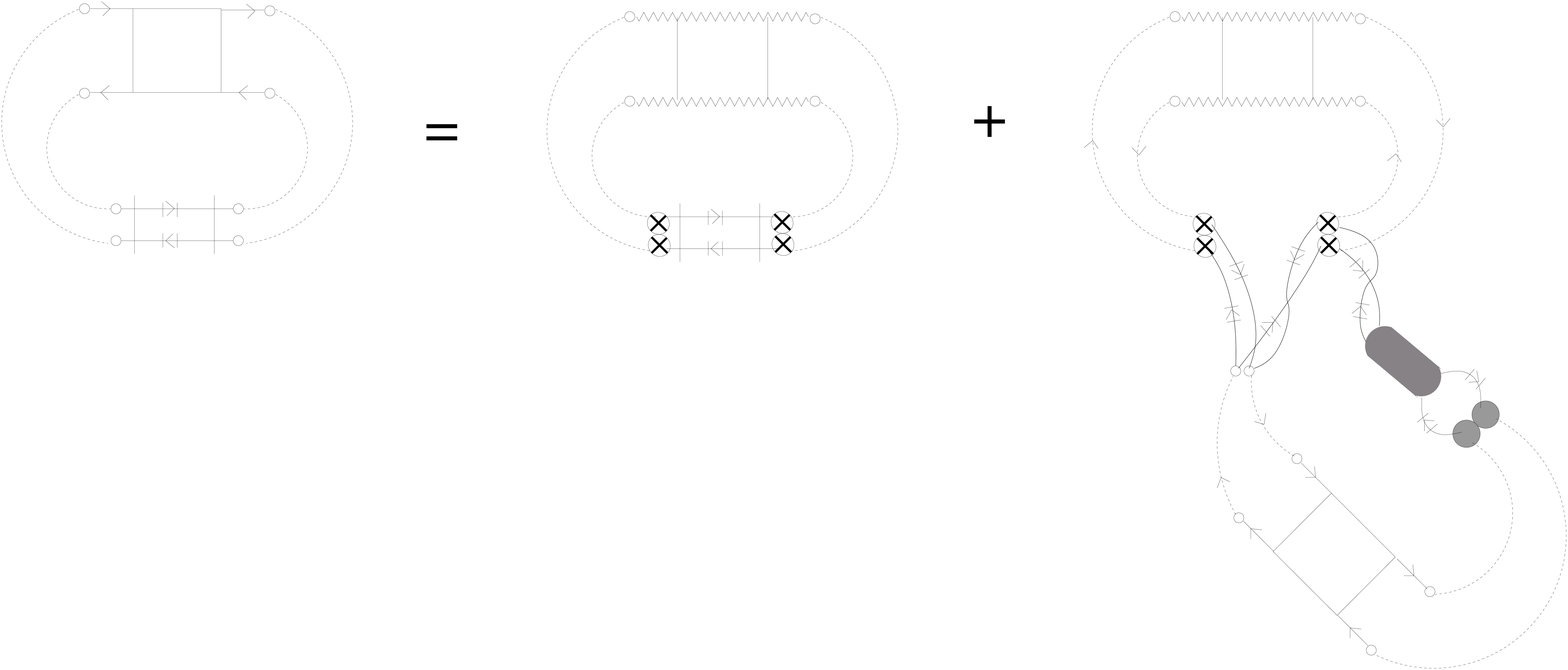}
  \end{center}
  \caption{The decomposition of the two-particle scattering matrix $\gm{TT}$ into
    irreducible propagations $\gm{\widetilde{TT}}$ and repeated visits to the
    local site.}
  \label{fig:betheTmatrix}
\end{figure}
In complete analogy to the  one-particle case  (see Eq.~\eqref{eq:Tmatrix})
the two-particle scattering matrix $\gm{TT}$ is build up 
from 
an irreducible scattering matrix $\gm{\widetilde{TT}}$. The latter   
describes the simultaneous propagation of two particles starting and ending at the 
same site without any intermediate returns to this specific site. 
This decomposition is shown  graphical in Fig.~\ref{fig:betheTmatrix}
which reads analytically
%\begin{widetext}
\begin{align}
  \label{eq:betheTTtildeTT}
  &\twoPart{TT}{i\nu_n}{i\o_1}{i\o_2}
  =\sum_{i\o_3} \Big[
  \onePart{\Lambda}{i\nu_n}{i\o_2}\,\delta_{\o_3,\o_2}
  \\ \notag
  &
  %\twoPart{\widetilde{TT}}{i\nu_n}{i\o_1}{i\o_2}\,\onePart{\Lambda}{i\nu_n}{i\o_1}
% \\\notag  &
  +
  \sum_{i\o_4} \twoPart{TT}{i\nu_n}{i\o_4}{i\o_2}\,
  \twoPartO{\tilde G}{2,\Lambda}{i\nu_n}{i\o_3}{i\o_4}
  \Big]\times
\\ \notag
  &
  \phantom{=====}\times\twoPart{\widetilde{TT}}{i\nu_n}{i\o_1}{i\o_3}\,\onePart{\Lambda}{i\nu_n}{i\o_1}
  .
\end{align}
%\end{widetext}
The structural difference of this equation to the one-particle case Eq.~\eqref{eq:Tmatrix} 
lies in the appearance of the factors $\gm{\Lambda}$ which account for the single-electron recurrence.  

At this point the identity
\begin{align}
  \label{eq:betheGtildeIdentity}
&  \sum_{i\o_3}\twoPart{TT}{i\nu_n}{i\o_3}{i\o_2}\,\twoPartO{\tilde{G}}{2,\Lambda}{i\nu_n}{i\o_1}{i\o_3}
  \\ \notag
  & =\sum_{i\o_3} \onePart{\Lambda}{i\nu_n}{i\o_2}\, \twoPart{\widetilde{TT}}{i\nu_n}{i\o_3}{i\o_2}\,
  \twoPartO{\chi}{\mathrm{loc}}{i\nu_n}{i\o_1}{i\o_3} 
\end{align}
and the equivalent form with the order of the matrix products 
reversed (as well as the order of the factors $\gm{\Lambda}$)
\begin{comment}
  \begin{align}
    \label{eq:betheGtildeIdentityRev}
    \sum_{i\o_3}\onePart{\Lambda}{i\nu_n}{i\o_2}\twoPartO{\tilde{G}}{2,\Lambda}{i\nu_n}{i\o_3}{i\o_2}
    \,\onePart{\Lambda}{i\nu_n}{i\o_3}^{-1}\twoPart{TT}{i\nu_n}{i\o_1}{i\o_3}
    \\\notag
    =\sum_{i\o_3}\twoPartO{\chi}{\mathrm{loc}}{i\nu_n}{i\o_3}{i\o_2} 
    \,\twoPart{\widetilde{TT}}{i\nu_n}{i\o_1}{i\o_3}\,\onePart{\Lambda}{i\nu_n}{i\o_1} 
\end{align}
\end{comment}
can be proven by comparing Eqs.~\eqref{eq:betheLocChi} and \eqref{eq:betheTTtildeTT}. 
These 
% are the two-particle versions of~\eqref{eq:GtildeIdentity}  and as in the one-particle case,
reflect the two possible ways to  decompose  the local Green function
and the effective local cumulant in terms of reducible and irreducible loops.

Utilizing the above identities in Eq.~\eqref{eq:betheLocChi}, the scattering matrix 
can be eliminated in favor of the irreducible scattering matrix 
and the local susceptibility,
\begin{align}
  \label{eq:betheLocImp}
  &\twoPartO{\chi}{\mathrm{loc}}{i\nu_n}{i\o_1}{i\o_2}
  =
  \\ \notag
&  \onePart{\Lambda}{i\nu_n}{i\o_2}\,\twoPartO{\tilde{G}}{2,\Lambda}{i\nu_n}{i\o_1}{i\o_2}
  \\\notag
  &
  +\sum_{i\o_3,i\o_4} 
  \twoPartO{\chi}{\mathrm{loc}}{i\nu_n}{i\o_3}{i\o_2} 
  \,\twoPart{\widetilde{TT}}{i\nu_n}{i\o_4}{i\o_3}
  \times
\\ \notag &\times
\onePart{\Lambda}{i\nu_n}{i\o_4} \,
  \twoPartO{\tilde G}{2,\Lambda}{i\nu_n}{i\o_1}{i\o_4}
  .
\end{align}

All equations derived so far incorporate inner frequency sums and thus have the
character of integral equations. 
The local susceptibility $\chi^{\mathrm{loc}}$ is determined from 
solution of the impurity model and then used 
as an input functions for Eq.~\eqref{eq:betheLocImp}. 
From this, ${\tilde G}^{2,\Lambda}$ is determined which then 
has to be used to solve  
the lattice equations, Eqs.~\eqref{eq:betheLattChiAlt} and \eqref{eq:betheLattSC}. 
Thus, closed explicit expressions for the momentum dependent susceptibilities  
cannot be obtained, in contrast to the single-particle case.

In order to complete the mapping onto the effective impurity model
we need to specify the processes to be incorporated into  the
effective medium. Therefore, it is beneficial to analyze the 
structure of this set of equations by switching to the 
Matsubara-matrix notation.
Equation~\eqref{eq:betheLocImp} can be solved in Matsubara-matrix 
space to give
\begin{align}
  \label{eq:betheLocImpInv}
  \gm{\hat \chi}^{\mathrm{loc}}_{i\nu} &=
  \Big[
    \left(      \hat{\gm{\Lambda}\gm{\tilde{G}}}^{2,\Lambda}_{i\nu}   \right)^{-1}
    - \gm{\hat{\widetilde{TT}}}_{i\nu}
    \Big]^{-1}
  .
\end{align}
This clearly demonstrates the resemblance to the one-particle equation~\eqref{eq:GlocG}.
The dressed local two-particle Green function is given by $\gm{\Lambda}\gm{\tilde{G}}^{2,\Lambda}$,
where the one-particle recurrence problem leads to the unexpected factor of $\gm\Lambda$,  
and $\gm{\widetilde{TT}}$ is the irreducible effective two-particle medium.
The effective local two-particle vertex can now be expressed in terms
of the local susceptibility function
\begin{comment}
  \begin{align}
    \label{eq:PiChiLoc}
    \gm{\Pi}^{\mathrm{amp}}&= 
    \left( \gm{\Lambda}\, \gm{\tilde{G}\tilde{G}}\right)^{-1}
    +  \gm{\chi}^{\mathrm{loc}} 
    \Big[
    \gm{\Lambda}\,\gm{\tilde{G}\tilde{G}}
    \big(
      1+\gm{\chi}^{\mathrm{loc}} \:\gm{\widetilde{TT}}
    \big)
    \gm{\Lambda}\,\gm{\tilde{G}\tilde{G}}
    \Big]^{-1}
  .
\end{align}
\end{comment}
and used to obtain an explicit form for the 
lattice susceptibility  in Matsubara-matrix space
\begin{align}
  \label{eq:lattSusMatsubara}
  \gm{\hat \chi}_{\q,i\nu_n}&=\left[-\gm{\hat P}_{\q,i\nu_n}^{-1}+{\gm{\hat \chi}^{\mathrm{loc}}_{i\nu_n}}^{-1}
    + \left( 
     \gm{\Lambda}\gm{\tilde{G}\tilde {G}}_{i\nu_n}
      \right)^{-1}+
      \gm{\hat{\widetilde{TT}}}_{i\nu_n}      
    \right]^{-1}
  .
\end{align}

\begin{comment}
Equations \eqref{eq:betheLocImpInv} and \eqref{eq:lattSusMatsubara}
clearly suggest the same interpretation as in
the one-particle case: when expressed in terms of physical  local 
averages, the effective local cumulant Green function and vertex 
incorporate a multitude of unphysical products connected by irreducible 
two-particle loops. 
These must be subtracted with the help of the irreducible scattering matrix
to yield a physical two-particle Green function. 
\end{comment}
   
As  the effective impurity is embedded into an noninteracting 
medium within the DMFT, the  irreducible two-particle medium  $\gm{\widetilde{TT}}$ 
can be specified in terms of the effective medium for 
the one-particle Green function, 
\begin{align}
  \label{eq:betheTtildeDmft}
  &\twoPart{\widetilde{TT}}{i\nu_n}{i\o_1}{i\o_2}= 
  \\ \notag
  &
  -\left(\um{\Gamma}(i\o_1)\otimes\um{\Gamma}(i\o_1+i\nu_n)\right)\,
  \onePart{\Lambda}{i\nu_n}{i\o_1}^{-1}\delta_{\o_1,\o_2}
  .
\end{align}
The negative sign reflects the fact that the effective medium 
is derived from the $\q$-summed particle-hole propagator, where 
a minus sign has to be included due to the closed fermion-loop.
The factor of $\um{\Lambda}^{-1}$  removes those processes
where only one electron intermediately re-visits the site, 
which are present in the product of the two one-particle 
media $\Gamma$.
This is necessary as 
$\gm{\widetilde{TT}}$ characterizes the amplitude where \textit{both} 
electrons must simultaneously leave the local site.

Notice that in this approximation the scattering matrix $\gm{TT}$ is not just 
the product of the uncorrelated one-particle scattering matrices, i.e.\
\begin{align}
  \label{eq:betheTDmft}
  \twoPart{TT}{i\nu_n}{i\o_1}{i\o_2}
  &\neq-\um{T}(i\o_1)\otimes\um{T}(i\o_1+i\nu_n)\delta_{\o_1,\o_2}
  ,
\end{align}
as it could be suspected na\"ively. The reason is that two-particle 
correlations are discarded only in the medium,
but locally all interaction vertices are retained.
The  Matsubara-matrix space  representation reveals
the structure of  $\gm{TT}$
\begin{align}
  \label{eq:betheTTstruct}
  \gm{TT}
  =  -\left(\um{T}\otimes\um{T}\right)\,
  \left[
    1+\gm{\Pi} 
    \left(\um{T}\otimes\um{T}\right)    
  \right]^{-1}
  .
\end{align}
%which can be derived by using Eqs.~\eqref{eq:betheTTtildeTT}, \eqref{eq:LambdaDef} and \eqref{eq:TmatrixTtilde}. 
This very instructive form
confirms the insight, that the uncorrelated one-particle loops get renormalized
by interaction vertices 
whenever the particles visit the local lattice site.

The explicit form \eqref{eq:betheTtildeDmft}  can be used to express the 
irreducible scattering matrix as
\begin{align}
  \label{eq:TtildeLCexpl}
    \gm{\widetilde{TT}}(i\nu_n|i\o_1,i\o_2)&=\Big[
  {\gm{P}^{\mathrm{loc}}(i\nu_n|i\o_1)}^{-1}
  \\ & \notag
  -\left( 
    \onePart{\Lambda}{i\nu_n}{i\o_1}\,\onePart{\tilde G\tilde G}{i\nu_n}{i\o_1}
    \right)^{-1}
    \Big]\delta_{\o_1,\o_2}
  ,
\end{align}
where the local particle-hole propagator is
\begin{align}
  \label{eq:PlocDef} 
  \gm{P}^{\mathrm{loc}}(i\nu_n|i\o_1)
  &=
  \frac{1}{N}\sum_{\q}\onePart{P}{\q,i\nu_n}{i\o_1}
%  =\frac{1}{N^2}\sum_{\q,\k}\gm{GG}(q,i\nu_n|\k|i\o_1)
  \\\notag 
  &=\um{G}^{\mathrm{loc}}(i\o_1)\otimes\um{G}^{\mathrm{loc}}(i\o_1+i\nu_n)
  % &  =\onePart{G^{\mathrm{loc}}G^{\mathrm{loc}}}{i\nu_n}{i\o_1}
  .
\end{align}

Inserting this into the Eq.~\eqref{eq:lattSusMatsubara} yields
the final form for the lattice susceptibility as stated in Eq.~\eqref{eq:MatsubaraDMFT}.

As stated in the main text, the  Matsubara-matrix
formalism is not  used  in this work. 
Instead, the analytic continuation of all Matsubara frequencies 
to the real axis is the principal goal of the above treatment.
This seems possible, as the local
susceptibility function $\gm{\chi}^{\mathrm{loc}}(i\nu_n|i\o_1,i\o_2)$
along with its analytical continuations to the real axis 
$\gm{\chi}^{\mathrm{loc}}(\nu_n\pm i0^+|\o_1\pm i0^+,\o_2\pm i0^+)$ 
could be obtained from the effective impurity model.
Equation~\eqref{eq:betheLocImp}, along with the irreducible scattering
matrix~\eqref{eq:betheTtildeDmft}, can be continued to the real axis
and the integral equation can be solved for the 
local two-particle Green functions $\gm{G}^{2,\Lambda}(\nu_n\pm i0^+|\o_1\pm i0^+,\o_2\pm i0^+)$. 
These are then used to determine the lattice susceptibility 
function at the real axis via Eqs.~\eqref{eq:betheLattChiAlt} and \eqref{eq:betheLattSC}.

In practice, this procedure is rather involved and more importantly 
the extraction of the analytically continued two-particle Green function
$\gm{\chi}^{\mathrm{loc}}(\nu_n\pm i0^+|\o_1\pm i0^+,\o_2\pm i0^+)$ 
from the impurity model is far from trivial and  strongly
depends on the impurity solver. It is straight 
forward (but very cumbersome) within semi-analytical 
approaches like ENCA, where one has a direct handle on the analytic structure 
of the this function. But, for example, in QMC or NRG 
it is conceptual not clear how extract two-particle quantities at the real
axis.

%%%%%%%%%%%%%%%%%%%%%%%%%%%%%%%%%%%%%%%%%%%%%%%%%%%%%%%%%%%%%%%%%%
% \section{Decoupling scheme}

In order to circumvent this involved procedure, we revert to an additional approximation.
The summations over inner Matsubara frequencies  
%Eqs~\eqref{eq:betheLattChi}, \eqref{eq:betheLattSC} and \eqref{eq:betheLocImp} 
are decoupled in a manner similar to the random phase approximation (RPA), 
i.e.\
\begin{align}
  \label{eq:RPAdecouple}
  &\sum_{i\o_3} \twoPart{A}{i\nu_n}{i\o_3}{i\o_2} \twoPart{B}{i\nu_n}{i\o_1}{i\o_3} 
\\ \notag & \phantom M  \Rightarrow 
  \twoPart{A}{i\nu_n}{i\o_1}{i\o_2}  \sum_{i\o_3}  \twoPart{B}{i\nu_n}{i\o_1}{i\o_3} 
.
\end{align}

With this, the inner frequency summations over $i\o_1$ and $i\o_2$ can be performed and
the equations can be solved explicity to yield the
final form for the susceptibility displayed in Eq.~\eqref{eq:susFinal}.

The local and lattice particle-hole propagators entering this equation 
can be calculated at the real axis  using standard techniques for the evaluation
of Matsubara sums,
\begin{align}
  \label{eq:PlocFinal}
  &\gm{P}^{\mathrm{loc}}(\nu)=\int\limits_{-\infty}^\infty\!d\o\: f(\o)
  \Big[
    \um{\rho}(\o)\otimes\um{G}(\o\!+\!\nu\!+\!i\delta)
    \\ & \notag
    \phantom{\int_{-\infty}^\infty\!d\o\: f(\o) \Big[}
    + \um{G}(\o\!-\!\nu\!-\!i\delta)\otimes\um{\rho}(\o)
    \Big]
    \\  \label{eq:PqFinal}
    &\gm{P}(\q,\nu)=
 \\ \notag
 &  %\phantom{\int\limits_{-\infty}^\infty\!d\o}
  \int\limits_{-\infty}^\infty\!d\o  \int\limits_{\mathrm{BZ}}\!\frac{d^D\k}{(2\pi)^D}f(\o)  \Big[
    \um{\rho}(\k,\o)\otimes\um{G}(\k\!+\!\q,\o\!+\!\nu\!+\!i\delta)
    \\ \notag    &\phantom{\int_{-\infty}^\infty\!d\o   MM}
    + \um{G}(\k,\o\!-\!\nu\!-\!i\delta)\otimes\um{\rho}(\k\!+\!\q,\o)
    \Big]
    ,
\end{align}
with the spectral functions
\begin{align}
  \label{eq:specFunG}
  \um{\rho}(\k,\o)&=-\frac{1}{2\pi i}\left(\um{G}(\k,\o\!+\!i\delta)-
    \um{G}(\k,\o\!-\!i\delta) \right)
  \\
  \um{\rho}(\o) &=      \int\limits_{\mathrm{BZ}}\!\frac{d^D\k}{(2\pi)^D}  \um{\rho}(\k,\o)
% =-\frac{1}{2\pi i}\left(\um{G}(\o\!+\!i\delta)-    \um{G}(\o\!-\!i\delta) \right) 
  .
\end{align}
All momentum integrals are over the whole Brillouin-Zone (BZ) in $D$ 
space dimensions.

% \bibliography{masterbib}

%\begin{comment}
  %
%\end{comment}

\end{document}